\documentclass[twocolumn]{aastex631}

\usepackage{graphicx}
\usepackage{txfonts}
\usepackage{longtable} 
\usepackage{gensymb}
\usepackage[verbose]{placeins}
\usepackage[utf8]{inputenc}
\usepackage[T1]{fontenc}
\DeclareUnicodeCharacter{2212}{-}

\begin{document} 
\title{VLASS-based survey of transition state galaxies and their relationship to compact peaked-spectrum radio sources}


\author[0000-0002-6741-9856]{Kunert-Bajraszewska, Magdalena}
\author[0009-0002-5915-4592]{Krauze, Aleksandra}
\affiliation{Institute of Astronomy, Faculty of Physics, Astronomy and Informatics, NCU, Grudziądzka 5/7, 87-100, Toruń, Poland}
\author[0000-0001-9324-6787]{Kimball, Amy E.}
\affiliation{National Radio Astronomy Observatory, 1011 Lopezville Road, Socorro, NM 87801, USA}
\author[0000-0002-7263-7540]{Stawarz, Łukasz}
\affiliation{Astronomical Observatory of the Jagiellonian University, Orla 171, 30-244 Kraków, Poland}
\author[0000-0003-3203-1613]{Kharb, Preeti}
\affiliation{National Centre for Radio Astrophysics (NCRA) - Tata Institute of Fundamental Research (TIFR), S. P. Pune University Campus, Post Bag 3, Ganeshkhind, 411007 Pune, India}
\author[0000-0003-2686-9241]{Stern, Daniel}
\affiliation{Jet Propulsion Laboratory, California Institute of Technology, 4800 Oak Grove Drive, Pasadena, CA 91109, USA}
\author[0000-0002-2557-5180]{Mooley, Kunal}
\affiliation{National Radio Astronomy Observatory, Socorro, NM 87801, USA; Cahill Center for Astronomy and Astrophysics, California Institute of Technology, Pasadena, CA 91125, USA; Indian Institute of Technology Kanpur, Kanpur 208016, U.P. India}
\author[0000-0003-1991-370X]{Nyland, Kristina}
\affiliation{U.S. Naval Research Laboratory, 4555 Overlook Avenue SW, Washington, DC 20375, USA}
\author[0000-0003-4323-0984]{Dorota Kozie\l -Wierzbowska}
\affiliation{Astronomical Observatory of Jagiellonian University, ul. Orla 171, 30-244 Kraków, Poland}


\begin{abstract}
We present multi-frequency and high-resolution studies of a sample of 24 radio transients sources discovered by comparing the NRAO VLA Sky Survey (NVSS) and  Very Large Array Sky Survey (VLASS) surveys. All of them are characterized by a significant increase in radio flux density over the last two decades. Their convex spectra, small sizes and high brightness temperatures are typical for young gigahertz-peaked spectrum (GPS) radio sources and indicative of an AGN buried in the host galaxy. On the other hand, they are much weaker than the archetypical GPS objects and their parsec-scale radio structures, although indicating the presence of young radio jets, are similar to radio-quiet AGNs like Seyfert and low-ionization nuclear emission-line region (LINER) galaxies. Based on the distribution of these objects in power$-$size ($P - D$) and peak frequency$-$size ($\nu_p − D$) diagrams, we suggest that after stabilizing their radio activity, some of the GHz-peaked radio transients (galaxies and quasars) will develop into radio-intermediate and radio-quiet (RI/RQ) quasars and low-frequency peaked-spectrum (PS) objects. We discuss several possible origins for the transient radio emission in our sources and conclude that changes in the accretion rate combined with low-power radio ejecta are the most probable cause. This is the scenario we also propose for one of our sources, 101841$−$13, which was independently identified as a candidate tidal disruption event (TDE) based on its infrared variability. However, we cannot exclude that 101841$−$13 or other sources in our sample are TDEs.

\end{abstract}


\section{Introduction}

Ultraviolet, optical and X-ray studies show that active galactic nuclei (AGNs) may go through short phases of dramatic variability. Recent reports, in particular, show that AGNs may brighten, fade, or change their spectral type, and some may do so even more than once \citep{Shapovalova, Denney2014, LaMassa15, MacLeod16, Stern2017, Ruan19}. Such {\it changing-look AGNs} show large changes in their X-ray and/or optical brightness as a result of either variable obscuration by an irregularly shaped torus \citep{Storchi-Bergmann}, a change of the accretion rate onto the central black hole \citep{Denney2014}, or thermal changes in the inner accretion disk \citep{Stern2018}. In the latter two cases, which are related to the source's activity state, sources can gain or lose broad emission lines in their UV/optical spectra and the name {\it changing-state AGNs} is sometimes used \citep{Graham2020}. It is beyond doubt that {\it changing-state AGNs} are extremely useful for studying the evolution of AGNs. They can also help us understand the radio jet production process. However, in the radio range, analogous wide-field time-domain studies were, until recently, very modest. New possibilities were provided by the Caltech-NRAO Stripe 82 Survey \citep[CNSS;][]{Mooley,Mooley2019} and the ongoing Very Large Array Sky Survey \citep[VLASS;][]{Lacy}.

The multi-epoch CNSS survey carried out with the Jansky VLA has facilitated an unbiased study of changing-state sources for the first time at radio wavelengths. It has resulted in the discovery of objects that appeared as new radio sources after $>5-20$ years of absence \citep{Mooley, MKB2020, Wolowska}. They are transient phenomena with respect to the Faint Images of the Radio Sky at Twenty Centimeters (FIRST) survey which, after the significant increase in radio luminosity, have been classified as radio-loud objects. Moreover, in the case of one of these sources, quasar 013815$+$00, the transition from radio-quiet to radio-loud phase coincided with changes in its accretion disk luminosity, which has been interpreted as a result of an enhancement in the supermassive black hole (SMBH) accretion rate \citep{MKB2020}. More detailed, multi-frequency studies of these switched-on radio sources revealed that they are characterized by convex spectra and compact morphologies, typical of the well-known class of gigahertz-peaked spectrum (GPS) sources which are indicative of the emergence of a newborn young jet \citep{Wolowska}. 
The recent detection of radio transient galaxies \citep{Zhang2022} and changing-state quasars \citep{Nyland} from the VLASS survey also supports this scenario.
This links {\it changing-state AGNs} and variable accretion onto SMBHs with episodic radio activity \citep{Reynolds, Kharb2006, Czerny, MKB10, Wolowska2017, Sullivan2024}. 

Radio sources with peaked spectra such as GPS and compact steep spectrum (CSS) objects are considered progenitors of large-scale radio sources with Fanaroff-Riley type I (FRI) or type II (FRII) morphology \citep{Fanaroff}. In this model of evolution, GPS and CSS objects grow continuously in size and change their spectra to eventually become massive galaxies \citep{Fanti, Redhead, Odea, Snellen2000, Odea20}. However, statistical studies have revealed that there is a significant excess of compact sources in comparison to powerful, fully developed, luminous radio-galaxies \citep{Odea, Sadler}. There is also recent evidence that some compact sources are in the fading phase \citep{MKB2005, MKB2006, Orienti2010, Callingham}. The proposed explanation for these observations is intermittent radio activity on timescales $\lesssim 10^4 - 10^5$ years, suggesting that many AGNs might be short-lived objects \citep{Reynolds}. This `incomplete' life cycle scenario may apply for both low-power sources \citep{MKB10, Slob} and high-luminosity sources \citep{Readhead2024}, whereby the physical mechanism behind such short term activity is still a topic of intense debate and may vary from source to source.
Among the proposed possibilities for initiating such variability are radiation pressure instabilities within the accretion disk \citep{Czerny} and tidal disruption events (TDEs) of giant branch stars with masses $\rm \gtrsim 1 M_{\odot}$ \citep{Readhead2024, Sullivan2024}.
Time-domain surveys of radio transients allow detection of this phenomenon and study its origin, as well as probe possible evolution of new-born radio jets.

In this paper we report on a sample of 24 slow radio transients discovered using the VLASS survey. They were undetected in the NRAO VLA Sky Survey carried out in $\sim1995$ \citep[NVSS;][]{Condon1998}, but discovered in VLASS to have brightened dramatically over the last $>$ 24 years. We present multi-frequency and high-resolution observations of these sources as well as discuss their nature. 

Throughout this paper we assume a concordance cosmology with $\rm H_0= 70~km~s^{−1}~Mpc^{−1}$, $\rm \Omega_M= 0.3$ and $\Omega_{\Lambda}= 0.7$. The radio spectral index $\alpha$ is defined as $S_{\nu} \propto \nu^{\alpha}$, where $S_{\nu}$ is the flux spectral density.

\section{Sample selection}
\label{sec:sample}
We used the observations from the first epoch of the VLASS\footnote{https://archive-new.nrao.edu/vlass/} survey and compared them with the NVSS survey, carried out at 1.4 GHz, in order to ﬁnd possible new radio sources (i.e., transient candidates) in VLASS.
We retained only those VLASS sources that: 1) are absent in the NVSS catalog (that is, they have a flux density below 2.5 mJy), 2) are point-like and have 3 GHz VLASS flux density $\geq$ 8 mJy (i.e. implied spectral index $\alpha>2$, between 1.4 GHz and 3 GHz), 3) are coincident within 2 arcsec with the nucleus of nearby galaxies having $r<$ 20 mag based on visual inspection of Pan-STARRS optical images.
The resulting radio source list of 24 objects is given in Table \ref{table1_basic}. 
Four of these objects have already been suggested as good candidates for transient sources based on cross-correlations of FIRST and VLASS observations \citep{Zhang2022}.

\begin{figure*}[ht]
   \centering
   \includegraphics[scale=0.81]{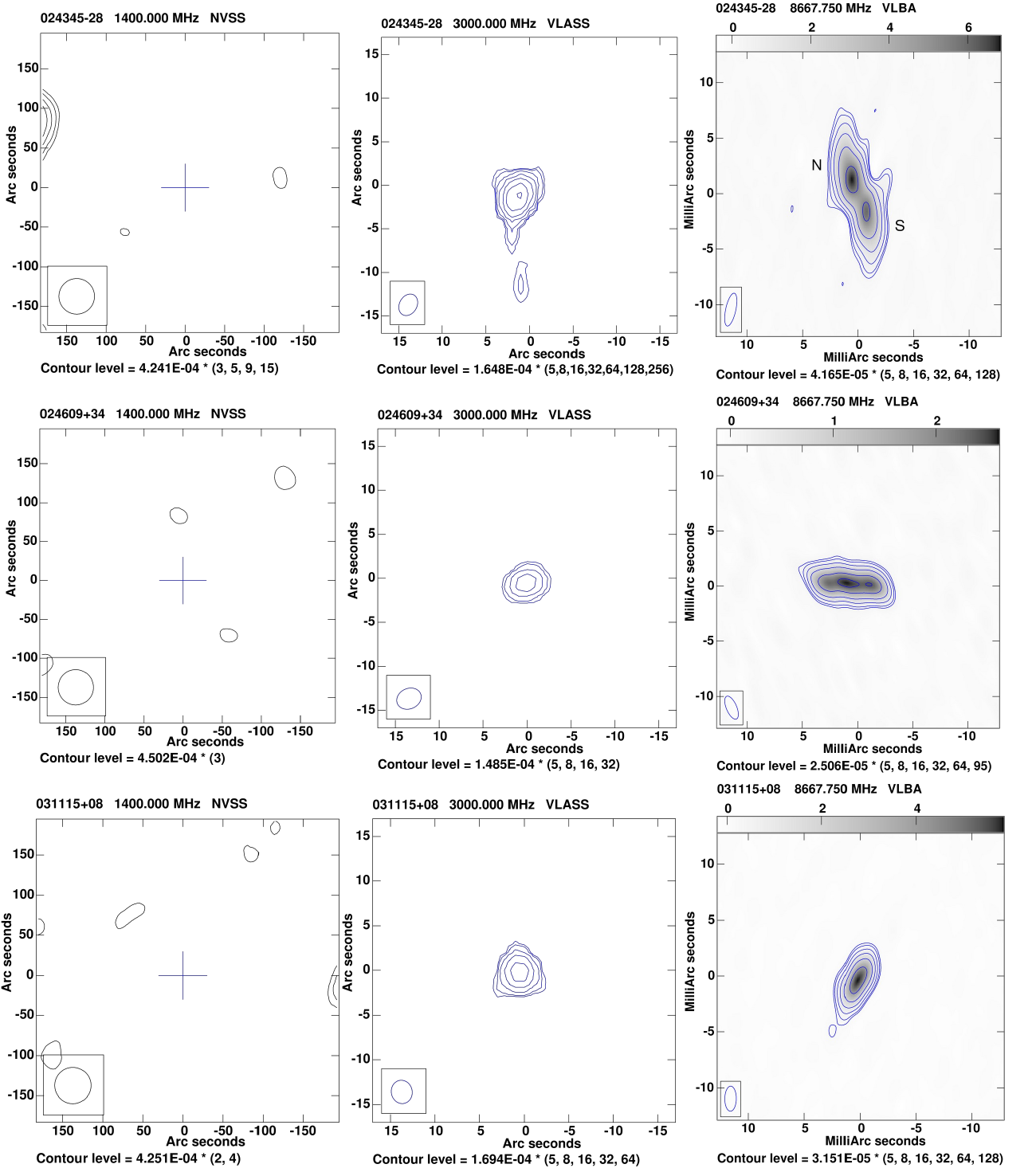}
\caption{High-frequency radio images for 024345$-$28, 024609$+$34 and 031115$+$08; from left to right: NVSS 1.4 GHz, VLASS 3 GHz (first epoch) and VLBA 8.7 GHz. 
The other 21 sources are presented in Appendix, Figure \ref{figure_appendix_images}.
} 
              \label{figure_images}%
    \end{figure*}

We then carefully inspected the NVSS images. For several sources, we found radio emission at the source position below the sensitivity limit of the NVSS catalog but above the $3\sigma$ noise level. We did a similar analysis for the subset of sources within the FIRST survey coverage. 
We used the JMFIT task in the NRAO AIPS\footnote{http://www.aips.nrao.edu} software to measure the integrated flux for each source in the NVSS, FIRST and VLASS images (Table \ref{table1_basic}). 
Thus far, VLASS has completed three epochs of observations as of October 2024.
The NVSS and VLASS images of the 24 selected sources are presented in Figure \ref{figure_images} and continued in Figure \ref{figure_appendix_images}.

It is important to note that the VLASS Quick Look images that we used to measure the flux density of our sources are not final data products, and therefore have a relatively high measurement uncertainty \citep{Lacy2020PASP}. Therefore, following the recommendations described in VLASS Memo \#13, we increased VLASS total flux densities by 10\% for all Epoch 1 observations and adopted a 10\% measurement uncertainty to the total flux densities for all VLASS observations. The resulting values are given in Table \ref{table1_basic}.

We have undertaken a multi-frequency follow-up campaign for these 24 transient galaxies, including radio studies at different resolution (this paper), as well as optical spectroscopic studies (forthcoming publication).

\begin{deluxetable*}{c c c l c c c c r r r r c l r}[th!]
\tabletypesize{\scriptsize}
\tablecaption{The sample of 24 transient radio sources.}
\tablehead{
& & &   &  NVSS & NVSS& FIRST& FIRST& \multicolumn{1}{c}{VLASS$_1$}& \multicolumn{1}{c}{VLASS$_2$}& \multicolumn{1}{c}{VLASS$_3$} &  &  &  &   \\  
Name & RA  & DEC & \multicolumn{1}{c}{\it z} & $\rm S_{1.4\,GHz}$& obs. & $\rm S_{1.4\,GHz}$& obs. & \multicolumn{1}{c}{$\rm S_{3\,GHz}$}& \multicolumn{1}{c}{$\rm S_{3\,GHz}$}  & \multicolumn{1}{c}{$\rm S_{3\,GHz}$} &\multicolumn{1}{c}{$\rm V_{3\,GHz}$}  &$\rm logL_{5\,GHz}$   & \multicolumn{1}{c}{LAS}     &\multicolumn{1}{c}{LLS}  \\
& h~m~s & $\degr$~$\arcmin$~$\arcsec$&   & [mJy]  & year & [mJy]  & year & \multicolumn{1}{c}{[mJy]}      & \multicolumn{1}{c}{[mJy]} & \multicolumn{1}{c}{[mJy]} & &$\rm [W~Hz^{-1}]$& \multicolumn{1}{c}{[mas] }  & \multicolumn{1}{c}{[pc]} \\
 (1) & (2) & (3) &\multicolumn{1}{c}{(4)}& (5) & (6) & (7) & (8) & \multicolumn{1}{c}{(9)} & \multicolumn{1}{c}{(10)}& \multicolumn{1}{c}{(11)} & \multicolumn{1}{c}{(12)}          & \multicolumn{1}{c}{(13)}   & \multicolumn{1}{c}{(14)} & \multicolumn{1}{c}{(15)}\\  
}
\startdata
024345$-$28&02 43 45.7&$-$28 40 39.4	&0.075	&$<$1.27   &1993 & $-$	        & $-$  &54.35$\pm$ 5.44& 59.60$\pm$ 5.96&55.53$\pm$ 5.56   & $-$0.05$\,\,$ &	23.81	        & 4.6$^d$	  & 6.5	\\
024609$+$34&02 46 09.3&$+$34 08 20.3	&0.131	&$<$1.35   &1993 & $-$	        & $-$  &8.75$\pm$ 0.87& 8.26$\pm$ 0.83 & \multicolumn{1}{c}{$-$}  & $-$0.07$^c$	 &	23.73	        & 5.4&	12.6 \\
031115$+$08&03 11 15.5&$+$08 51 32.6	&0.050 &$<$1.28   &1993 & $-$	        & $-$  &15.19$\pm$ 1.52&19.25$\pm$ 1.93&20.67$\pm$ 2.07 & 0.05$\,\,$ & 22.99	& 2.7&	2.6	\\
053509$+$83&05 35 09.9&$+$83 45 41.3	&0.165	&$<$1.15   &1993 & $-$	        & $-$  &24.45$\pm$ 2.45&26.85$\pm$ 2.69&23.89$\pm$ 2.39 & $-$0.04$\,\,$&24.22	        &5.3$^d$	  &	15.0 \\
064001$+$28&06 40 01.8&$+$28 03 07.1	&0.016	&1.62$\pm$ 0.81$^b$&	1993& $-$&$-$  &19.06$\pm$ 1.91&22.20$\pm$ 2.22&\multicolumn{1}{c}{$-$} & $-$0.02$^c$	 &	21.92	&	4.3$^d$&	1.4\\
070837$+$32&07 08 37.8&$+$32 04 20.6	&0.275	&$<$1.42   &1993 & $-$	        & $-$  &9.56$\pm$ 0.96 &5.71$\pm$ 0.57&\multicolumn{1}{c}{$-$} & 0.60$^c$	 &	23.53	&	4.9	&	20.5	\\
071829$+$59&07 18 29.3&$+$59 42 52.0	&0.229	&1.59$\pm$ 0.63&1993&2.34$\pm$ 0.45	&2001  &19.86$\pm$ 1.99&15.79$\pm$ 1.58&17.26$\pm$ 1.73 & 0.01$\,\,$ &24.52	&	2.3	&	8.4	\\
095141$+$37&09 51 41.6&$+$37 03 34.6	&0.236	&$<$1.13&1993&$<$0.42&	1994&	10.59$\pm$ 1.06	&10.70$\pm$ 1.07	&\multicolumn{1}{c}{$-$} & $-$0.04$^c$	 &	24.28	&	1.6	&	6.0	\\
101841$-$13&10 18 41.4&$-$13 00 06.5	&0.030	&1.96$\pm$ 1.06	&1995&$-$&$-$	&11.30$\pm$ 1.13	&6.28$\pm$ 0.63	&1.44$\pm$ 0.14	& 0.73$\,\,$ &	21.38	&	4.8	&	2.9	\\
105035$-$07&10 50 35.7&$-$07 45 02.1	&0.122	&2.03$\pm$ 0.77	&1995	&1.43$\pm$ 0.43	&2001	&8.92$\pm$ 0.89	&3.73$\pm$ 0.37	&\multicolumn{1}{c}{$-$} & 0.41$^c$	 &23.03	&	1.9	&	4.2	\\
110239$-$06&11 02 39.2&$-$06 11 50.0	&1.393$^a$&2.87$\pm$ 1.43&1995	&0.84$\pm$ 0.43	&2001	&35.57$\pm$ 3.56	&31.35$\pm$ 3.14	&\multicolumn{1}{c}{$-$} & 0.01$^c$	 &26.47&	6.5$^d$&	54.8	\\
112940$+$39&11 29 40.0&$+$39 00 46.5	&0.288	&$<$1.04&1993	&$<$0.41&1994	&12.75$\pm$ 1.27	&10.83$\pm$ 1.08	&\multicolumn{1}{c}{$-$} & $-$0.02$^c$	 &24.50	&	0.4&	1.7	\\
114101$+$10&11 41 01.7&$+$10 28 21.4	&0.072	&$\geq$1.25	&1995	&1.36$\pm$ 0.46	&2000	&10.88$\pm$ 1.09	&14.84$\pm$ 1.48	&9.48$\pm$ 0.95 & 0.12$\,\,$ &23.00	&	5.1$^d$&	7.0	\\
121619$+$12&12 16 19.6&$+$12 40 27.7	&0.059	&$<$1.17   &1995&2.40$\pm$ 0.46	&2000	&10.43$\pm$ 1.04	&9.37$\pm$ 0.94	&\multicolumn{1}{c}{$-$} & 0.01$^c$	 &	22.85	&	2.1	&	2.4	\\
130400$-$11&13 04 00.7&$-$11 58 57.5	&0.139	&1.44$\pm$ 0.58	&1995	&$-$	&$-$	&9.38$\pm$ 0.94 &8.08$\pm$ 0.81	&\multicolumn{1}{c}{$-$} & $-$0.03$\,\,$ &	23.57 &	3.6	&	8.8	\\
150415$+$28&15 04 15.9&$+$28 29 47.9	&0.058	&1.95$\pm$ 0.97	&1995	&$<$0.60	&1993	&9.75$\pm$ 0.97 &10.11$\pm$ 1.01	&7.65$\pm$ 0.77& 0.04$\,\,$ &22.76&	4.2	&	4.7	\\
155847$+$14&15 58 47.7&$+$14 12 13.4	&0.035	&$<$1.53	&1995	&$\geq$0.70	&2000	&32.10$\pm$ 3.21	&27.53$\pm$ 2.75	&\multicolumn{1}{c}{$-$} & $-$0.02$^c$	 &22.88	&3.4	&	2.4	\\
164607$+$42&16 46 07.0&$+$42 27 37.5	&0.050	&$<$1.18	&1995	&$<$0.40	&1995	&11.01$\pm$ 1.10	&9.46$\pm$ 0.95	&\multicolumn{1}{c}{$-$} & 0.21$^c$	 &22.41&0.9	&	0.9	\\
180940$+$24&18 09 40.5&$+$24 36 40.9	&0.016	&$<$1.13	&1995	&$-$	&$-$	&9.16$\pm$ 0.92	&6.77$\pm$ 0.68	&\multicolumn{1}{c}{$-$} & 0.22$^c$	 &21.30&	3.1	&	1.0	\\
183415$+$61&18 34 15.4&$+$61 24 00.6	&0.223	&$<$1.40	&1995	&$-$	&$-$	&11.56$\pm$ 1.16	&11.48$\pm$ 1.15	&14.07$\pm$ 1.41 	& 0.00$\,\,$ &24.24&	2.6	&	9.3	\\
195335$-$04&19 53 35.2&$-$04 52 03.5	&0.046	&1.80$\pm$ 0.95	&1993	&$-$	&$-$	&18.63$\pm$ 1.86	&18.28$\pm$ 1.83	&\multicolumn{1}{c}{$-$} & 0.07$^c$	 &	22.75&	1.2	&	1.1	\\
203909$-$30&20 39 09.1&$-$30 45 20.4	&0.049	&$<$1.18	&1993	&$-$	&$-$	&9.88$\pm$ 0.99	&6.20$\pm$ 0.62	&\multicolumn{1}{c}{$-$} & 0.13$\,\,$ &	22.46	&	8.3	&	8.0	\\
223933$-$22&22 39 33.2&$-$22 31 26.8	&0.119	&$<$1.22	&1993	&$-$	&$-$	&18.71$\pm$ 1.87	&15.08$\pm$ 1.51	&\multicolumn{1}{c}{$-$} & 0.01$\,\,$ &	23.80	&	3.1$^d$&	6.7	\\
233058$+$10&23 30 58.9&$+$10 00 30.2	&0.040	&$<$1.62	&1996	&$-$	&$-$	&13.44$\pm$ 1.34	&6.35$\pm$ 0.64	&3.44$\pm$ 0.34 & 0.52$\,\,$ &21.98&	1.8	&	1.4	\\
\enddata
\vspace{0.1 in}
\tablecomments{
The columns are marked as follows: (1) source name; (2,3) VLA coordinates in J2000.0; (4) spectroscopic redshift, $^a$ indicates the only quasar on this list; (5) NVSS 1.4 GHz flux density measurement or the 3$\sigma$ noise level at the location of the source, $^b$ indicates that the value applies to component A only; (6) NVSS observation year; (7) FIRST 1.4 GHz flux density measurement or the 3$\sigma$ noise level at the location of the source, '$-$' means an area outside the survey coverage; (8) FIRST observation year; (9, 10, 11) VLASS 3 GHz flux density measurement: epoch 1 (2017 - 2019), epoch 2 (2020 - 2022), epoch (2023); (12) variability ratio at 3\,GHz, described in section \ref{sec:other_source_parameters}, $^c$ indicates that in the absence of data from the VLASS$_3$, we took into account our VLA observations (2022) if they differed significantly in time from the VLASS$_2$ observations; (13)
log of the k-corrected radio luminosity at 5 GHz calculated based on the VLA flux density measurement performed in 2021 and 2022 as described in section \ref{sec:observations}; (14) Largest Angular Size as the deconvolved major axis of the source or separation between the outermost component peaks measured in the 8.7 GHz VLBA image, $^d$ indicates sources with multiple VLBA components; (15) Largest Linear Size.}
\label{table1_basic}
\end{deluxetable*}

\section{Observations and calculations}
\label{sec:observations}
\subsection{VLA observations}

We conducted multi-frequency VLA observations of the entire sample of 24 transient sources 
in two parts under project codes: AK1064 (2.5 hours) and AK1084 (14.7 hours). The former was carried out for three sources (024345$-$28, 031115$+$08 and 203909$-$30) in B configuration in September 2021 using five receivers covering the spectral range from 1000 to 16884 MHz (L, S, C, X, Ku). Observations for the remaining 21 objects (project AK1084) were performed in C configuration using four receivers (L, S, C, X) in the period from October to December 2022. 
The observing setup was the correlator with 16 (L, S and Ku-band) or 32 (C and X-band) spectral windows and 64 2-MHz-wide channels. 
We used 3C48 and 3C147 as the primary flux density calibrators and phase calibrators chosen from the VLA calibrator manual were also observed.
Detailed calibration and processing of the VLA data was carried out using CASA\footnote{http://casa.nrao.edu} software.
In order to obtain broad spectral coverage while maintaining high signal-to-noise ratio, $\leq$8 adjacent spectral windows were averaged in each band to measure average flux densities, resulting in four measurements of a flux density for the C- and X-bands, and two measurements for the L-, S- and Ku-bands for most objects. The final measurements with error estimations are shown in Figure \ref{figure_spectral_modeling} and listed in Table \ref{tableA1_vla_flux}.

\subsection{VLBA observations}
VLBA X-band observations were conducted in the period from August to December 2021 under the observation code BK240. The sources were observed individually or in blocks of two sources along with the phase calibrator and fringe-finder. The integration time for a single source varied from 3 to 4 hours. The total time assigned to this project was 87 hours. Data reduction (including editing, amplitude calibration, instrumental phase corrections and fringe-fitting) was performed with the standard procedure using the NRAO AIPS software.
After this stage the AIPS task IMAGR was used to produce the final total intensity images, which are presented in Figure \ref{figure_images} and continued in Figure \ref{figure_appendix_images}. Most of the sources turned out to be very compact at 8.7 GHz. 
Only in the case of six sources did observations reveal more than one radio component and for these objects the angular size is calculated as the separation between the outermost component peaks measured in the 8.7 GHz VLBA image. In other cases, the largest angular size (LAS) is the same as the size of the deconvolved major axis of the Gaussian fit to the single source obtained using the task JMFIT (Table \ref{table1_basic}). More detailed information about individual sources and their components is presented in Table \ref{table_vlba}.

\subsection{GMRT observations}
To complement the spectra at low frequencies, sub-GHz observations with the upgraded GMRT in Band-3 (250-500 MHz) and Band-4 (550-850 MHz) were carried out in the period from June to October 2021, as project 40\_089. The total allocated time was 48 hours. However, due to technical problems, high quality data was collected for only some of the sources.
The observations were then analyzed using the CASA-6 compatible version of the CAsa Pipeline-cum-Toolkit for Upgraded GMRT data REduction (CAPTURE)\footnote{https://github.com/ruta-k/CAPTURE-CASA6} developed at NCRA \citep{Kale}.
The GMRT flux density measurements are presented in Table \ref{tableA2_lowfreq_flux}.

\subsection{LOFAR observations}
We searched for archival data at 144 MHz from the LOw-Frequency ARray (LOFAR) Two-metre Sky Survey \citep[LoTSS;][]{Shimwell} and found detections for four sources from our sample. 
LOFAR images for these four sources are shown in Figure \ref{figure_appendix_lofar} and flux density measurements taken from the catalog LoTSS Data Release 2 (LoTSS$-$DR2) are presented in Table \ref{tableA2_lowfreq_flux}. These observations were made in 2015 (source 112940$+$39) and 2019 (sources 095141$+$37, 150415$+$28 and 164607$+$42).

\subsection{Optical observations}
Using the Low-Resolution Imaging Spectrograph (LRIS) mounted on the Keck I telescope
and Double Spectrograph (DBSP) mounted on the Palomar 200-inch Hale Telescope, we have obtained high quality spectroscopic observations of our sample. Publication of these data, along with interpretation and discussion, will be presented elsewhere. 
An exception is the brief discussion of optical observations of 101841$-$13 presented in Appendix~G. For the remaining objects we only report the redshifts determined based on these observations (Table \ref{table1_basic}), and note that only one of the sources, 110239$-$06, is classified as a quasar.

\subsection{Spectral modeling}
\label{Spectral_modeling}
In order to characterize the broadband radio spectra of our peaked-spectrum sources, the following generic curved model \citep{Snellen} was used:

\begin{equation}
\mathrm{S_{\nu}=\frac{S_{p}}{(1-e^{-1})} \times \left(\frac{\nu}{\nu_{p}}\right) ^{\alpha_\mathrm{thick}} \times \left(1-e^{-\left(\frac{\nu}{\nu_{p}}\right)^{\alpha_\mathrm{thin}-\alpha_\mathrm{thick}}}\right)}
\label{equation_powerlaw}
\end{equation}
where $\rm \alpha_{thick}$ is the optically thick spectral index, $\rm \alpha_{thin}$ is the optically thin spectral index, and $\rm S_{p}$ and $\rm \nu_{p}$ are the peak flux density and peak frequency, respectively. However, this model could not be meaningfully applied to two sources, 070837$+$32 and 105035$-$07, because of an overall `upturn' of their radio continua. In these two cases we have therefore estimated spectral indices based on flux density measurements straddling the minimum in their spectra. We provide further discussion and interpretation of the spectra in Section \ref{sec:shape of the spectrum}.
The obtained values for each source are given in Table \ref{table_spectral_modeling} and the spectra are presented in Appendix C, Figure \ref{figure_spectral_modeling}.

\subsection{Other source parameters}
\label{sec:other_source_parameters}

The variability index is estimated based on the 3 GHz VLASS observations using the formula adopted from \citet{Aller}:

\begin{equation}
\mathrm{V=\frac{(S-\sigma)_{\rm max} - (S+\sigma)_{\rm min}}{(S-\sigma)_{\rm max} + (S+\sigma)_{\rm min}}}
    \label{equation_variability}
\end{equation}
where $\rm S_{max}$, $\rm S_{min}$, $\rm \sigma_{max}$, $\rm \sigma_{min}$ are the limiting values of flux densities and their associated measurement errors, respectively. The results of our calculations are presented in Table \ref{table1_basic}.

The k-corrected spectral luminosities of our sources are calculated as:
\begin{equation}
\mathrm{L_{\nu}=4\pi D^2_L S_{\nu}(1+{\it z})^{-(1+\alpha)}}
 \label{equation_luminosity}
\end{equation}
where $\rm D_L$ is the luminosity distance, $\rm S_{\nu}$ is the flux density at a given frequency, $\it z$ is the redshift and $\rm \alpha$ is a thin or thick spectral index, depending on the spectral shape at a given frequency.

The brightness temperature $\rm T_B$ of a given radio component of a source is calculated using the formula derived in Appendix A in the following form:
{\small
\begin{equation}
\mathrm{T_{B}=1.38 \times 10^{12} (1+{\it z})\, \left(\frac{S_{\rm comp}}{{\rm Jy}}\right)\, \left(\frac {\theta_{\rm min}}{{\rm mas}}\right)^{-1}\,  \left(\frac{\theta_{\rm maj}}{{\rm mas}}\right)^{-1}\, \left(\frac{\nu}{{\rm GHz}}\right)^{-2}\quad {\rm K}\,},
\end{equation}
}
where $\rm \theta_{maj}$ and $\rm \theta_{min}$ are the deconvolved major and minor component axes, and $\rm S_{comp}$ is the flux density of a given component at the observed frequency $\nu$. The results of these calculations are presented in Appendix E, Table \ref{table_vlba}.

\begin{figure}[t]
   \centering
   \includegraphics[scale=0.52]{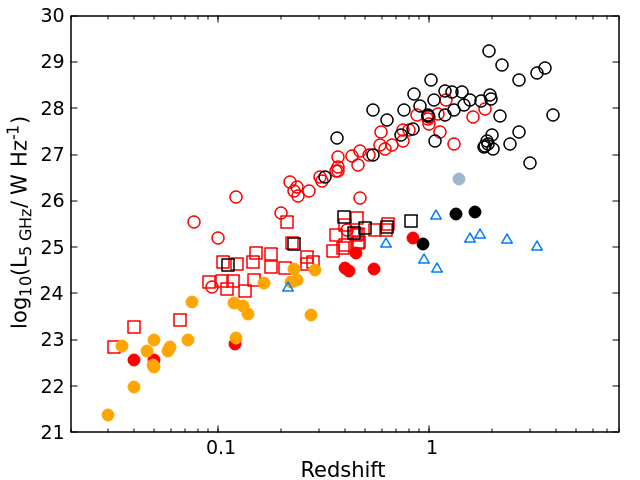}
\caption{5 GHz luminosity vs. redshift diagram. The transient sources from this work are marked with filled yellow (galaxies) and gray (quasar) circles. The open red circles (galaxies) and open black circles (quasars) indicate CSS and GPS sources taken from \citet{Odea2}, \citet{Snellen}, \citet{devries}, \citet{Stranghellini}, and \citet{Fanti} as collected by \citet{Keim} into one sample. The red and black squares indicate galaxies and quasars from the LLC sample \citep{MKB10}. The filled red circles (galaxies) and filled black circles (quasars) indicate radio transients discovered in the CNSS survey by \citet{Wolowska}. 
The open blue triangles indicate transient quasars from \citet{Nyland}. }
              \label{figure_lumred}
    \end{figure}

\section{Results and Discussion}
\label{sec:discussion}
All 24 objects presented in this work are transients with respect to the NVSS survey.
They were undetected in NVSS (1993–1996) at the sensitivity level of 2.5 mJy at 1.4 GHz but discovered in the VLASS first epoch observations (2017–2019) to have brightened signiﬁcantly (Table \ref{table1_basic} and Figure \ref{figure_images} and \ref{figure_appendix_images}). 
However, a close inspection of the NVSS radio images reveals the presence of low level radio emission from several sources. The presence of this early emission was also confirmed by the $\geq 3$ times more sensitive FIRST survey, whenever such observations were available. 
For the sources with no low level emission in the NVSS images, but for which FIRST observations are also available, we found a range of results. Three sources, 095141$+$37, 112940$+$39 and 164607$+$42, were observed by FIRST within one year of the NVSS observations; in all three cases, there was no low-level emission in the FIRST data. Two sources, 121619$+$12 and 155847$+$14, were observed by FIRST five years after the NVSS observations; in both cases, the sources were detected by FIRST. Finally, 150415$+$28 was undetected by FIRST but was faintly detected by NVSS two years later.
Nevertheless, what we observe in these sources is a significant increase in flux density after some period of very low-level emission, or even its absence. 
The increase in flux density varies for individual sources and in the whole sample ranges from 3 to even 50. 

We have compared the redshift and luminosity distribution of our sources to the sample of strong GPS and CSS objects collected by \citet{Keim}, the low-luminosity compact (LLC) sources defined by \citet{MKB10}, and radio transients discovered using the CNSS survey \citep{Wolowska} and VLASS first epoch observations \citep{Nyland}, where the latter is limited to quasars. This comparison is provided in Figure \ref{figure_lumred}. It shows that transient radio sources identified in this study have, on average, lower radio luminosities at 5 GHz compared to archetypical GPS, CSS and LLC sources. Other than the sole quasar in this sample, 110239$-$06, the 5 GHz luminosities are below $\rm 10^{25}\,W\,Hz^{-1}$ in all cases.
Their redshift distribution is limited to the local universe and does not exceed $z = 0.3$, which is likely the result of the adopted selection criteria (in particular, coincidence of the radio transient with an $r<20$\,mag galaxy). The exception here is, again, the only quasar in our sample. Importantly, however, the selection criteria used in this study do not bias toward limiting the source power. Therefore, the fact that among our sources there are no transients with radio power exceeding $\rm 10^{25}\,W\,Hz^{-1}$ at 5 GHz reflects a real observed physical limit, which can be connected with the different life cycle of these sources.  
It is also worth noting that relatively low radio luminosities are also characteristic of other radio transients, both galaxies and quasars, discovered by \citet{Wolowska} and \citet{Nyland} by comparing the FIRST survey to the CNSS and VLASS surveys, respectively. 
It has been already postulated by \citet{MKB10}, based on the studies of LLC objects, that low-power radio sources can be more short-lived than high-luminosity objects.

\begin{deluxetable}{l r r c r c}
\tabletypesize{\scriptsize}
\tablecaption{Results of spectral modelling.}
\tablehead{
\multicolumn{1}{c}{Name} & \multicolumn{1}{c}{$S_p$} & \multicolumn{1}{c}{$\nu_p$} & $\alpha_{thin}$ & \multicolumn{1}{c}{$\alpha_{thick}$}& $\rm \Delta \alpha$ \\
      & \multicolumn{1}{c}{[mJy]} & \multicolumn{1}{c}{[GHz]}   &      &    &   \\
\multicolumn{1}{c}{(1)} & \multicolumn{1}{c}{(2)}   &  \multicolumn{1}{c}{(3)}    & (4)            &   \multicolumn{1}{c}{(5)}    & (6)     \\    
}
\startdata
024345$-$28 & 63.23 ± 1.22 & 2.08 ± 0.33 & -0.69 ± 0.06 & 0.86 ± 0.13 & 1.55\\
024609$+$34 & 14.23 ± 0.28 & 5.41 ± 0.35 & -0.53 ± 0.05 & 1.63 ± 0.12 & 2.16\\
031115$+$08 & 19.71 ± 0.39 & 1.51 ± 0.09 & -0.46 ± 0.03 & 1.59 ± 0.11 & 2.05\\
053509$+$83 & 22.44 ± 0.27 & 1.52 ± 0.03 & -0.28 ± 0.01 & 2.66 ± 0.15 & 2.95\\
064001$+$28 & 33.12 ± 0.37 & 1.01 ± 0.09 & -0.77 ± 0.03 & 1.28 ± 0.23 & 2.16\\
071829$+$59 & 22.58 ± 0.87 & 2.66 ± 0.17 & -0.52 ± 0.06 & 2.58 ± 0.29 & 3.10\\
095141$+$37 & 12.75 ± 0.19 & 3.42 ± 0.11 & -0.62 ± 0.03 & 2.01 ± 0.09 & 2.63\\
101841$-$13 & 2.50 ± 0.07 & 1.03 ± 0.16 & -0.75 ± 0.06 & 2.02 ± 0.65 & 2.77\\
110239$-$06 & 33.22 ± 0.33 & 3.89 ± 0.20 & -0.74 ± 0.05 & 1.58 ± 0.12 & 2.32\\
112940$+$39 & 11.97 ± 0.37 & 2.79 ± 0.11 & -0.25 ± 0.03 & 2.75 ± 0.24 & 3.00\\
114101$+$10 & 10.22 ± 0.19 & 2.72 ± 0.18 & -0.94 ± 0.06 & 1.54 ± 0.15 & 2.48\\
121619$+$12 & 12.04 ± 0.32 & 1.25 ± 0.12 & -0.57 ± 0.03 & 1.89 ± 0.27 & 2.46\\
130400$-$11 & 8.40 ± 0.11 & 3.79 ± 0.22 & -0.94 ± 0.53 & 1.05 ± 0.07 & 1.99\\
150415$+$28 & 7.47 ± 0.36 & 4.82 ± 0.73 & -1.01 ± 0.13 & 0.39 ± 0.04 & 1.40\\
155847$+$14 & 30.85 ± 0.70 & 2.65 ± 0.11 & -0.82 ± 0.05 & 2.74 ± 0.24 & 3.56\\
164607$+$42 & 9.16 ± 0.23 & 1.36 ± 0.19 & -0.85 ± 0.04 & 0.91 ± 0.55 & 1.77\\
180940$+$24 & 7.79* & 1.50* & -0.74 ± 0.01 & -0.40* & $-$\\
183415$+$61 & 13.83 ± 0.54 & 4.33 ± 0.64 & -0.39 ± 0.07 & 1.20 ± 0.21 & 1.60\\
195335$-$04 & 13.24 ± 0.59 & 4.27 ± 0.41 & -1.25 ± 0.08 & 0.53 ± 0.10 & 1.78\\
203909$-$30 & 7.00* & 3.50* & -1.33 ± 0.06 & 0.25* & $-$\\
223933$-$22 & 18.25 ± 0.37 & 10.79 ± 0.26 & -1.85 ± 0.20 & 0.51 ± 0.03 & 2.35 \\
233058$+$10 & 4.82 ± 0.83 & 2.70 ± 0.55 & -1.54 ± 0.11 & 0.78 ± 0.57 & 2.33\\
\hline
070837$+$32$\dagger$&  $-$  & $-$ & 0.66 $\pm$ 0.10 & -0.31 $\pm$ 0.30 & $-$\\
105035$-$07$\dagger$&  $-$  & $-$ & 0.76 $\pm$ 0.30 & -0.30 $\pm$ 0.28 & $-$\\
\enddata
\vspace{0.1 in}
\tablecomments{The columns are marked as follows: (1) source name; (2) peak ﬂux density; (3) observed frequency of the peak (spectral turnover); (4) spectral index of the optically thin part of the radio spectrum; (5) spectral index of the optically thick part of the radio spectrum; (6) spectral curvature defined as $\Delta \alpha = \alpha_{thick} - \alpha_{thin}$.
* - indicates values that have been ﬁxed to obtain a proper ﬁt, $\dagger$ - for these sources values of the indices have been calculated based on flux densities straddling the minimum in the spectrum.}
\label{table_spectral_modeling}
\end{deluxetable}

\begin{figure}[ht]
   \includegraphics[scale=0.3]{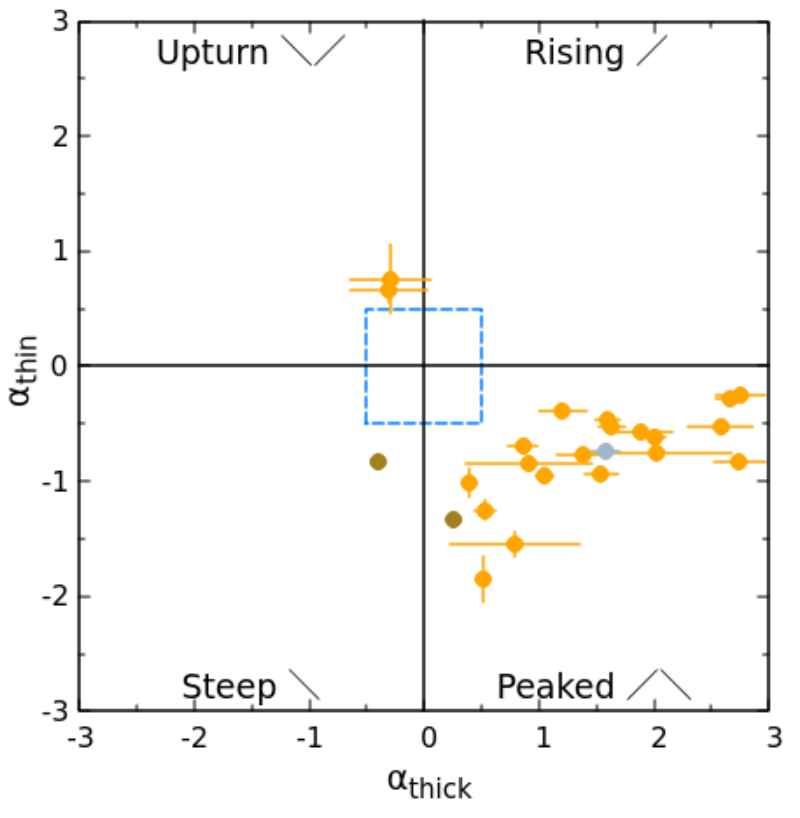}
\caption{Radio color-color diagram for our sample. The blue square corresponds to the spectral index range from $-$0.5 to 0.5 (“blazar box”). Galaxies are marked with yellow circles, and the only quasar in the sample is shown in gray. 
The two galaxies, 180940$+$24 and 203909$-$30, with fixed values for their spectral indices indicated by brown circles.
Details on the calculation of $\alpha_{thin}$ and $\alpha_{thick}$ are provided in Section \ref{Spectral_modeling} and Table \ref{table_spectral_modeling}. }
              \label{figure_alpha}%
    \end{figure}

\subsection{Shape of the spectrum}
\label{sec:shape of the spectrum}
\begin{figure*}[t]
   \centering
   \includegraphics[scale=0.53]{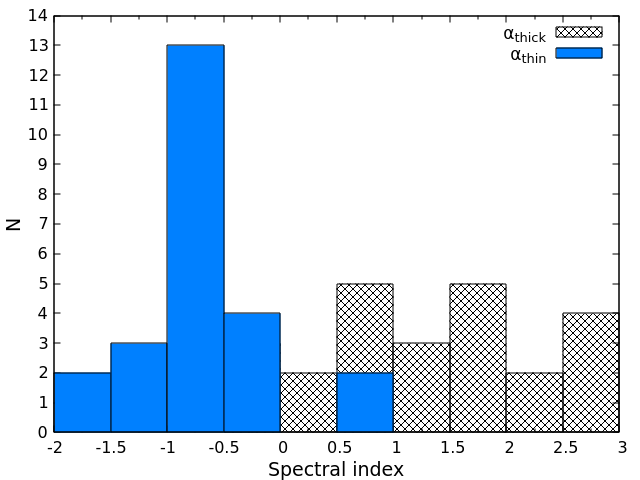}
   \includegraphics[scale=0.53]{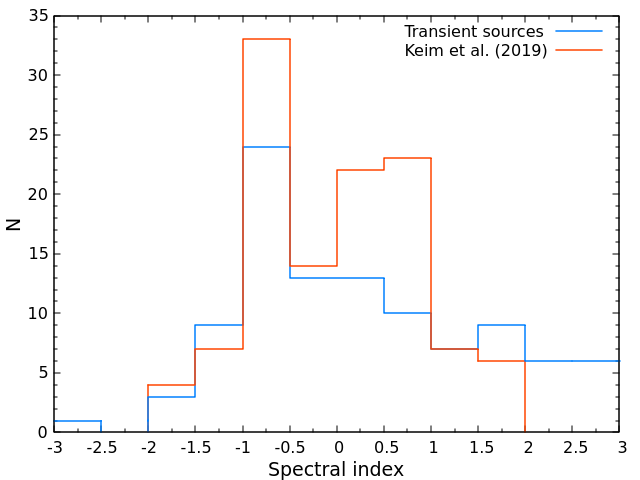}
\caption{(Left): Distribution of spectral indices for our sample of 24 transient sources. (Right): Comparison of the distribution of spectral indices for the combined sample of archetypal GPS and CSS sources from \citet{Keim} and the combined sample of radio transients from this work and from \citet{Wolowska}. }
              \label{figure_histogram_index}%
    \end{figure*}

Using the modified power law model, we determined the characteristic features of the radio spectra of our objects, i.e. the frequency and flux density of the peak and the spectral indices on both sides of the peak. The fitted spectra are presented in Figure \ref{figure_spectral_modeling} and the obtained values are summarized in Table \ref{table_spectral_modeling}. However, in the case of four sources, the procedure yielded spurious results.
Figure \ref{figure_alpha} presents the spectral-type distribution in the convenient form of a radio color-color diagram in which $\alpha_{thin}$ is plotted against $\alpha_{thick}$. The majority of our objects reside in the lower right quadrant of this figure, indicating convex spectra peaking at or above 1 GHz and corresponding spectral indices of $\alpha_{thick} \geq 0.25$ and $\alpha_{thin} \leq −0.25$. 
In the case of two sources, 070837$+$32 and 105035$-$07, the shape of their spectra was classified as `upturn', which means falling at low frequencies and rising at higher frequencies with spectral minima in the range between 4.5 and 5 GHz. Another four objects, 031115$+$08, 053509$+$83, 112940$+$39 and 183415$+$61, have values of $\alpha_{thin} > −0.5$, which formally defines them as flat-spectrum sources. Finally, the sources 180940$-$24 and 203909$-$30 do not have formal maxima in the observed range (1$-$10 GHz) and the peak frequency of 164607$+$42 is uncertain.
The distribution of spectral indices for the whole sample is presented in the left panel of Figure \ref{figure_histogram_index}. 
According to the generally accepted unified scheme for radio–loud AGNs \citep{Barthel,Urry}, spectra with high–frequency flattening or upturn shapes occur when emission from the AGN compact core begins to dominate at higher frequencies over the more extended and steeper-spectrum component prominent at lower frequencies \citep{Tucci}. The necessary condition here is a small angle between the relativistic jet and the line of sight. For example, in the case of flat-spectrum radio quasars, this angle is $< 10\degree$ \citep{Ghisellini}. As the value of this angle increases, we observe quasars with steep spectra and, for even larger jet inclinations, radio galaxies. However, as discussed by \citet{Snellen1998}, if a newborn radio source is oriented towards the observer, Doppler boosting of the inner radio component will also result in a one-sided jet morphology and a variable flat spectrum. 
Therefore, what defines an object as a GPS source is a peak in the spectrum around GHz frequencies and significant spectral curvature, $\rm \Delta \alpha > 0.6$ (Table \ref{table_spectral_modeling}), i.e. the difference between the spectral indices below and above the peak \citep{devries}. All our sources in the lower right quadrant of Figure \ref{figure_alpha} meet these criteria. 

It is also worth noting that several of our sources have highly absorbed spectra with $\alpha_{thick} > 2$ or even exceed the canonical value of 2.5 expected in the case of a uniform synchrotron self-absorbed (SSA) source, similar to the transient sources discovered by \citet{Wolowska}. Figure \ref{figure_histogram_index} shows that compared to GPS and CSS sources, compact radio transients have steeper spectra in the optically thick segment of their spectra, probably as a result of SSA,
but we cannot exclude the existence of a thermal plasma of different opacity along the line of sight around the source, as proposed by \citet{Bicknell}.
Note that uncertainties in the spectral index, especially for the optically thick part of the spectrum, are very large in some cases and result from too few measurements at low frequencies. This, in turn, prevents us from modeling the radio spectrum more precisely and testing whether SSA or free–free absorption (FFA) is responsible for the observed curvature.

\subsection{Radio morphology}
Most of our sources, although resolved in VLBA observations at 8.7 GHz, remained single objects with sizes in the range of $\rm 0.9 - 20.5$ pc. Several of them have a slightly extended structure, and in the case of six objects, we identify multiple radio components, some of which are probably radio jets. 
In Table \ref{table_vlba} we propose a morphological classification for all sources. Single (S) indicates a point-like object, unresolved or slightly resolved. Core-jet (Cj) indicates a source with bright central component and much weaker elongated, one-sided emission (i.e. a probable jet). Double-lobed (D) and triple (T) indicate objects consisting of two or three components of similar brightness, respectively. Finally, extended (E) indicates a structure consisting of a bright center with radio emission on both sides.
We calculated the brightness temperature of the components that we believe may contain the radio core and obtained values in the range of $\rm 10^6 - 10^{9}\,K$ (Table \ref{table_vlba}) at 8.7 GHz, indicating an AGN origin for their radio emission \citep{Condon1991}. However, in many cases this is a rough estimate since we are restricted by the resolution of our observations.
These values do not exceed the inverse Compton brightness temperature limit of $\rm 10^{12}\,K$ which means that relativistic beaming is not formally required.
Comparing this characterization of the radio structures of our sources with other groups of objects, we note that similar parsec-scale features have been observed in radio-quiet AGN like Seyfert or LINER galaxies \citep[e.g.,][]{Baldi2018,Kharb2021}, and not often in radio-loud AGNs \citep[e.g.,][]{Lister2009,Lister2018}. 

\begin{figure*}[t!]
\centering  
\includegraphics[scale=0.43]{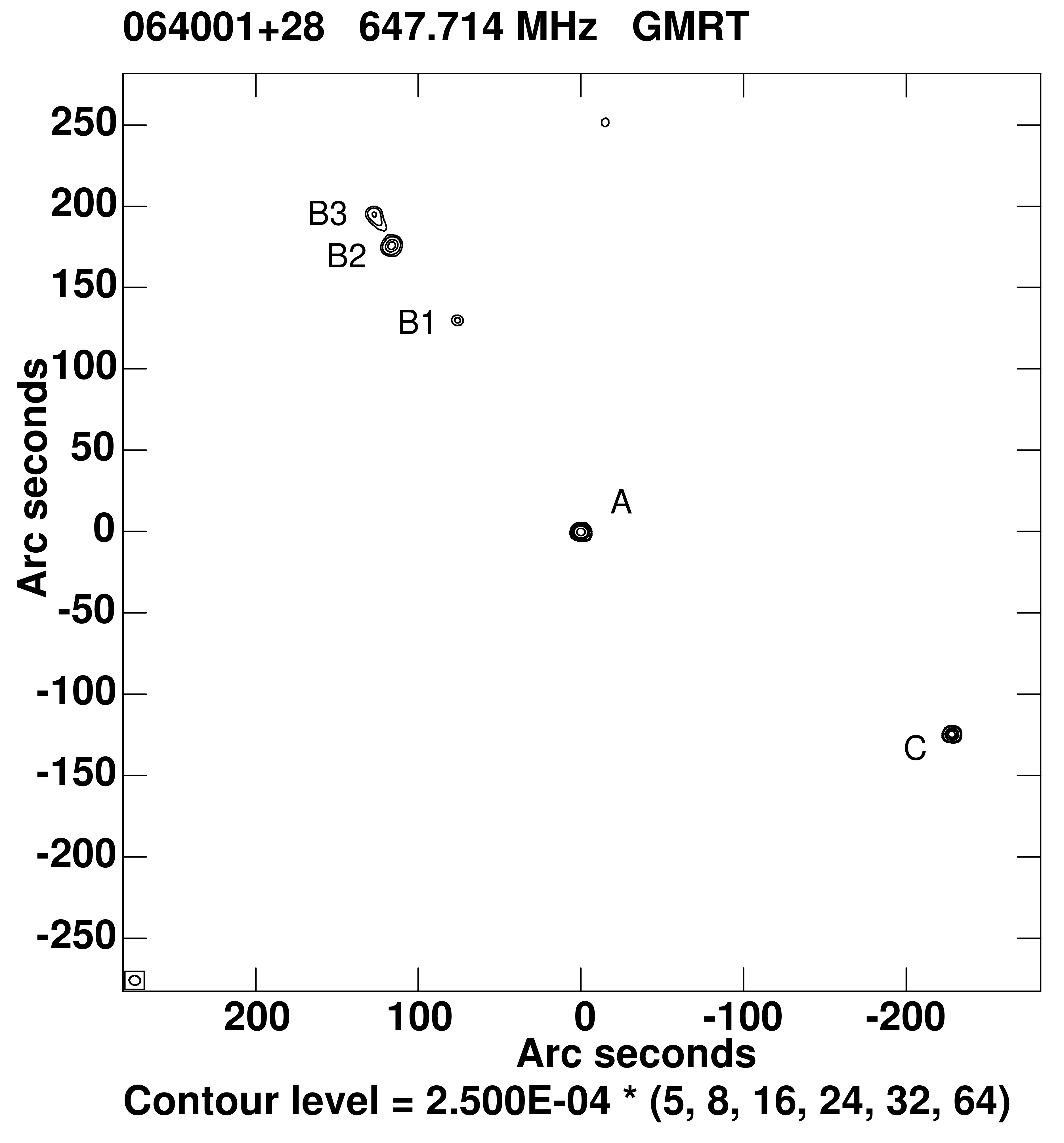}
\includegraphics[scale=0.43]{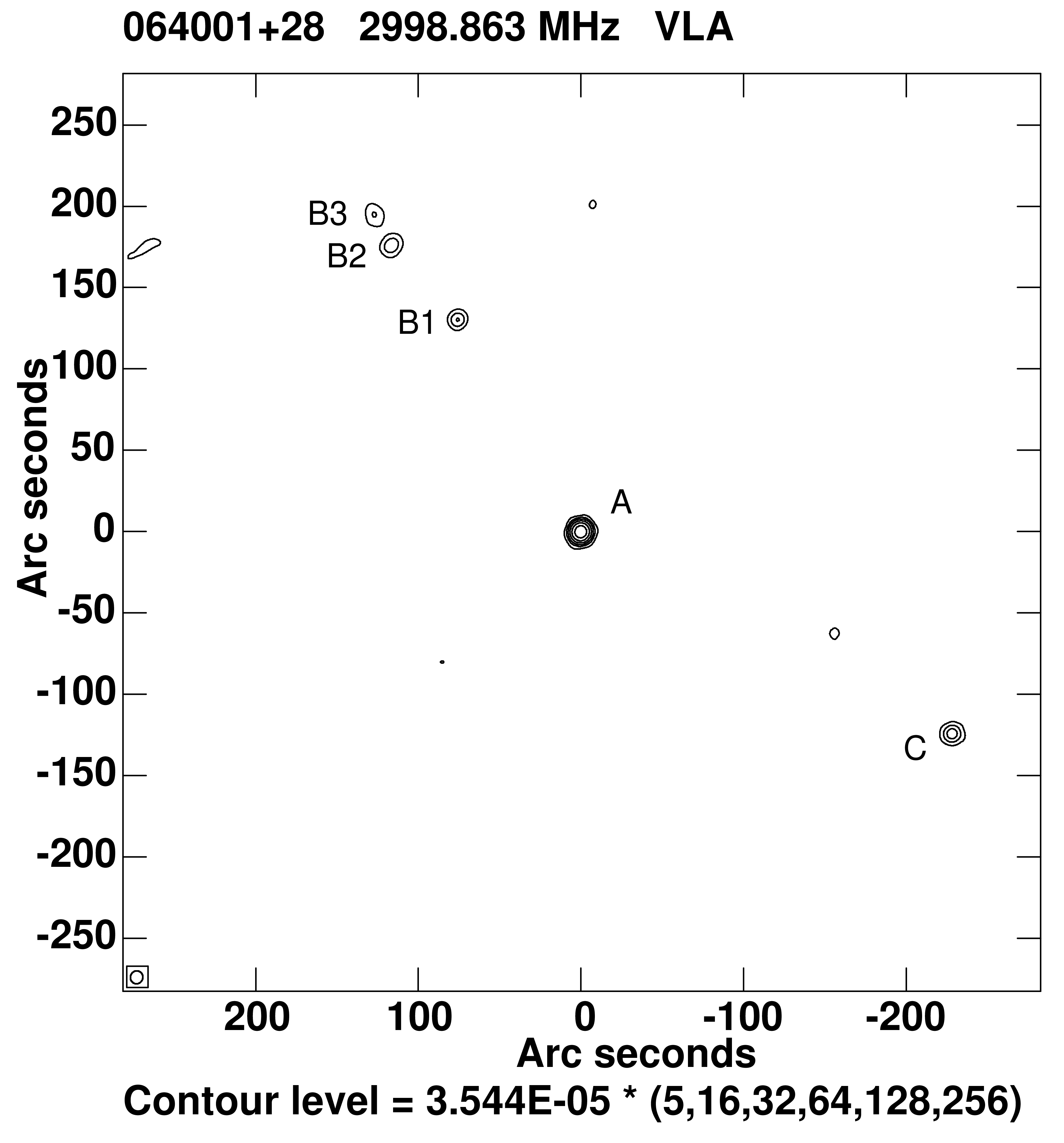}
\includegraphics[scale=0.43]{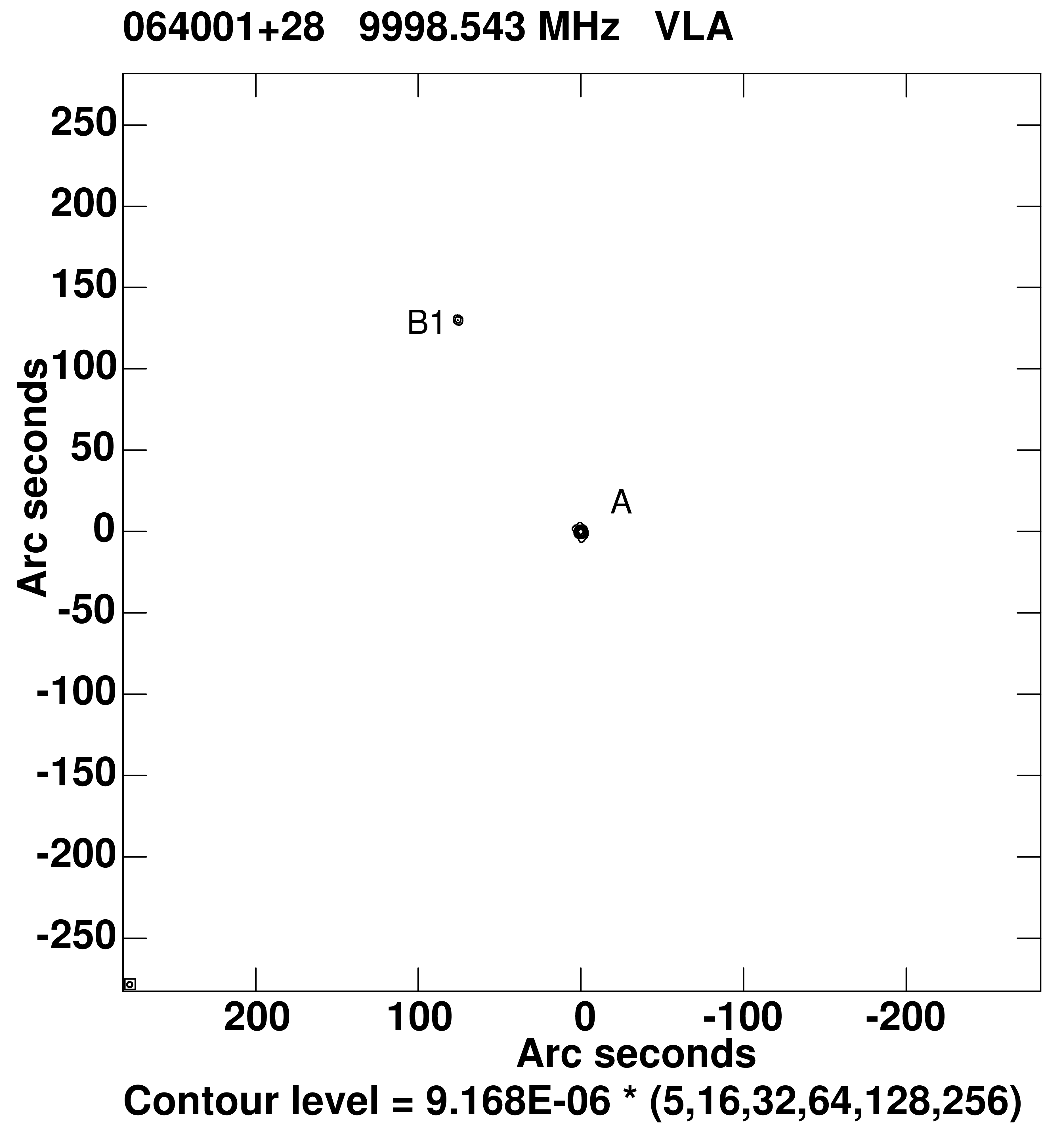}
\includegraphics[scale=0.37]{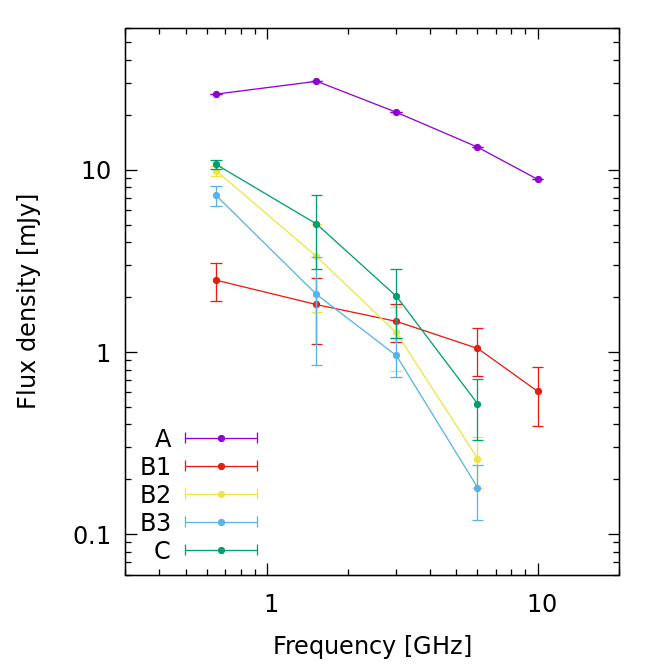}
\caption{The GMRT image and selected VLA images showing source 064001$+$28 (marked as A) and additional components/sources marked as B1- B3 and C. The lower right panel shows measurements of their flux densities, with measurements of component A added for comparison. Note, however, that a more detailed spectrum of component A with a fitted model is in Figure \ref{figure_spectral_modeling}.}  
\label{figure_appendix_0640}
    \end{figure*}

\begin{figure*}[t!]
\centering  
\includegraphics[scale=0.4]{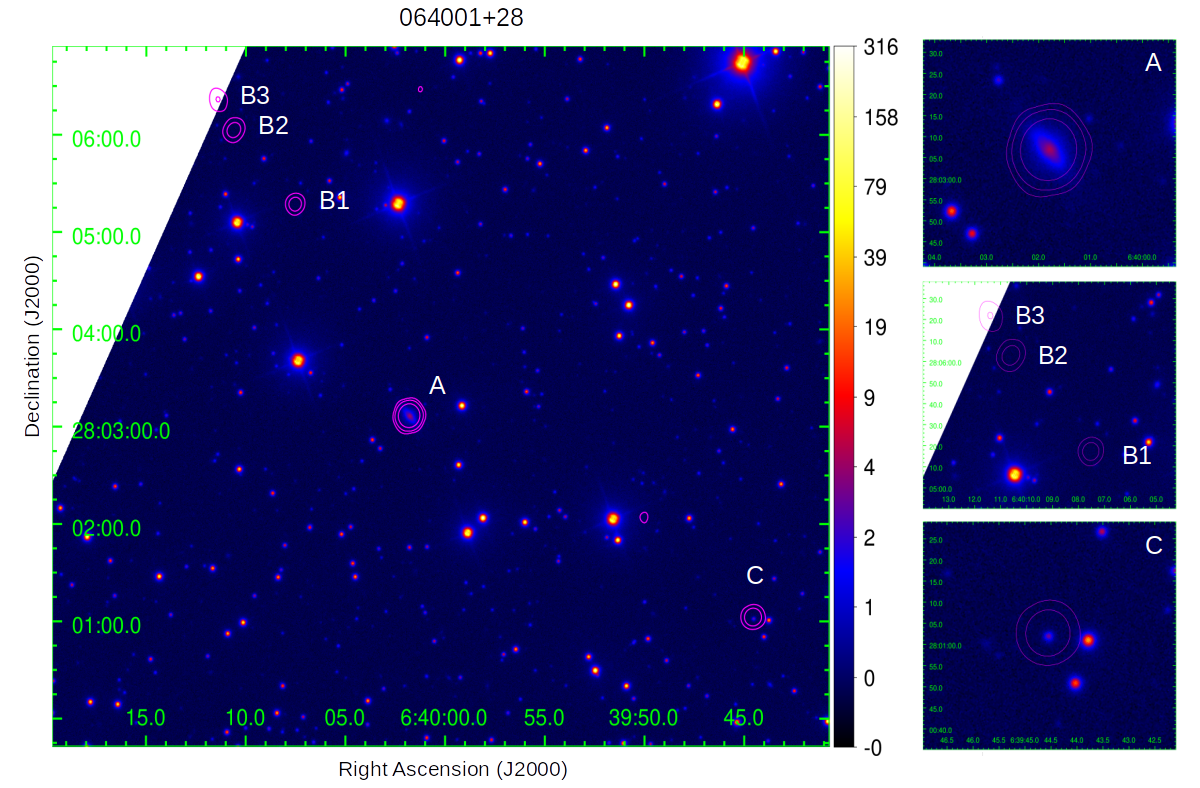}
\caption{The VLA S-band image (magenta contours) overlaid on the SDSS $i$-band image of 064001$+$28. The presented contours have the following values: $\rm 1.7, 5, 20 \times 10^{-4}\, Jy\, beam^{-1}$.}  
\label{figure_appendix_0640_sdss}
    \end{figure*}

Assuming that the entire VLBA structures of our sources were formed after the radio brightening, we estimated the expansion velocities of the newly formed outflows. The obtained results are in a very wide range of $0.1c - 8.9c$ (Table \ref{table_vlba}) and indicate that in some objects this velocity may not be relativistic. However, this is only a very rough estimate of the lower limit of this velocity, assuming that the entire radio structure developed in the time that elapsed between VLBA and NVSS/FIRST observations, i.e. 20-28 years. In reality, however, this time may be much shorter. Additionally, the linear size of the sources unresolved in VLBA observations measured by us is only its upper limit, which also affects the uncertainty of the estimated result.
For comparison the multi-epoch observations of compact radio sources made it possible to measure the radio lobes (or hot spots) proper motion at $\sim0.19c$ \citep{Polatidis, Gugliucci2005} while the jet velocity is much higher and may be superluminal and reach even $\sim 7c$ \citep{An2010, Lister2013}. In the case of Seyferts and LINERS or radio-quiet quasars both sub-relativistic jet speeds in the range of $0.01c − 0.5c$ \citep{Middelberg2004, Wang2023} and relativistic outflows are detected \citep{Kharb2024}. And finally small expansion velocities of the order of $0.02 - 0.15c$ are also estimated for thermal outflows \citep{Anderson2020, Cendes2024} which result from tidal disruption events (TDEs). Further discussion of possible scenarios for the origin of the radio emission in our sources is given in section \ref{sec_origin}.

In the case of five sources from our sample, the presence of radio emission on scales up to tens of kiloparsecs is plausible. One example is 064001$+$28 (marked as A) where the NVSS map at 1.4 GHz shows additional components, B and C, at distances of $\sim 70$ and $\sim 85$ kpc, respectively (Figure \ref{figure_appendix_images}). For component B, its elongated radio structure and collinear position with 064001$+$28 (A) suggests a physical link.
Meanwhile, high-resolution imaging at 8.7 GHz of component A reveals probable core-jet structure perpendicular to the line defined by components A and B. This could mean that the source we identify as 064001$+$28 is in fact the central component of a larger source in which we observed a restart of radio activity along with a change in the direction of the jet propagation. This interpretation seems to be supported by our new GMRT and multi-band VLA observations, in which we identify all potential components of the 064001$+$28 source with better resolution (Figure \ref{figure_appendix_0640}). These observations show that the flux density of objects B and C decreases with frequency, with object B consisting of three components. The value of the spectral index obtained by fitting with a standard non-thermal power-law model shows that the spectra of components C, B2, and B3 are very steep. In the case of component B1, the value is on the border between flat and steep spectra and suggests a more compact component (Figure \ref{figure_appendix_0640} and Table \ref{table_0640_comps_flux} in Appendix F). Additionally, Figure \ref{figure_appendix_0640_sdss} presents VLA radio contours of all identified components superimposed on the SDSS optical image, which shows that only components A and C have optical counterparts. The above findings lead us to interpret object B as the expiring jet/lobe of the source A. Assuming a moderate value for the lobe proper motion in compact AGNs of 0.19$\it c$ \citep{Polatidis}, we obtain a rough estimate of the age of this potential A$-$B kpc structure of $\rm \sim 1.2 \times 10^6$ years. This is in agreement with previously postulated interpretations of some compact radio sources as objects that are unable to develop large-scale radio structures \citep{Reynolds, Czerny, MKB10}.
Detection of several candidate dying compact sources \citep{MKB2006, Orienti2010, Callingham} also supports this scenario. 
However, finding objects in which both phases of source activity, old and new, are visible at the same time is very difficult, especially in the case of compact objects. According to models, after source activity ceases, the radio plasma remains visible for less than half of the jets' active phase due to the rapid decay in brightness caused by radiative energy losses and adiabatic expansion \citep{Reynolds}. 
If component B is indeed part of source 064001$+$28, it may mean that the previous phase of radio activity ceased in this source $\lesssim 6 \times 10^5$ years ago.

The next four sources are detected at very low frequencies, i.e. in LOFAR low-resolution (6\arcsec) observations at 144 MHz, as single objects (Figure \ref{figure_appendix_lofar}, Appendix D). However, the flux densities of three of them (095141$+$37, 112940$+$39, and 164607$+$42) do not match the spectral model fitted to the VLA and GMRT measurements and suggest the presence of an additional radio emission outside the very center of the source (Figure \ref{figure_spectral_modeling}). 
In the case of a fourth object, 150415$+$28, the LOFAR image corresponds to the position of the central component of a triple structure, the flux density fits well into the entire radio spectrum. 
The remaining two components, located symmetrically on both sides of the center, are actually separate sources that have SDSS optical counterparts (Figure \ref{figure_appendix_lofar}, Appendix D). The high-resolution LOFAR images of these sources are necessary to identify the potential new components. Another possible interpretation is that this excess radio emission at low frequencies has an origin other than AGN activity.

\subsection{Variability}
\label{sec:variability}
\begin{figure*}[t]
   \centering
   \includegraphics[scale=0.36]{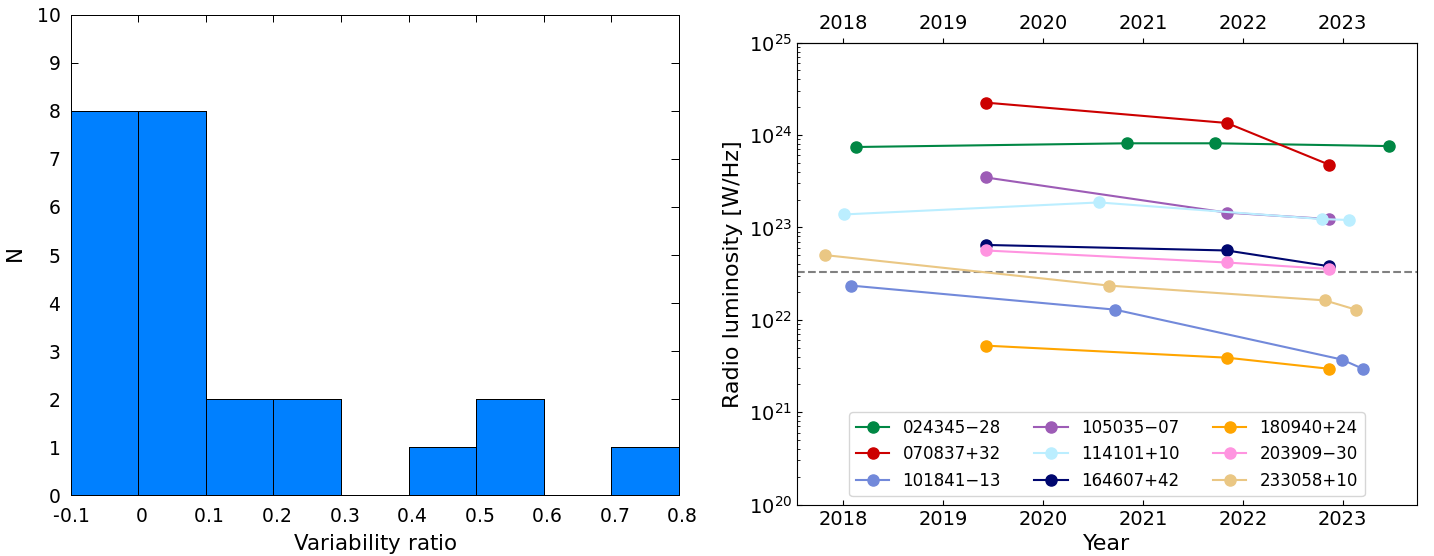}
\caption{(Left): Distribution of variability ratio at 3 GHz for our sample. (Right): The 3 GHz light curves for selected sources. The gray dashed line indicates the recalculated value of  $\rm \nu L_{\nu,GHz}= 10^{39}\,erg\,s^{−1}$ below which most known thermal TDEs occur \citep{Cendes2024}.}
              \label{figure_histogram_variability}%
    \end{figure*}

The multi-epoch VLASS observations give us the opportunity to examine the variability of transient sources after the onset of their activity. However, since the measurement of source flux density based on VLASS Quick Look images is subject to relatively large uncertainties (as described in section \ref{sec:sample}), the value of the variability index is only a rough estimate allowing us to identify the most significant changes with high certainty. 
We calculated the variability index at 3 GHz ($\rm V_{3\,GHz}$) using the standard formula (Equation \ref{equation_variability}), and these results are presented in Table \ref{table1_basic} and Figure \ref{figure_histogram_variability}. In cases where no third epoch of VLASS data was yet available, we used our own VLA observations from 2022 for that epoch.
Negative values of $\rm V_{3\,GHz}$ correspond to cases where the estimated error is greater than the observed scatter of the data

Only eight sources exceed the value of $\rm V_ {3\,GHz} = $0.10, which makes them potential candidates for variable objects. 
For four objects, 070837$+$32, 101841$-$13, 105035$-$07, and 233058$+$10, the change in flux density is very significant, $\rm V_{3\,GHz} > 0.30$, and indicates a large, systematic decline from epoch to epoch. Figure \ref{figure_histogram_variability} presents the 3 GHz light curves for these eight variable objects and a non-variable source 024345$-$28 as an example (see also the discussion in section \ref{sec:tde}).
For comparison, the long-term monitoring of AGNs at GHz frequencies performed by \citet{Sotnikova2022} revealed average values of the variability indices to be 0.15 and 0.17 at 2.3 GHz for flat-spectrum radio quasars and blazars, respectively. Both indices are higher at 22.3 GHz, with values of 0.28 and 0.30, respectively. Summarizing the long-term monitoring of peaked-spectrum sources described so far in the literature, a significant fraction (between 50 and 70\%) have a peaked spectrum temporarily, in the flaring state, particularly quasars \citep{Torniainen, Mingaliev2012, Sotnikova}. In contrast, genuine GPS sources are characterized by less variability, and maintain their convex spectral shape for many years. Recently, \citet{Wolowska} showed the same regularity for slow radio transients discovered in the CNSS survey. Multi-epoch VLA spectra showed that GPS quasars transform over time into flat-spectrum sources while galaxies maintain the convex shape of their spectra. However, it should be clarified here that a certain change of the spectrum, even for so-called genuine GPS sources, is observed as a consequence of the development of their radio structure. When compact components of the source expand we observe a shift of the peak to lower frequencies and variations in optical depth and thus also in flux density. Moreover, such a spectral change in very young objects happens in a short time, over just a few years \citep{Orienti2008}. For the sources studied here, the vast majority of which are galaxies, we conclude that they are characterized by a typical convex spectra shape. This shape is present even though several years have passed since the burst of their radio activity. Of course, further broadband monitoring of these sources would be advisable to confirm the shape and evolution of their radio spectra, because observing the variability of GPS sources only at one frequency does not give the full picture. Similarly, the potential variability of some of our sources at 3 GHz will only be confirmed when the final VLASS data products are available.
At the moment, however, it seems that the variability at 3 GHz observed for some of our sources could be the result of rapid adiabatic expansion of a young radio structure. This applies to source 114101$+$10 with a clear core-jet structure and up-down flux density fluctuations, and source 203909$−$30 with an elongated structure suggesting a jet outflow. This is also a probable explanation for the two-component shape of the spectra of 070837$+$32 and 105035$-$07, in which the peak visible at very high frequencies, $>$10 GHz, indicates the dominance of a bright compact core. While the low frequency peak is the result of a large drop in the flux density of these sources at 3 GHz and indicates an expanding outflow.

The next four objects, 101841$-$13, 164607$+$42, 180940$+$24 and 233058$+$10, are very compact sources, single, and have a convex or steep spectrum. The probable evolution of their spectra is a progressive decrease in flux density in a wide frequency range and a shift of the peak to lower and lower frequencies, eventually even below 1 GHz. In the case of one of these sources, 101841$-$13, its flux density has returned to the level before the radio brightening. The remaining objects have no earlier detection of radio emission, but the decrease in their radio flux density over several years is significant and amounts to 42--74\%.

\begin{figure*}[t]
   \centering
   \includegraphics[scale=0.6]{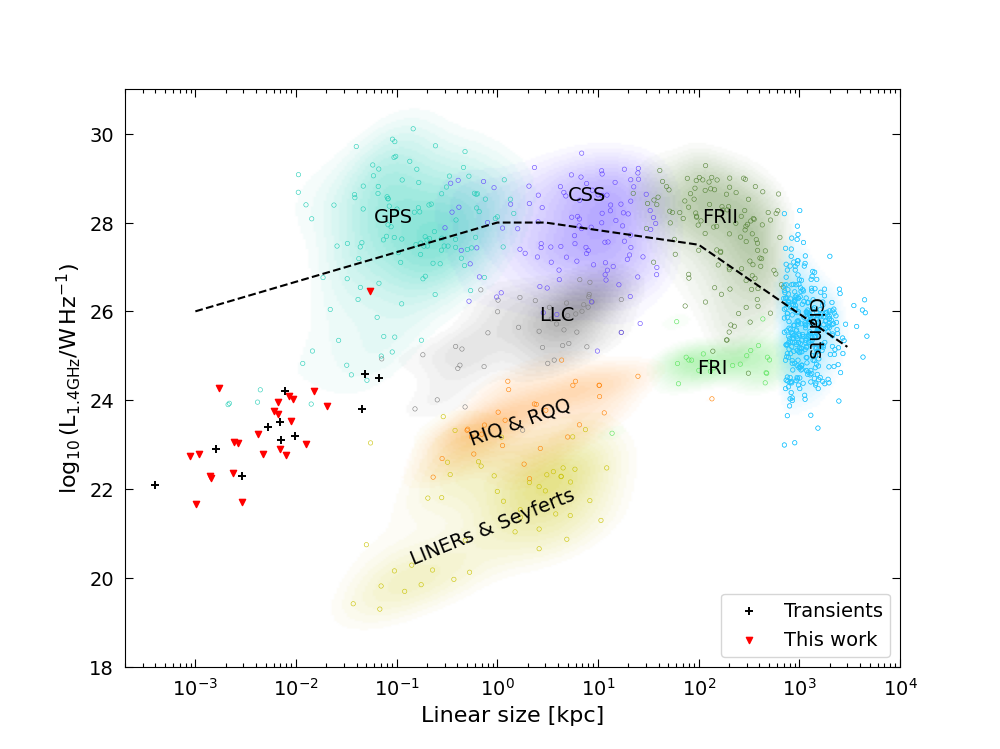}
\caption{The radio power vs. linear size ($P - D$) diagram adapted from \citet{MKB10}, including various types of radio-selected AGNs. The plots of GPS, CSS, LLC, FR\,I and FR\,II sources are taken from \citet{MKB10}. The sample of giant radio galaxies are adapted from \citet{Kuzmicz2018}. The radio-intermediate and radio-quiet quasars (RIQ \& RQQ) are taken from \citet{Kukula1998} and \citet{Jarvis2019}. 
LINERs and Seyferts galaxies are from \citet{Gallimore2006}, \citet{Baldi2018} and \citet{Singh2013}.
The transient sources are taken from \citet{Wolowska} and from this work, and are indicated with black crosses and red triangles, respectively. The black dashed line shows a possible evolutionary track for high power objects as described in \citet{An}.}
              \label{figure_P_D_diagram}%
    \end{figure*}

\subsection{Radio power $-$ linear size diagram }
Recent studies of radio-loud (RL) and radio-quiet (RQ) quasars suggest that the appropriate measure of radio loudness is the radio spectral luminosity $\rm L_{\nu}$ rather than radio ﬂux density or radio–optical ﬂux density ratio {\it R} \citep{Kimball2011, Kellerman2016}. This work suggests a luminosity separating RL and RQ of $\rm log_{10}[L_{6\,GHz}/ W\,Hz^{-1}]\sim 23 $ at low redshifts. Below this limit, the star-forming galaxy component contributes significantly to the radio emission or is even its sole source. More detailed observations of radio-quiet quasars lead to the conclusion that the population of these objects is a mixture of mainly starburst-powered and jet-powered galaxies \citep{Silpa, McCaffrey2022}, or they may be undergoing periods of increased radio activity during which they are able to launch a relativistic jet \citep{Silpa2021}.
If we apply this division criterion to our sample, we find that even after the significant increase in radio flux density that allowed us to detect these objects, half of them can be considered as radio-quiet sources (Table \ref{table1_basic}). Note, however, that the level of star formation-related radio emission in galaxies studied here may be different from that characterizing quasar hosts as investigated by \citet{Kellerman2016}. 

In order to quantify our comparison to various AGN populations, we investigate the relationship between radio luminosity and physical size for our sources in Figure \ref{figure_P_D_diagram} using the updated version of the radio power $-$ linear size ($P - D$) diagram from \citet{MKB10}. The parametrization of radio sources in terms of their radio power and linear size serves as a powerful probe of major evolutionary trends. We indicate with the red dashed line the evolutionary track for high power radio sources, connecting traditional radio-loud AGN classes as proposed by many authors \citep{Fanti, Redhead, Odea}. In the case of low luminosity radio sources, the evolutionary
paths are less clear, and a range of variants are discussed in the literature, including
further development of a source into a low luminosity stagnated FR-I-like source \citep{An}, or an episodic, short-lived outburst of radio activity which ceases after $\sim 10^5$ years \citep{Reynolds, Snellen2000, Czerny, MKB10}. In the latter case, we will observe a weakening radio structure of medium size, clearly visible only at low radio frequencies.

Analyzing the ($P - D$) diagram, we see that the transient sources from our work and those described by \citet{Wolowska}, both galaxies and quasars, are similar in power to the low luminosity compact (LLC) sources, radio-intermediate and radio-quiet quasars (RIQ/RQQ) and Seyfert galaxies. 
In terms of projected physical size alone, we see that the transient sources are more compact and overlap with the most compact GPS objects. Therefore, the radio properties of transient sources place them on the ($P - D$) diagram at the beginning of the evolutionary path, which suggests their further development towards larger LLC and RIQ/RQQ sources, and perhaps Seyferts. Some of our sources, however, may be here only for a while.

The information about the spectral properties of LLC and RIQ/RQQ objects is very limited. However, it is known that their radio spectra are steep and peak between 1.4 GHz and 150 MHz in the case of RIQ/RQQs \citep{Jarvis2019} and well below 150 MHz in the case of LLC sources \citep{MKB10}. To better determine the spectral shape of LLC objects, observations at low radio frequencies are needed, which will soon be provided by the LOFAR survey. 
The radio spectra of Seyfert galaxies are both flat and steep, with the vast majority of the latter in a wide frequency range from 150 MHz to 1.4 GHz \citep{Kharb2016, Singh2018}.
Recently, two new large groups of objects with peak frequencies between 72 MHz and 1.4 GHz were identified by \citet{Callingham17} and \citet{Slob} using the GaLactic and Extragalactic All-sky Murchison Wideﬁeld Array (GLEAM) survey and LOFAR surveys. The values of their radio power at 1.4 GHz lies within a wide range, $\rm 10^{22} - 10^{27}\, W\,Hz^{-1}$, and high-resolution observations are needed to know their exact physical sizes. Interestingly, however, there is an observed relationship between radio size and the frequency of peak emission in the radio SEDs of compact radio galaxies \citep{Odea, Orienti, Sotnikova}, which we use to further investigate the evolutionary scenario.

\begin{figure*}[t]
   \centering
   \includegraphics[scale=0.55]{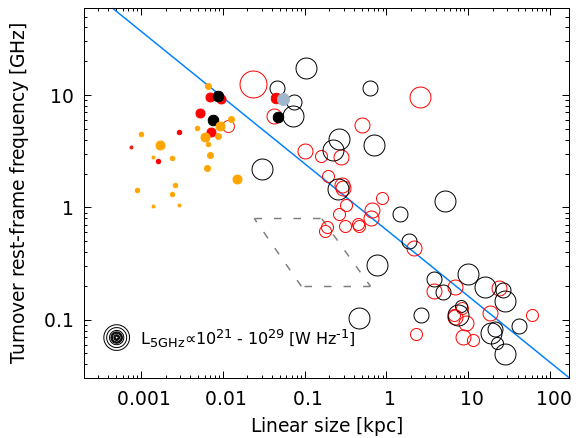}
\caption{The intrinsic turnover frequency vs. linear size diagram. The transient sources from this work are marked with yellow (galaxies) and gray (quasar) circles. The open red circles (galaxies) and open black circles (quasars) indicate CSS and GPS sources taken from \citet{Odea2}, \citet{Snellen}, \citet{devries}, \citet{Stranghellini}, and \citet{Fanti} collected by \citet{Keim} in one sample. The filled red circles (galaxies) and filled black circles (quasars) indicate radio transients discovered in the CNSS survey by \citet{Wolowska}. The sizes of the circles correspond to their radio luminosity at restframe 5GHz. The blue solid line indicates the linear relationship log~$\nu_p = -0.21 - 0.59\times \mathrm{log~LLS}$, found by \citet{Orienti}. The gray diamond indicates the range of values for peaked-spectrum sources from \citet{Callingham17}.}
              \label{figure_turnover}%
    \end{figure*}

\subsection{The peak frequency $-$ linear size relationship}
\label{sec:peak_size}
The peak in the radio spectrum, the most characteristic feature of GPS and CSS sources, is a function of source size and therefore age. The continuous distribution of young AGNs along the frequency of the peak (turnover frequency) $-$ linear size ($\nu_p − D$) plane \citep{Odea, Orienti, Sotnikova} suggests a source development path (Figure \ref{figure_turnover}). At the same time, there is an ongoing discussion in the literature about the mechanism that causes this turnover in the spectrum. Several studies indicate that SSA is able to correctly reproduce the $\nu_p − D$ relation \citep{Snellen2000, Jeyakumar}. The particular evolutionary models in this framework are moreover based on the assumption of minimum energy content, corresponding to a near equipartition of energy between the radiating particles and the magnetic field. There are many examples of individual sources for which the magnetic field values calculated from measurements indeed agree well with those obtained by assuming equipartition \citep{Orienti2008}. On the other hand, we also have examples of objects in which the magnetic field values and the shape of the broad-band radio spectrum indicate a different absorption mechanism, namely FFA caused by a thermal plasma \citep{Callingham, Keim}.
Using the equipartition magnetic field $\rm B_{eq}$ formula (Equation \ref{equation:magnetic_field_eq}), the derivation of which is presented in Appendix \ref{section:magnetic_field}, we calculate that $\rm B_{eq}$ for our sources is of the order of several dozen mG (Table \ref{table_vlba}), i.e. typical for compact sources ($\rm < 1\,kpc$) with SSA-related peak frequencies within the GHz range.

In the $\nu_p − D$ graph (Figure \ref{figure_turnover}), our sources occupy the same area as transient sources from \citet{Wolowska}, which, according to the accepted interpretation, means that they are at an early stage of their life cycle. On the other hand, some transient objects, mainly those with lower radio luminosities, show some deviation from this relationship. This, as discussed in \citet{Wolowska}, may imply a different (parallel) path of radio source evolution and a limit to the size of the source. Radio source evolution consistent with a parallel path would mean that weak radio sources of larger sizes ($\rm > 1\,kpc$) would have a peak in their spectrum below the rest-frame frequency of $\sim 200$ MHz. Parallel development paths depending on jet power are discussed in \citet{Jeyakumar} in the context of the SSA model. As noted previously, \citet{Callingham17} and \citet{Slob}, using low-frequency radio surveys, found sources that display spectral peaks between 72 MHz and 1.4 GHz, and speculated that these are low-frequency analogs of archetypal GPS and CSS objects. \citet{Slob} then suggested that some of these sources, especially low-power ones, may not evolve into large-scale AGNs. We have marked the location of low-frequency peaked-spectrum sources from the sample of \citet{Callingham17} on the $\nu_p − D$ plane, estimating the sizes of these objects assuming equipartition. For this we used the equation \ref{eq:eq_size} and the following median spectral properties of the peaked-spectrum \citet{Callingham17} sample: $z=0.98$, upper and lower limits of observed peak frequency with values of 100 and 380 MHz, and low and high values of flux density equal to 0.16 and 10 Jy, respectively. The estimated values of equipartition angular size, $\theta_{eq}$, are in the range of 11 - 80 mas and 3 - 21 mas for peak frequencies of 100 and 380 MHz, respectively. Based on these estimates, we hypothesize that our low power transient sources with GPS spectra may be progenitors of low-frequency, low-luminosity peaked-spectrum objects (Figure \ref{figure_turnover}).

\subsection{Origin of the radio transient emission}
\label{sec_origin}
The study of the radio variability of our objects is limited due to the method of their selection and the sparse cadence of observations. Therefore, we have at our disposal the NVSS/FIRST observations and the first epoch of VLASS observations, which are separated by more than two decades, two or three VLASS observation epochs with a three-year cadence and our single VLA measurement. Additionally, as already mentioned, the available VLASS data are subject to large errors, which greatly limits the study of variability on a time scale of several years. Taking this into account, we observe a significant increase in the flux density of all 24 sources, $>200\%$, on a scale of several decades, and variability over the next few years for eight of them (section \ref{sec:variability} and Figure \ref{figure_histogram_variability}). Such long-term radio variability may have various physical origins, including extrinsic variability induced by scintillation, intrinsic variability of a radio-quiet AGN, changes in the accretion onto an SMBH due to instability in an accretion disk or tidal disruption of a star, connected or not with the young, expanding radio jet. Moreover, the physical explanation for the observed transient emission might not be the same in all sources and it might be a combination of more than one mechanism.  The following subsections discuss these mechanisms.

\subsubsection{Variability induced by interstellar scintillation}
Extrinsic variability is caused by refractive or diffractive interstellar scintillation and observed at megahertz frequencies on time scales of months-years and days-weeks, respectively \citep{Rickett1986}. Such variability is observed when a radio wave emitted by a compact, point-like source or point-like feature, whose emission dominates the spectrum, passes through fluctuations in the density of the plasma of the interstellar medium (ISM). Recently, \citet{Ross2021} has shown that variability at low frequencies ($<$1 GHz) is more prevalent in peaked-spectrum sources compared to typical radio galaxies and in the case where we do not observe a change in the shape of the radio spectrum in the source, it may be caused by scintillation. \citet{Nyland} estimated that flux density fluctuations caused by refractive scintillation at frequencies above 1 GHz are of order $\sim 30\% – 40\%$.
In the case of our objects, we do not have multi-epoch observations of their radio spectra or monitoring of variability at lower frequencies. Nevertheless, our estimates show that after the radio brightening, these sources typically show relatively low variability at 3 GHz, of order 20\% - 40\% from epoch to epoch, implying that scintillation is a possible explanation for this behavior. However, the scale of the flux density increase observed for the radio brightening phenomenon itself as well as the scale of the flux density decay for the most variable sources is much higher. This allows us to rule out refractive scintillation as the origin of the observed radio transitions, as well as the significant variability seen in some sources.

\subsubsection{Emission mechanisms in radio-quiet AGNs}
As mentioned in the previous sections, half the sources in our sample, even after radio brightening, have radio luminosities typical of radio-quiet sources, although in the case of many of them this value is close to our adopted limit of $\rm log_{10}[L_{6\,GHz}/ W\,Hz^{-1}]\sim 23 $ \citep{Kellerman2016}.
While radio-loud AGNs have the ability to launch powerful relativistic jets, which are their main source of radio emission, in the case of radio-quiet AGNs, the possible mechanisms generating radio emission are more numerous and more diverse. They include star formation (SF), accretion disc winds, coronal disc emission, low-power jets or a combination of them \citep{Panessa2019}. However, it should be remembered that the division into radio-loud and radio-quiet sources based only on radio emission is not sharp and unambiguous. Therefore, comprehensive multi-frequency and high-resolution observations are essential to correctly classify sources. Our sample are radio-compact objects that in VLBA observations at 8.7 GHz remain either single or are partially resolved with jet-like components, similar to what is observed in LINERs and Seyfert galaxies. Neither of these morphologies is preferred among either the radio-quiet or the radio-loud objects in our sample. Further, the majority of these sources are characterized by a GPS-like spectrum, high brightness temperature and magnetic fields of order tens of mG, typical for newborn radio sources with kinematically young radio jets. 
Therefore, the above properties allow us to exclude as a potential source of radio emission in our sources such phenomena as AGN winds and thermal free-free emission, both of which are characterized by a flat spectrum and low-brightness temperature ($\rm T_{B} < 10^5\,K$), as well as SF processes, which are characterized by a constant steep radio slope from megahertz to gigahertz frequencies, low surface brightness, and, most importantly, a lack of radio variability.

\subsubsection{Interplay between the jet and the BLR clouds}
Very recently strong high-frequency radio variability has been reported in radio-weak narrow-line Seyfert~1 (NLS1) galaxies \citep{Lahteenmaki, Jarvela2024}. No satisfactory explanation for such behavior has yet been found. However, the lack of a typical peaked radio spectrum in these sources has led to the rejection of the scenario of young jet formation. Instead, the authors suggest that the recorded variability may be the result of some kind of interplay between the jet and the broad line region (BLR) clouds \citep{Jarvela2024}. According to this scenario, the jets of these sources are still within the BLR. They are completely free-free absorbed but from time to time they are exposed by the movement of ionized gas clouds, which manifests itself in the appearance of a radio flare. The variability timescale of these NLS1 galaxies is days and the observed frequency is very high, 37 GHz. This is therefore a different situation from our sample, and it is not clear whether similar scenario, e.g. with longer variability times, could be applicable to them. 
However, it is worth noting that some of our objects showed low levels of radio emission before radio brightening, which could indicate the existence of a persistent jet.

\subsubsection{Changes in the accretion process}
Another possible explanation for the transient behavior detected in our sources is intermittent activity of the nucleus caused by a radiation pressure instability in the accretion disk \citep{Czerny}. This mechanism can operate in both radio-loud and radio-quiet sources. According to this model, an accretion disk alternates between the two basic states which are thermally stable: (1) an outburst state combined with a jet ejection, and (2) a period of quiescence at a low accretion rate. 
The duration of the active period and quiescence scales with the mass of the black hole. For the typical black hole masses of radio-loud AGNs ($\rm 10^7 - 10^8\, M_{\odot}$) the expected outburst durations are $10^4 - 10^5$ years. 
For the typical black hole masses of NLS1 galaxies ($\rm 10^6 - 10^7\, M_{\odot}$), the outburst duration is 40 - 440 years and the time of the activity decay is even shorter, with the prediction that NLS1 objects should fade and transition to a low luminosity AGN state in 4 - 40 years. 

We estimate that our sources started their active period at the earliest $\sim$20 years ago, which is manifested by the appearance of a new, synchrotron radio outflow, similar to the case of other radio transients detected so far \citep{Nyland, MKB2020, Wolowska}. Additionally, this is associated with a significant increase in the source brightness in the initial phase of development, which is consistent with the evolutionary model of radio objects proposed by \citet{An}. On the other hand, the observed low variability of the studied objects after the radio brightening suggests that the radio activity stabilized at a higher level and is continuing. Further monitoring of the flux densities of these sources over a wide range of frequencies will allow testing of this scenario. At the moment, however, this seems the most likely explanation for the radio properties of majority of our sources. Moreover, \citet{MKB2020} suggested that a small change in the accretion rate of $\sim 30\%$ might be sufficient to trigger such low power jet activity. This in turn may explain the fact that we did not find significant amplitude variability in the optical light curves of our sources observed by Catalina Sky Survey \citep[CRTS;][]{Drake} and Zwicky Transient Survey \citep[ZTF;][]{Graham2019}, and in their mid-IR light curves from NEOWISE survey \citep{Mainzer2014}. The exception here are four sources whose behavior in the radio and infrared wavelengths allows for a different interpretation, which we discuss in the next section.

\subsubsection{Contamination from Tidal Disruption Events}
\label{sec:tde}
Tidal disruption events (TDEs) occur when a star passes too close to an SMBH \citep{vanVelzen2011,Alexander2020} and is consequently ripped apart. The stellar material is then partially accreted onto the SMBH leading to a multi-wavelength flare, which is also visible at radio wavelengths in $\sim 40\%$ of TDEs \citep{Cendes2024}. The vast majority of TDEs are thermal, non-relativistic outflows with $\rm \nu L_{\nu,GHz} \lesssim 10^{39}\,erg\,s^{−1}$ and the mechanism generating the radio emission is poorly understood \citep{Alexander2020}. However, in a few cases, a relativistic, non-thermal jet with luminosity $\rm >10^{40}\,erg\,s^{−1}$ has been observed to be launched as a result of a TDE \citep{Eftekhari2018}. The timescale of TDEs are usually in the range of tens to hundreds of days, but longer events have been observed in the group of so-called radio-selected TDEs \citep{Somal2024}. This is comparable to the radio luminosity changes observed for our eight objects (Figure \ref{figure_histogram_variability}). 
Nevertheless, even at their minimum luminosity at 3 GHz, most are brighter than known thermal TDEs \citep[e.g.,][]{Cendes2024}. Additionally, as discussed above, other radio properties for four of them (070837$+$32, 105035$−$07, 114101$+$10 and 203909$-$30) make the TDE scenario an unlikely interpretation for those sources. 

However, for the remaining four sources, 101841$-$13, 164607$+$42, 180940$+$24 and 233058$+$10, the TDE scenario can be considered. Their luminosities, the shapes of the radio spectrum and their magnetic field strengths are comparable to radio-selected TDEs \citep{Somal2024}. At the same time, these radio properties may suggest the formation of a small, low-power, probably non-relativistic radio jet in three of them. The exception is 180940$+$24, which does not have a turnover in its radio spectrum and also has a diffuse, low-luminosity radio structure. Furthermore, the WISE infrared light curve of 180940$+$24 shows a long-term decay since 2010 which can be interpreted as the dust-heated echo  of the central optical flare, which would have have occurred earlier and could potentially have been caused by a stellar tidal disruption event. 
However, the available optical data do not allow confirmation of any significant optical variability in this source.

The situation is still different for 101841$-$13, which shows a distinct infrared flare which recently led to its classification as a mid-infrared-selected TDE \citep{Masterson2024}. Modeling the infrared light curve indicates a potential tidal disruption in 2015, which could also be the onset of the radio brightening tracked by VLASS starting in 2017. This is similar to previously discovered TDE phenomena in which an infrared flare is accompanied by radio emission from a relativistic jet \citep{Mattila2018}. However, as discussed earlier, weak radio emission from 101841$-$13 was already visible in 1995 in the NVSS 1.4 GHz data. Its low level may indicate a star formation origin or the presence of a quiescent AGN core, or a combination of the two. 
Spectroscopic observations of 101841$-$13 made by us in 2022, i.e. after the extinction of the infrared flare and at the end of the radio flare, clearly classify the source as a Seyfert galaxy. 101841$-$13 has no obvious broad emission lines in its spectrum, but the narrow emission line flux ratios imply photoionization by an AGN. 
This could be interpreted as a temporary increase in ionization levels due to the postulated infrared-detected TDE. However, an archival 6dF optical spectrum of 101841$-$13 taken prior to the infrared and radio brightening shows significant [O\,III] and [N\,II] emission, indicating the presence of an AGN buried in the host galaxy \citep{Jones2004, Jones2009}.
Therefore, our proposed alternative interpretation of the infrared variability of 101841$-$13 associates it with AGN optical variability being reprocessed by the dusty torus. The optical variability may, in turn, be the result of an increased accretion rate, similar to what is expected in changing-state AGNs, or could be associated with a TDE in an AGN. An additional effect of the accretion rate change is the launch of a radio jet-like outflow by this pre-existing AGN. 
Interestingly, spectroscopic observations made exactly one year later by \citet{Masterson2024} show changes in line luminosities that no longer allow such a clear classification of this source, which argues that the source underwent a short-lived accretion enhancement. In their BPT diagrams, 101841$-$13 appears as a Seyfert, composite or star-forming galaxy. Ultimately, a combination of these two phenomena, a TDE and a very low-level accretion process, is not ruled out.
We present our spectroscopic observations of $101841-$13 along with a comparison to the 6dF spectrum in Appendix~G.

The remaining two sources, 164607$+$42 and 233058$+$10, do not show any indications of optical or infrared flares. 

Finally, in the context of TDEs, we will comment on 024345$−$28 whose 3~GHz light curve is also shown in Figure \ref{figure_histogram_variability}. This source increased in brightness by a factor of $\sim 50$ between the NVSS (1993) and VLASS (2018), which is the largest flux change in our sample, but the source does not show any significant radio flux density changes after the brightening event. The properties of 024345$−$28 are discussed by \citep{Somal2023}, who hypothesize that this source is either a slowly evolving TDE that produced a synchrotron jet or is a young radio source that was initiated by an enhancement in the accretion rate. We suggest that the radio properties of 024345$−$28 presented here, including the double radio structure seen in the high-resolution data and the convex shape of the synchrotron spectrum, favor the latter explanation.

In conclusion, we cannot definitively rule out that the radio brightness changes of 101841$-$13, 164607$+$42, 180940$+$24 and 233058$+$10 are caused by TDEs. However, radio properties such as weak radio emission from a single component and rapidly decaying flux densities make us consider the TDE phenomenon to be the most probable interpretation for 180940$+$24 and 233058$+$10.

\subsection{Implications}
Recent studies of radio transient phenomena have yielded estimates of the period of occurrence of a new episode of radio activity in AGNs at $\sim 10^5$ years \citep{Mooley, Nyland}. This is a timescale previously proposed by theoretical and dynamical models of the evolution of radio sources, which were motivated by the overrepresentation of young, compact radio objects in population studies \citep{Odea, Reynolds, Czerny, MKB10, An}. Also, recent dynamical modeling of LOFAR radio source populations in the HETDEX field \citep{Hardcastle} and in the Lockman Hole \citep{Shabala2020} show the need for a large fraction of short-lived objects. Moreover, the duration of the active radio phase may scale with jet power, such that the low-power objects are shorter-lived and may stay compact for most of their lifetime \citep{Hardcastle, Capetti, MKB2020, Wolowska, Kharb2021}. Such episodic jet activity is likely the result of changes in the accretion rate of material onto the SMBH regulated by the radiative or mechanical feedback from the AGN. However, as shown in \citet{Wolowska} and in this article, deep multi-epoch and multi-frequency follow-up observations are needed to finally correctly classify sources because some contamination from the TDE phenomenon is possible. 
Based on this work, we estimate the TDE contamination rate to be 8\% -- 17\% and the duration of TDE radio emission to be at least several years. On the other hand, a TDE was recently discussed as the possible cause of episodic radio activity lasting even thousands of years \citep{Readhead2024, Sullivan2024}. Such a short period of radio activity, i.e., compared to large-scale radio galaxies, allows for the development of a small, jetted radio structure with a size of $<$ 500~pc. These sources therefore remain compact throughout their life and can be called young only in a very early phase of their development (also see discussion in section \ref{sec:peak_size}). 
The study of radio transient objects fits very well into understanding the formation and evolution of radio sources and black hole – galaxy co-evolution. It remains largely unexplored whether there are other physical conditions required for jet production rather than changes in the accretion rate. This requires further, regular radio time-domain observations of large areas of the sky, pushing to lower luminosities and monitoring of objects that have already been discovered as radio transients.

\section{Summary and Conclusions}
The radio transient sources presented in this article were discovered by comparing the NVSS and VLASS surveys. The entire sample contains 24 sources, one of which was identified as a quasar and the remaining ones as galaxies. All these objects, except the quasar, belong to the local Universe and have redshifts ${ z < 0.3}$. We obtained multi-frequency and high resolution radio observations of this sample using instruments such as VLA, VLBA, GMRT and LOFAR, the conclusions of which are presented below.
   \begin{itemize}
      \item The studied sources are characterized by a significant increase in their radio flux density after some period of absence or marginal activity. The characteristics of the radio emission found in the vast majority of these objects, such as the convex shape of their spectra, small size and high brightness temperature (not exceeding however the Compton catastrophe), are typical for young GPS-type radio sources and indicative of an AGN buried in the host galaxy.

      \item High-resolution VLBA observations reveal that about half of our objects have a more extended, symmetrical structure or more than one component with an indication of young radio jets, similar to what is observed in LINERs or Seyfert galaxies. Additionally, lower frequency radio observations indicate the presence of radio emission on kiloparsec scales in a few objects. We suggest this may be remnants of previous jet activity, which we estimate to be $\lesssim 10^6$ years old.
       
      \item In terms of radio luminosity, the transient sources are much weaker than archetypal GPS objects and will probably undergo a different life cycle than high luminosity radio sources. Based on the distribution of these objects in power$-$size ($P - D$) and peak frequency$-$size ($\nu_p − D$) diagrams, we suggest that the low-luminosity GHz-peaked radio transients (galaxies and quasars) will develop into RI/RQ quasars and low-frequency peaked-spectrum objects, as recently identified at MHz frequencies.

      \item Most transient sources probably have low brightness variability once they turn on. However, for about 30\% of the sources in our sample, the changes in radio flux density are larger and have a decreasing nature.
      In the case of one source, 101841$−$13, the analysis of three epochs of 3\,GHz VLASS observations conducted in the period 2017 - 2023 reveals that it returned to its pre-brightening radio luminosity level. 
      These radio changes are likely associated with the prominent infrared flare recorded in WISE observations of 101841$−$13, based on which it has been classified as a mid-infrared-selected TDE. In contrast, our analysis of its radio and optical properties prior to the WISE and radio flares indicate the presence of an AGN in the host galaxy. We therefore speculate that the radio transient behavior of 101841$−$13 may be the result of a short-lived intrinsic accretion enhancement.

      \item We conclude that in the vast majority of our sources the radio transient emission event is probably caused by intrinsic changes in accretion properties,
      similar to what is observed in changing-state AGNs, lasting several decades or less.
      However, we cannot rule out the possibility that a few of our sources may belong to the rare class of radio-emitting TDEs. We estimate the TDE contamination of our sample to be less than 17\%.
    
   \end{itemize}

\begin{acknowledgements}
We thank the anonymous referee for helpful suggestions that
led to improvement of the paper.

The National Radio Astronomy Observatory is a facility of the National Science Foundation operated under cooperative agreement by Associated Universities, Inc. We thank the staff of the VLBA and VLA for carrying out these observations in their usual efficient manner.

We thank the staff of the GMRT who have made these observations possible. The GMRT is run by the National Centre for Radio Astrophysics of the Tata Institute of Fundamental Research. 
We thank Ruta Kale and Ishwara-Chandra C. H. for their help with data reduction and using the CAPTURE software.

We thank Matthew Graham, Thomas Connor and George Djorgovski for help in making optical observations using the Keck and Palomar telescopes. 

MKB acknowledges support from the `National Science Centre, Poland' under grant no. 2017/26/E/ST9/00216. The work of DS was carried out at the Jet Propulsion Laboratory, California Institute of Technology, under a contract with NASA. PK acknowledges the support of the Department of Atomic Energy, Government of India, under the project 12-R\&D-TFR-5.02-0700.
\end{acknowledgements}

\software{CASA \citep{McMullin}, AIPS \citep{vanMoorsel}, IRAF \citep{Tody86,Tody93}, STARLIGHT \citep{CidFernandes2005,Mateus2006}.}

\bibliographystyle{aasjournal}
\bibliography{main.bib}

\appendix
\section{Magnetic field}
\label{section:magnetic_field}

For the energy spectrum of radio-emitting electrons, we assume a single power-law
\begin{equation}
\mathcal{N}_e(\gamma) = N_0 \, \gamma^{-p} \quad {\rm for} \quad \gamma_{min} \leq \gamma \leq \gamma_{max} \, ,
\end{equation}
with $p\simeq 2$ and $\gamma_{max} \gg \gamma_{min}$, so that the electron energy density
\begin{equation}
U_e \equiv \int m_e c^2 \gamma \, \mathcal{N}_e(\gamma) \, d\gamma = m_e c^2 N_0 \, \ln(\gamma_{max}/\gamma_{min}) \simeq 10  \, m_e c^2 N_0 \, .
\end{equation}
We also define the equipartition ratio as
\begin{equation}
\eta_{eq} \equiv \frac{U_e}{U_B} \simeq \frac{10  \, m_e c^2 N_0 }{B^2/8\pi} \, ,
\label{eq}
\end{equation}
where $B$ is the magnetic field intensity, and $U_B$ is the magnetic field energy density.

For such a power-law electron energy distribution, the synchrotron emissivity as measured in the source rest frame (denoted by primes),
\begin{equation}
j'_{\nu'} = {1 \over 4 \pi} \, \int d \gamma \, P\!(\nu', \gamma) \,  \mathcal{N}_e\!(\gamma)  \, ,
\end{equation}
where $P(\nu', \gamma)$ is the spectral density of the synchrotron power emitted by a single electron, reads as
\begin{equation}
j_{\nu'} = a_j\!(p) {e^3 \over m_e c^2} \left({3 e \over 4 \pi m_e c}\right)^{(p-1)/2} N_0 B^{(p+1)/2} {\nu'}^{-(p-1)/2} \simeq  0.1 \, \left({3 e^7 \over 4 \pi m_e^3 c^5}\right)^{1/2} N_0 B^{3/2} {\nu'}^{-1/2} \, ,
\label{em}
\end{equation}
where
\begin{equation}
a_j\!(p) = {2^{(p-1)/2} \, \sqrt{3} \over 8 \sqrt{\pi} \, (p+1)} \, {\Gamma\left({3p - 1 \over 12}\right) \, \Gamma\left({3p + 19 \over 12}\right) \, \Gamma\left({p + 5 \over 4}\right) \over \Gamma\left({p +7 \over 4}\right)} \, ,
\end{equation}
and $a_j\!(2) \simeq 0.1$. Meanwhile, the SSA absorption coefficient \citep{Ghisellini1991},
\begin{equation}
\alpha'_{\nu'} = {c^2 \over 8 \pi m_e c^2 {\nu'}^2} \, \int d \gamma \, \gamma^2 \, P\!(\nu', \gamma) \, {d \over d \gamma}\!\left[ \gamma^{-2} \, \mathcal{N}_e\!(\gamma) \right] \, ,
\end{equation}
reads as
\begin{equation}
\alpha'_{\nu' } = a_{\alpha}\!(p) {e^3 \over m_e^2 c^2} \left({3 e \over 4 \pi m_e c}\right)^{p/2} N_0 \, B^{(p+2)/2} \, {\nu'}^{-(p+4)/2} \simeq 0.15 \, {3 e^4 \over 4 \pi m_e^3 c^3} N_0 B^2 {\nu'}^{-3} \, ,
\label{ab}
\end{equation}
where
\begin{equation}
a_{\alpha}\!(p) = {2^{p/2} \, \sqrt{3} \over 16 \sqrt{\pi}} \, {\Gamma\left({3p + 2 \over 12}\right) \, \Gamma\left({3p + 22 \over 12}\right) \, \Gamma\left({p + 6 \over 4}\right) \over \Gamma\left({p + 8 \over 4}\right)} \, ,
\end{equation}
and $a_{\alpha}\!(2) \simeq 0.15$. 

The source-frame peak frequency in the self-absorbed synchrotron spectrum, $\nu'_p$, corresponds to the SSA optical depth of the order of unity, 
\begin{equation}
\tau_{\nu'_p} \equiv \alpha'_{\nu'_p} \, R \simeq 1 \, ,
\label{tau}
\end{equation}
where $R$ is the linear size of the spherically symmetric (by assumption) source. Hence, from equation \ref{eq} and \ref{ab} it follows that:
\begin{equation}
B \simeq \left(\frac{320 \pi^2 \, m_e^4 c^5 \, {\nu'_p}^3}{3 e^4 \, \eta_{eq} \, R}\right)^{1/4} \sim 18 \, \eta_{eq}^{-1/4} (1+z)^{3/4} \left(\frac{R}{{\rm pc}}\right)^{-1/4} \left(\frac{\nu_{p}}{\rm GHz}\right)^{3/4} \,\, {\rm mG} \, ,
\label{equation:magnetic_field_eq}
\end{equation}
where $\nu_{p} = \nu'_p / (1+z)$ is the SSA-related peak frequency as measured in the observer frame. Note a very weak dependence of the resulting magnetic field intensity on the source linear size, $\propto R^{-1/4}$, and not particularly strong dependence on the observed peak frequency, $\propto {\nu_{p}}^{3/4}$. Also, the magnetic field intensity in the above expression does not depend explicitly on the observed flux, but instead on the equipartition ratio, and even this scaling is rather weak, namely $\propto \eta_{eq}^{-1/4}$. As a result, for astrophysical sources in approximate (by even orders of magnitude!) equipartition, $\eta_{eq} \sim 1$, with linear sizes between 1\,pc and 1\,kpc, and the SSA-related peak frequencies within the GHz range, the magnetic field intensity should always be of the order of $B \sim 10$\,mG.

At frequencies below the spectral peak, i.e. in the optically thick segment of the spectrum where $\tau_{\nu'} \gg 1$, the synchrotron specific intensity
\begin{equation}
 I'_{\nu'} = \frac{j'_{\nu'}}{\alpha'_{\nu'}} = a_I\!(p) \, \left({3 e \over 4 \pi m_{\rm e}^3 c }\right)^{-1/2} \, B^{-1/2} \, {\nu'}^{5/2} \quad,
\end{equation}
\noindent
where $a_I(p) \equiv a_j(p) / a_{\alpha}(p)$, and $a_I(2) \simeq 0.7$. Meanwhile, the observed flux density is
\begin{equation}
S_{\nu} = \int d\Omega \, I_{\nu} \quad,
\end{equation}
\noindent
where $d\Omega = dA / d_A^2$ for the angular diameter distance $d_A$. Introducing next the angular size of the source as $\theta = \sqrt{dA} / d_A$, and taking into account that
\begin{equation}
I_{\nu} = (1+z)^{-3} \, I'_{\nu'} \quad,
\end{equation}
\noindent
we finally obtain
\begin{equation}
S_{\nu} = (1+z)^{-3} \, \theta^2 \, I'_{\nu'} \quad.
\end{equation}
\noindent
Hence, for the optically thick segment of the synchrotron continuum and $p=2$, one has
\begin{equation}
S_{\nu} \sim 0.73 \, \left(1+z\right)^{-1/2} \, \left({\nu \over {\rm GHz}}\right)^{5/2} \, \left({B \over {\rm mG}}\right)^{-1/2} \, \left({\theta \over {\rm mas}}\right)^{2} \,\, {\rm [Jy]} \, ,
\label{eq:flux}
\end{equation}
which recovers the well-known formula. Note also that the brightness temperature, $T_B$, is defined through the relation 
\begin{equation}
T_B = {c^2 I'_{\nu'} \over 2 k \,  {\nu'}^2} \quad,
\end{equation}
\noindent
so that for a spherically symmetric source
\begin{equation}
    T_B \sim 1.38 \times 10^{12} \, (1+z) \, \left(\frac{S_{\nu}}{{\rm Jy}}\right) \, \left(\frac{\theta}{{\rm mas}}\right)^{-2} \, \left(\frac{\nu}{{\rm GHz}}\right)^{-2} \quad {\rm K} \, ,
\end{equation}

Finally, we note that the equipartition magnetic field as given in equation\,\ref{equation:magnetic_field_eq} with $\eta_{eq} = 1$, along with the flux density relation as given in equation\,\ref{eq:flux} for the peak frequency, $S_p \equiv S_{\nu_p}$, and also the physical source size $R = \theta \, d_A$, recovers the relation B3 from \citet{Readhead2021} for $p=2$, i.e. for the spectral index of the optically-thin segment of the synchrotron continuum $\alpha_{\rm thin} = -(p-1)/2=-0.5$, namely
\begin{equation}
    \theta_{eq} \sim 2.1 \, (1+z)^{7/17} \, \left(\frac{S_{p}}{{\rm Jy}}\right)^{8/17} \, \left(\frac{d_A}{{\rm Gpc}}\right)^{-1/17} \, \left(\frac{\nu_p}{{\rm GHz}}\right)^{-1} \quad {\rm mas} \, .
    \label{eq:eq_size}
\end{equation}

\newpage
\section{Flux density measurements}
Flux density measurements of our sources from our follow-up observations performed with GMRT (Table \ref{tableA2_lowfreq_flux}) and VLA (Table \ref{tableA1_vla_flux}). Additionally, Table \ref {tableA2_lowfreq_flux} includes flux densities from the LOFAR survey for four sources \citep{Shimwell}. It is important to note that all sources are unresolved at the angular resolution of GMRT and VLA. However, this is probably not the case for LOFAR observations (see discussion in Section \ref{sec:discussion}).

\begin{deluxetable}{c| c| c c c c}[h!]
\tabletypesize{\normalsize}
\tablecaption{Flux density measurements at low frequencies from LOFAR and GMRT.}
\tablehead{
 Name & \multicolumn{1}{c|}{LOFAR} & \multicolumn{4}{c}{GMRT} \\
 & \multicolumn{1}{c|}{144 MHz} & 402 MHz& 609 MHz& 648 MHz&687 MHz\\
   & \multicolumn{1}{c|}{[mJy]} & [mJy] & \multicolumn{1}{c}{[mJy]}& \multicolumn{1}{c}{[mJy]}& \multicolumn{1}{c}{[mJy]} }
\startdata
024345$-$28 & $-$& 20.91$\pm$2.97&$-$ & 37.83$\pm$2.53& $-$ \\    
024609$+$34 & $-$&        $-$       & $-$& 0.77$\pm$0.32 & $-$\\
031115$+$08 & $-$& 3.09$\pm$1.23 & $-$& 8.71$\pm$1.26 & $-$\\
053509$+$83 & $-$& 2.00$\pm$0.93 & $-$& 3.42$\pm$0.63 &$-$ \\
064001$+$28 & $-$&      $-$         & 25.47$\pm$2.25  & $-$& 26.72$\pm$2.28 \\  
071829$+$59 & $-$&         $-$      & $-$& 2.85$\pm$0.61 & $-$\\
095141$+$37 & 0.90$\pm$0.10& $-$ & $-$& 0.44$\pm$0.27 & $-$\\ 
101841$-$13 & $-$&       $-$        & $-$& 1.52$\pm$0.45 & $-$\\  
105035$-$07 & $-$&       $-$        & $-$& 1.73$\pm$0.31 & $-$\\ 
112940$+$39 & 1.60$\pm$0.20& $-$& $-$& $-$& $-$\\
114101$+$10 & $-$&        $-$       & $-$& 1.36$\pm$0.57 & $-$\\ 
121619$+$12 & $-$&       $-$        & $-$& 5.45$\pm$1.56 & $-$\\ 
130400$-$11 & $-$&        $-$       & $-$& 1.99$\pm$0.81 & $-$\\ 
150415$+$28 & 3.00$\pm$0.30$\dagger$& $-$ & $-$&        $-$       & $-$\\ 
155847$+$14 & $-$&        $-$       &$-$ & 0.88$\pm$0.33 & $-$\\ 
164607$+$42 & 8.80$\pm$0.70& $-$ & $-$&     $-$          & $-$\\ 
\enddata
\vspace{0.1 in}
\tablecomments{$\dagger$ - The value of the flux density refers to the central source visible on the LOFAR image (Figure \ref{figure_appendix_lofar}). The flux densities of the other sources in  150415$+$28 are as follows:  0.7$\pm$0.2 (east) and 2.8$\pm$0.4 (west).}
\label{tableA2_lowfreq_flux}
\end{deluxetable}

\movetabledown=50mm
\begin{rotatetable}
\begin{deluxetable}{c| c c| c c| c c c c| c c c c| c c} [ht!]
\tabletypesize{\scriptsize}
\tablecaption{VLA flux density measurements.}
\tablehead{
Name & \multicolumn{2}{l|}{L-band} & \multicolumn{2}{l|}{S-band} & \multicolumn{4}{l|}{C-band} & \multicolumn{4}{l|}{X-band} & \multicolumn{2}{l}{Ku-band}
}
\startdata
024345$-$28 & 1.26& 1.78& 2.50& 3.50& 4.49& 5.51& 6.49& 7.51& 8.48& 9.51& 10.49& 11.51& 13.50& 14.50\\
            &62.77$\pm$2.58 &63.65$\pm$2.34 &60.93$\pm$1.05 &56.55$\pm$1.63 &51.37$\pm$2.41 &47.52$\pm$3.00 &43.49$\pm$2.10 &40.09$\pm$3.36 &37.62$\pm$2.45 &35.30$\pm$2.15 &32.50$\pm$2.41 &30.70$\pm$2.95 &24.00$\pm$2.10 &22.60$\pm$1.80 \\
            \hline
024609$+$34 & & & 2.50& 3.50& 4.49& 5.51& 6.49& 7.51& 8.48& 9.51& 10.49& 11.51& & \\
            & & & 5.52$\pm$1.55& 9.97$\pm$1.38 & 13.53$\pm$0.96 & 14.64$\pm$ & 14.84$\pm$0.25 & 14.98$\pm$0.18 &14.86$\pm$0.14 & 14.51$\pm$0.13 & 14.24$\pm$0.18 & 13.72$\pm$0.22 & & \\
            \hline
031115$+$08 & 1.26& 1.78& 2.50& 3.50& 4.49& 5.51& 6.49& 7.51& 8.48& 9.51& 10.49& 11.51& 13.50& 14.50\\
            & 17.52$\pm$1.05& 21.13$\pm$0.69 & 20.49$\pm$0.24 & 19.00$\pm$0.39 & 17.90$\pm$0.21 &16.63$\pm$0.18 &15.85$\pm$0.15 & 15.00$\pm$0.31 & 14.19$\pm$0.13 & 13.37$\pm$0.12 &12.64$\pm$0.15 & 12.05$\pm$0.12 & 10.83$\pm$0.18 & 10.44$\pm$0.18\\
            \hline
053509$+$83 & 1.26& 1.78& 2.50& 3.50& 4.49& 5.51& 6.49& 7.51& 8.48& 9.51& 10.49& 11.51& & \\
            & 18.22$\pm$0.87& 24.93$\pm$0.48 & 27.60$\pm$0.17 & 26.81$\pm$0.17 & 25.44$\pm$0.35 & 24.27$\pm$0.39 & 23.35$\pm$0.34 & 22.59$\pm$0.27 & 21.40$\pm$1.10 & 21.30$\pm$1.00 & 19.70$\pm$1.10 & 19.60$\pm$1.20 & & \\
            \hline
064001$+$28 & 1.26& 1.78& 2.50& 3.50& 4.49& 5.51& 6.49& 7.51& 8.48& 9.51& 10.49& 11.51& & \\ 
            & 33.65$\pm$0.60& 28.74$\pm$0.45 & 24.05$\pm$0.27 &19.41$\pm$0.21 &15.88$\pm$0.19 &14.12$\pm$0.19&12.54$\pm$0.14 &11.25$\pm$0.14 &10.12$\pm$0.19 &9.24$\pm$0.13 & 8.47$\pm$0.16 &7.80$\pm$0.17 & & \\
            \hline
070837$+$32 & & 1.90& 2.69& 3.50& 4.49& 5.45& 6.49& 7.51& 8.48& 9.51& 10.49& 11.51& & \\ 
            & & 2.61$\pm$0.60&2.10$\pm$0.20 &1.97$\pm$0.11 &2.00$\pm$0.13 &2.12$\pm$0.14 &2.52$\pm$0.11 &2.79$\pm$0.11 &3.01$\pm$0.21 &3.24$\pm$0.09 &3.54$\pm$0.09 &3.71$\pm$0.10 & & \\
            \hline
071829$+$59 & 1.42& 1.78& 2.50& 3.50& 4.49& 5.51& 6.49& 7.51& 8.48& 9.51& 10.49& 11.51& & \\
            & 6.43$\pm$0.93&12.28$\pm$0.36 &21.56$\pm$0.23 &25.31$\pm$0.13 &24.73$\pm$0.17 &23.29$\pm$0.19 &21.10$\pm$0.14& 19.98$\pm$0.12 &19.16$\pm$0.17 &18.30$\pm$0.13 &17.57$\pm$0.17 &16.78$\pm$0.19 & & \\
            \hline
095141$+$37 & 1.26& 1.78& 2.50& 3.50& 4.49& 5.51& 6.49& 7.51& 8.48& 9.51& 10.49& 11.51& & \\ 
            & 2.79$\pm$1.23&5.39$\pm$0.75 &9.72$\pm$0.54 &12.84$\pm$0.25 &13.33$\pm$0.15 &12.94$\pm$0.18 &12.55$\pm$0.15 &11.88$\pm$0.12 &10.99$\pm$0.15 &10.41$\pm$0.15 &9.62$\pm$0.15 &9.17$\pm$0.20 & & \\
            \hline
101841$-$13 & & 1.61& 2.69& 3.30& 4.49& 5.51& 6.49& 7.51& 9.00& 11.00& & & & \\  
            & & 2.44$\pm$0.69&1.85$\pm$0.45 &1.67$\pm$0.30 &1.28$\pm$0.39 &1.02$\pm$0.21 &0.93$\pm$0.20 &0.87$\pm$0.22 &0.78$\pm$0.18 &0.75$\pm$0.18 & & & & \\
            \hline
105035$-$07 & & 1.87& 2.50& 3.37& 4.62& 5.51& 6.49& 7.51& 8.48& 9.51& 10.49& 11.51 & & \\ 
            & & 4.10$\pm$0.69&3.49$\pm$0.39 &3.31$\pm$0.33 &3.13$\pm$0.49 &3.39$\pm$0.45 &3.84$\pm$0.54 &4.22$\pm$0.46 &4.62$\pm$0.66 &5.00$\pm$0.78 &5.40$\pm$0.60 &6.27$\pm$0.75 & & \\
            \hline
110239$-$06 & 1.30& 1.90& 2.69& 3.31& 4.49& 5.51& 6.49& 7.51& 8.48& 9.51& 10.49& 11.51& & \\ 
            & 9.70$\pm$1.92&16.01$\pm$1.87 &28.77$\pm$2.85 &31.21$\pm$0.71 &33.69$\pm$0.30 &32.98$\pm$0.54 &30.84$\pm$0.51 &29.21$\pm$0.67 &27.20$\pm$0.77 &25.99$\pm$0.96 &23.48$\pm$1.35 &22.22$\pm$1.44 & & \\
            \hline
112940$+$39 & & 1.87& 2.50& 3.50& 4.49& 5.51& 6.49& 7.51& 8.48& 9.51& 10.49& 11.51& & \\
            & & 6.15$\pm$0.60&10.43$\pm$0.30 &14.04$\pm$0.20 &15.11$\pm$0.17 &15.19$\pm$0.13 &14.46$\pm$0.15 &14.20$\pm$0.15 &14.06$\pm$0.16 &13.94$\pm$0.12 &13.46$\pm$0.12 &13.28$\pm$0.15 & & \\
            \hline
114101$+$10 & 1.26& 1.78& 2.50& 3.50& 4.49& 5.51& 6.49& 7.51& 8.48& 9.51& 10.49& 11.51& & \\ 
            & 5.21$\pm$1.77&7.97$\pm$0.91 &9.91$\pm$0.60 &9.61$\pm$0.36 &8.83$\pm$0.20 &8.04$\pm$0.16 &7.00$\pm$0.16 &6.09$\pm$0.25 &5.28$\pm$0.25 &4.74$\pm$0.22 &4.48$\pm$0.36 &3.72$\pm$0.26 & & \\
            \hline
121619$+$12 & 1.42& 1.84& 2.50& 3.43& 4.55& 5.44& 6.49& 7.51& 8.48& 9.51& 10.55& 11.44& & \\ 
            & 12.83$\pm$0.95&12.47$\pm$0.75 &11.16$\pm$0.30 &10.34$\pm$0.16 &9.16$\pm$0.27 &8.77$\pm$0.28 &7.32$\pm$0.10 &6.68$\pm$0.29 &6.14$\pm$0.09 &5.73$\pm$0.07 &5.34$\pm$0.09 &5.24$\pm$0.22 & & \\
            \hline
130400$-$11 & & 1.87& 3.00& & 4.49& 5.51& 6.49& 7.51& 8.48& 9.51& 10.49& 11.51& & \\ 
            & & 6.39$\pm$1.59&8.11$\pm$0.84 & &7.97$\pm$0.23 &7.69$\pm$0.16 &6.82$\pm$0.11 &6.15$\pm$0.23 &5.68$\pm$0.12 &5.20$\pm$0.08 &4.66$\pm$0.09 &4.28$\pm$0.11 & & \\
            \hline
150415$+$28 & & & 2.80& 3.50& & & 6.49 & 7.38 & 8.48& 9.51& 10.49& 11.51& & \\ 
            & & & 8.25$\pm$1.42&8.55$\pm$1.23 & & &6.19$\pm$0.26 &5.86$\pm$0.10 &5.38$\pm$0.12 &5.04$\pm$0.10 &4.54$\pm$0.07 &4.33$\pm$0.07 & & \\
            \hline
155847$+$14 & 1.39& 1.78& 2.50& 3.50& 4.49& 5.51& 6.49& 7.51& 8.48& 9.51& 10.49& 11.51& & \\ 
            & 7.74$\pm$1.55&16.45$\pm$0.39 &29.81$\pm$0.54 &31.25$\pm$0.51 &29.21$\pm$0.36 &26.65$\pm$0.21 &23.72$\pm$0.15 &20.97$\pm$0.36 &18.73$\pm$0.13 &16.88$\pm$0.12 &15.00$\pm$0.09 &13.57$\pm$0.18 & & \\
            \hline
164607$+$42 & 1.30& 1.78& 2.50& 3.50& 4.49& 5.51& 6.49& 7.51& 8.48& 9.51& 10.49& 11.51& & \\ 
            & 9.12$\pm$0.54&8.82$\pm$0.51 &7.02$\pm$0.61 &5.82$\pm$0.65 &4.84$\pm$0.50 &4.21$\pm$0.12 &3.72$\pm$0.09 &3.31$\pm$0.09 &2.99$\pm$0.13 &2.78$\pm$0.12 &2.52$\pm$0.07 &2.27$\pm$0.09 & & \\
            \hline
180940$+$24 & & 1.94& 2.50& 3.50& 4.49& 5.51& 6.49& 7.51& 8.48& 9.51& 10.49& 11.51& & \\
            & & 6.84$\pm$1.24&5.50$\pm$0.63 &4.49$\pm$0.66 &3.81$\pm$0.17 &3.21$\pm$0.17 &2.83$\pm$0.16 &2.43$\pm$0.17 &2.22$\pm$0.20 &2.08$\pm$0.20 &1.97$\pm$0.07 &1.80$\pm$0.10 & & \\
            \hline
183415$+$61 & & & 2.50& 3.50& 4.49& 5.51& 6.49& 7.51& 8.48& 9.51& 10.49& 11.51& & \\
            & & & 10.15$\pm$2.40&12.98$\pm$1.89& 13.98$\pm$0.90 &14.44$\pm$0.19 &14.54$\pm$0.15 &14.40$\pm$0.90 &14.49$\pm$1.46 &14.06$\pm$1.42 &13.73$\pm$1.40 &13.36$\pm$1.35 & & \\
            \hline
195335$-$04 & 1.39& 1.81& 2.69& 3.31& 4.49& 5.51& 6.49& 7.51& 8.48& 9.51& 10.49& 11.51& & \\ 
            & 11.56$\pm$3.11&13.17$\pm$1.53 &14.38$\pm$2.16 &14.95$\pm$2.24 &12.81$\pm$1.05 &11.40$\pm$0.66 &9.59$\pm$0.22 &8.53$\pm$0.10 &7.68$\pm$0.13 &7.18$\pm$0.12 &6.04$\pm$0.10 &5.69$\pm$0.12 & & \\
            \hline
203909$-$30 & 1.39& 1.84& 2.50& 3.50& 4.49& 5.51& 6.49& 7.51& 8.48& 9.51& 10.49& 11.51& 13.50& 14.50\\
            & 9.55$\pm$1.82&9.01$\pm$1.04 &7.53$\pm$0.66 &6.51$\pm$0.63 &5.73$\pm$0.39 &5.03$\pm$0.36 &4.21$\pm$0.55 &3.60$\pm$0.51 &3.11$\pm$0.36 &2.57$\pm$0.30 &2.28$\pm$0.21 &2.08$\pm$0.27&1.65$\pm$0.21 &1.17$\pm$0.19 \\
            \hline
223933$-$22 & 1.36& 1.78& 2.50& 3.50& 4.49& 5.51& 6.49& 7.51& 8.48& 9.51& 10.49& 11.51& & \\  
            & 11.68$\pm$2.70&12.45$\pm$1.21 &13.86$\pm$0.39 &15.79$\pm$0.45 &18.50$\pm$0.31 &20.46$\pm$0.31 &21.58$\pm$0.36 &21.74$\pm$0.30 &20.64$\pm$0.81 &19.78$\pm$1.17 &18.88$\pm$0.91 &17.85$\pm$2.01 & & \\
            \hline
233058$+$10 & 1.39& 1.81& 2.69& 3.31& 4.49& 5.51& 6.49& 7.51& 8.48& 9.51& 10.49& 11.51& & \\
            & 4.63$\pm$0.87&4.94$\pm$0.69 &6.14$\pm$1.81 &3.96$\pm$1.82 &2.94$\pm$0.72 &2.36$\pm$0.45 &1.86$\pm$0.13 &1.47$\pm$0.09 &1.27$\pm$0.07 &1.05$\pm$0.08 &1.00$\pm$0.09 &0.78$\pm$0.06 & & \\            
\enddata
\vspace{0.1 in}
\tablecomments{There are two rows of values for each source. The first row is the frequency in GHz and the second is the flux density in mJy.}
\label{tableA1_vla_flux}
\end{deluxetable}
\end{rotatetable}

\clearpage
\section{Broadband radio spectra}

\begin{figure}[hbt!]
   \centering
   \includegraphics[scale=0.28]{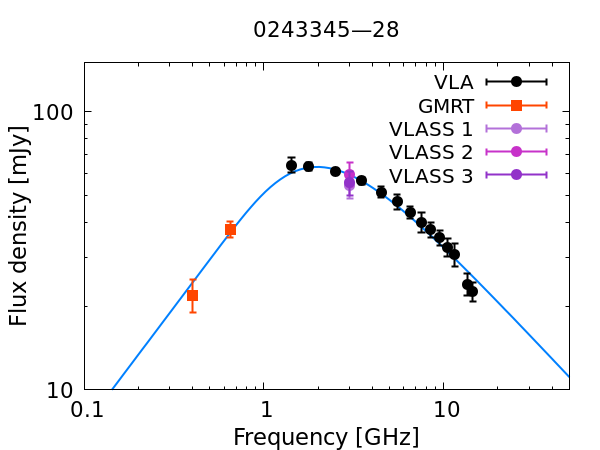}
   \includegraphics[scale=0.28]{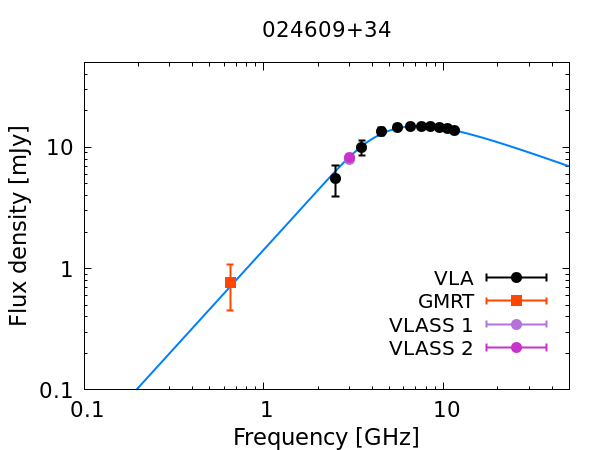}
   \includegraphics[scale=0.28]{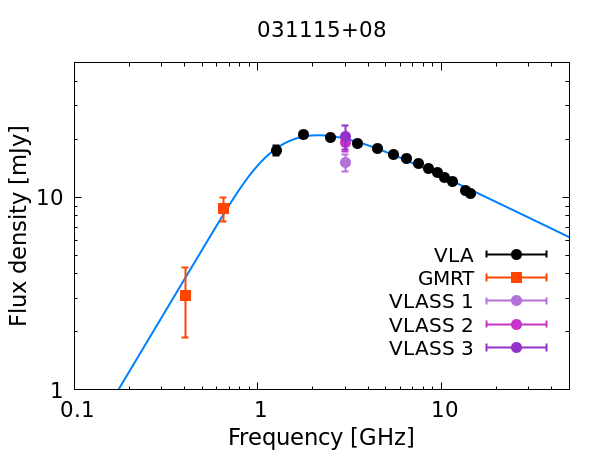}
   \includegraphics[scale=0.28]{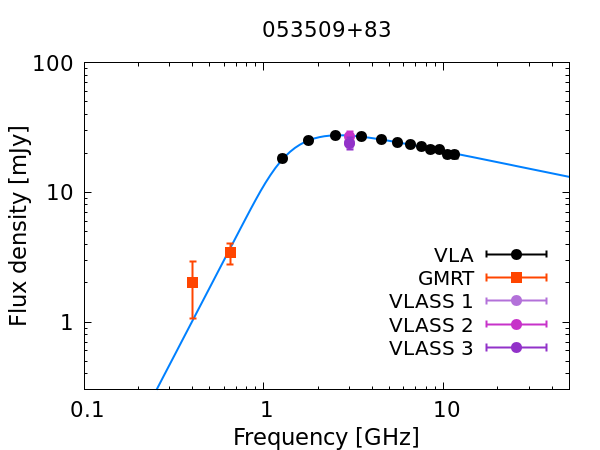}
   \includegraphics[scale=0.28]{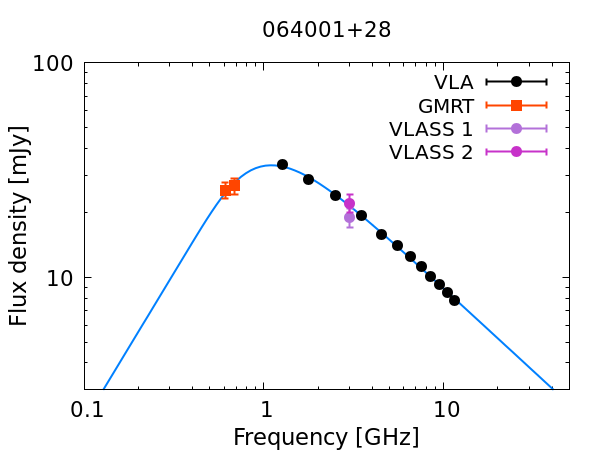}
   \includegraphics[scale=0.28]{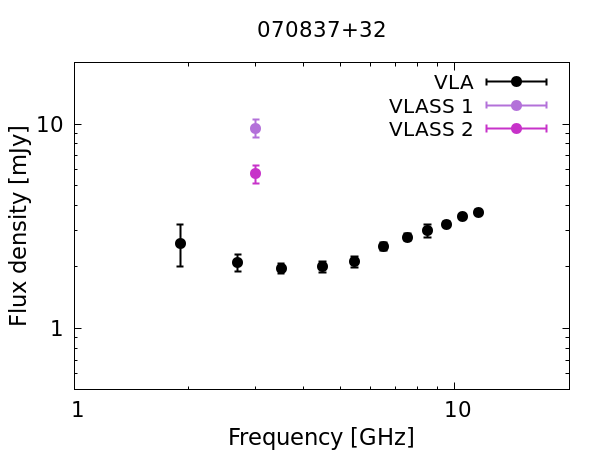}
   \includegraphics[scale=0.28]{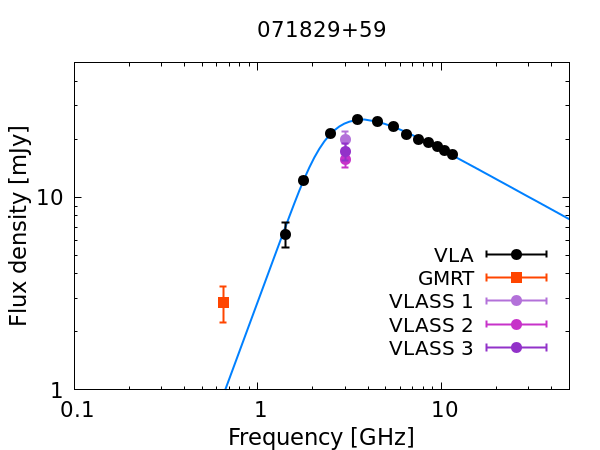}
   \includegraphics[scale=0.28]{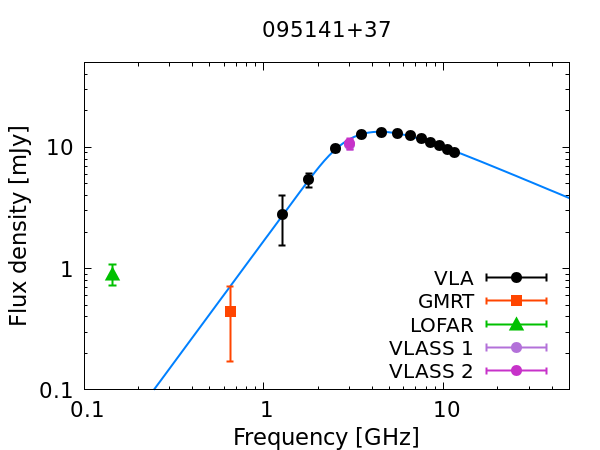}
   \includegraphics[scale=0.28]{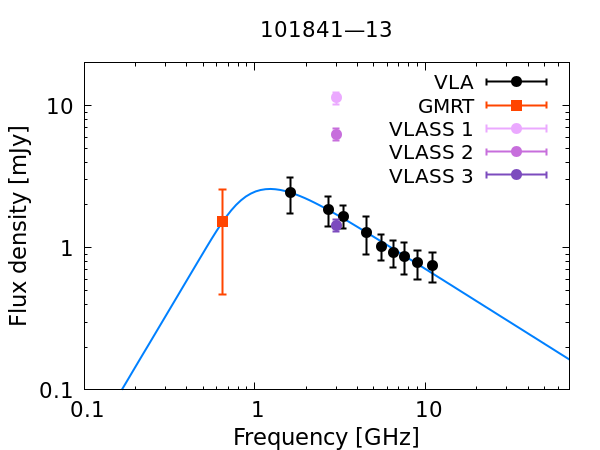}
   \includegraphics[scale=0.28]{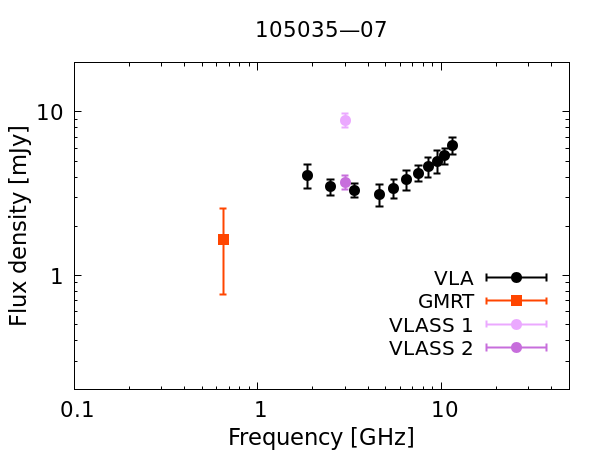}
   \includegraphics[scale=0.28]{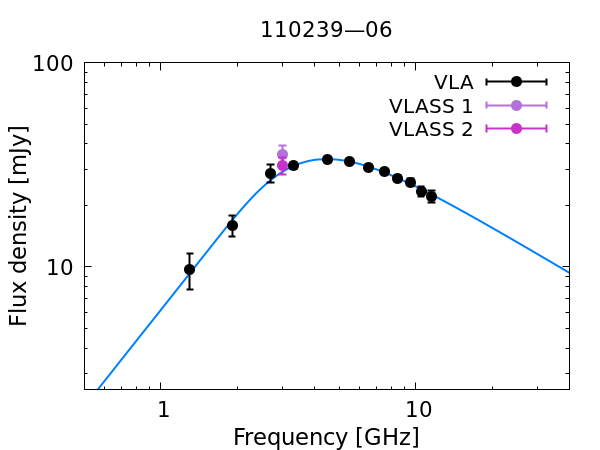}
   \includegraphics[scale=0.28]{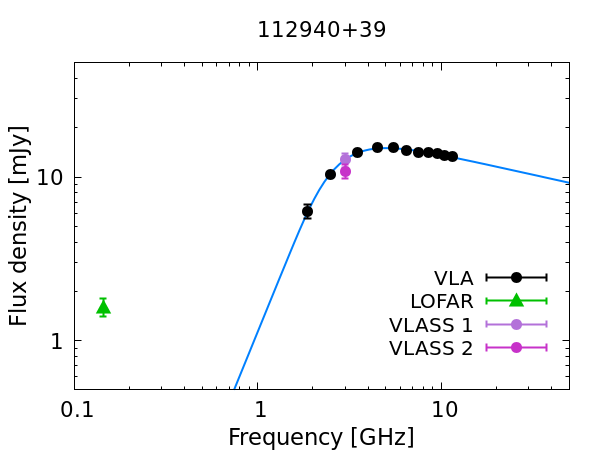}
    \end{figure}

   \begin{figure}[h!]
   \centering  
   \includegraphics[scale=0.28]{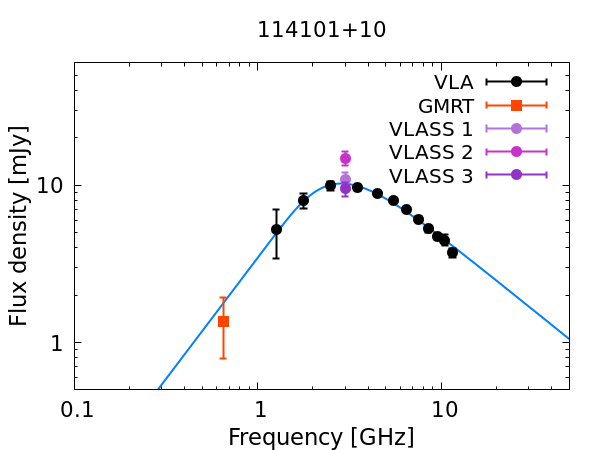}
   \includegraphics[scale=0.28]{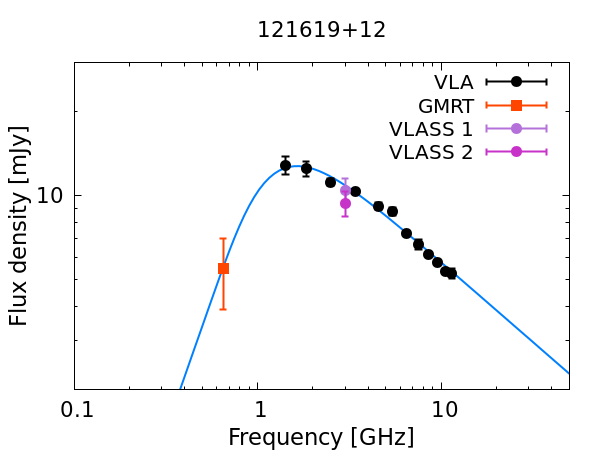}
   \includegraphics[scale=0.28]{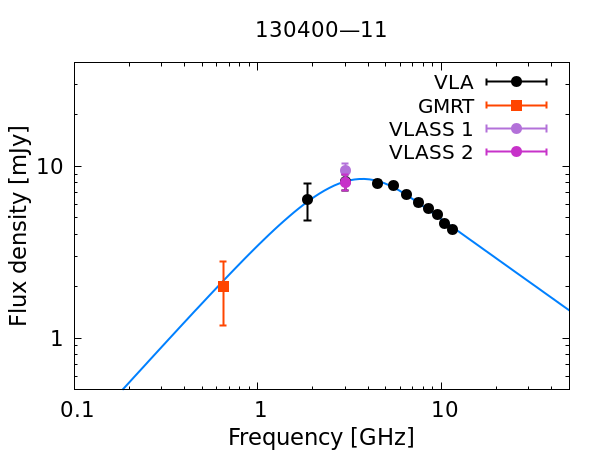}
   \includegraphics[scale=0.28]{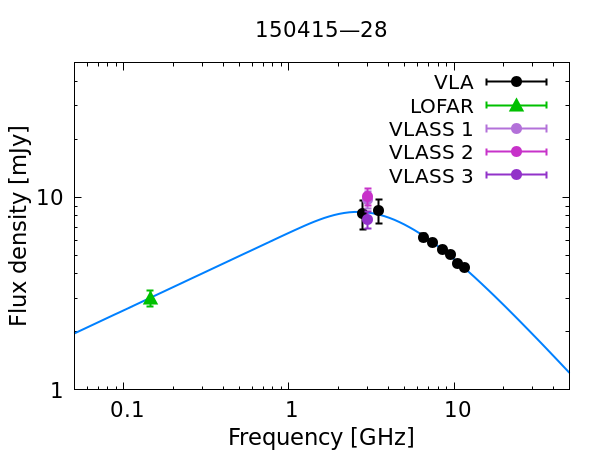}
   \includegraphics[scale=0.28]{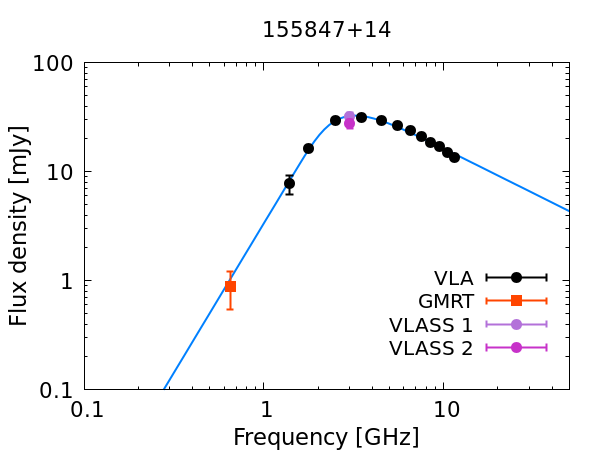}
   \includegraphics[scale=0.28]{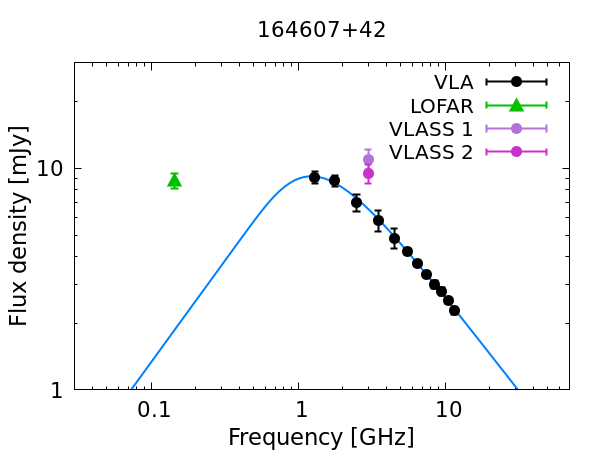}
   \includegraphics[scale=0.28]{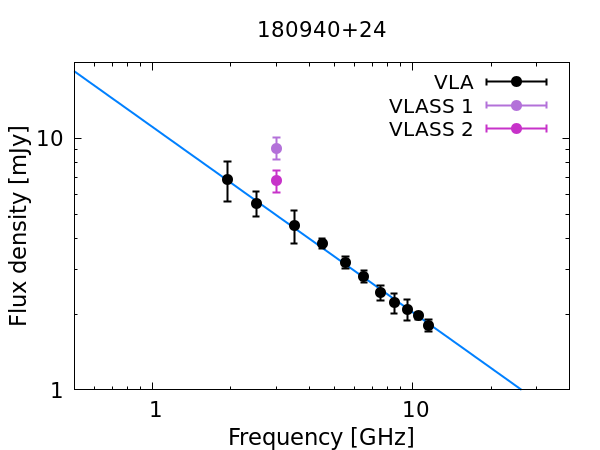}
   \includegraphics[scale=0.28]{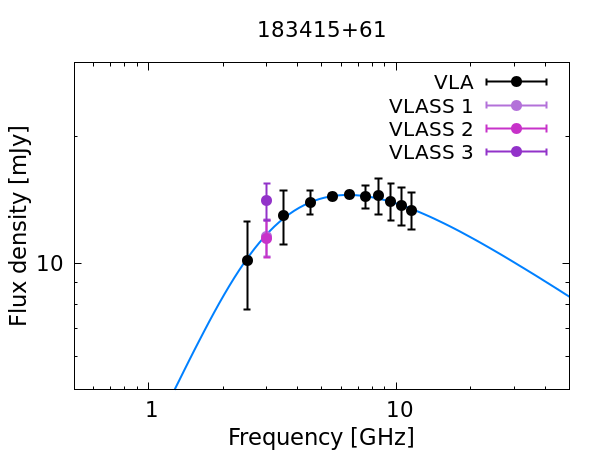}
   \includegraphics[scale=0.28]{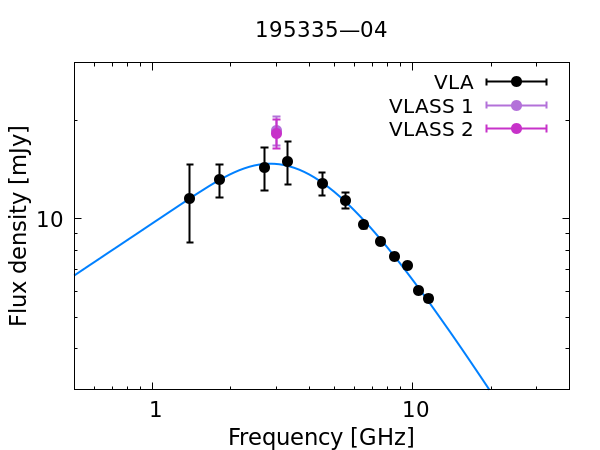}
   \includegraphics[scale=0.28]{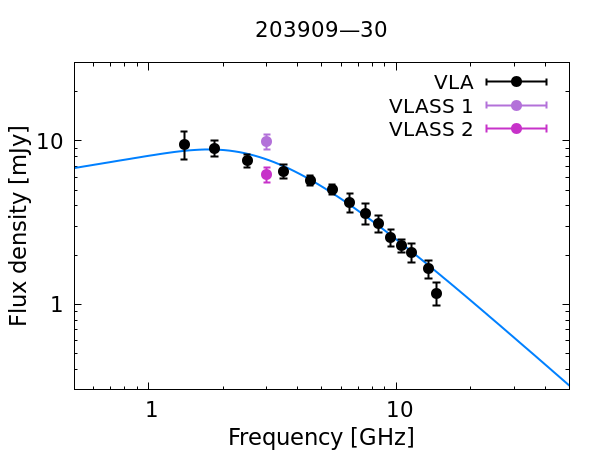}
   \includegraphics[scale=0.28]{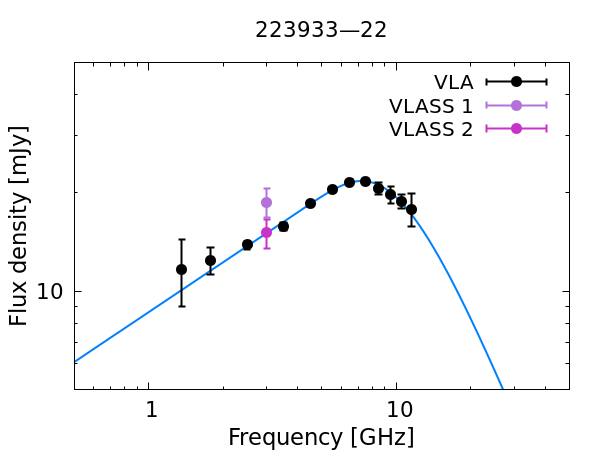}
   \includegraphics[scale=0.28]{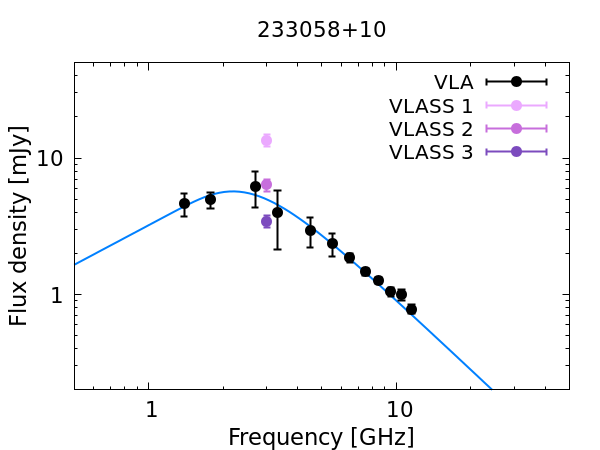}
\caption{The radio SEDs between 144 MHz and 16.9 GHz. The solid line shows the best-fit to the data with the modified power-law model (Equation \ref{equation_powerlaw}).}
 \label{figure_spectral_modeling}
    \end{figure}

\clearpage
\section{LOFAR radio images}
\begin{figure*}[h!]
\centering  
\includegraphics[scale=0.35]{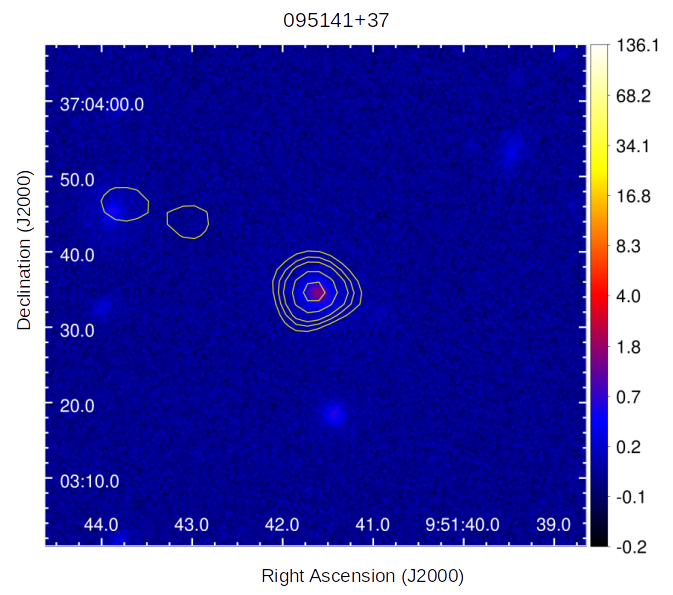}
\includegraphics[scale=0.35]{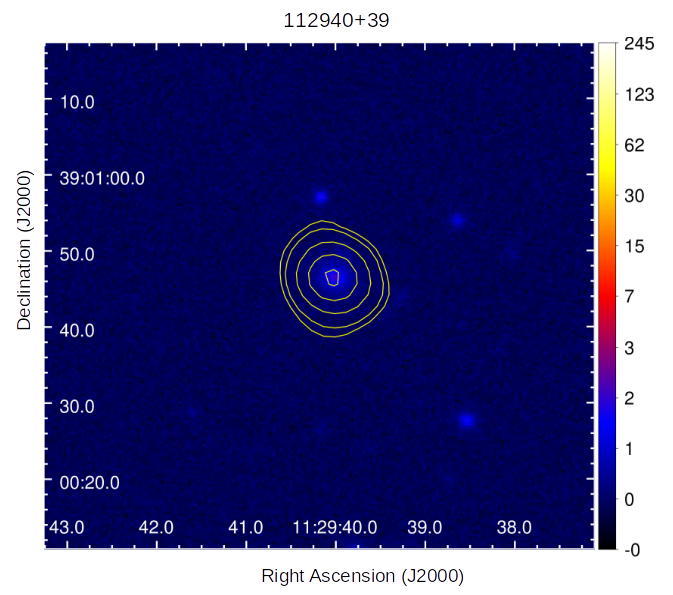}
\includegraphics[scale=0.35]{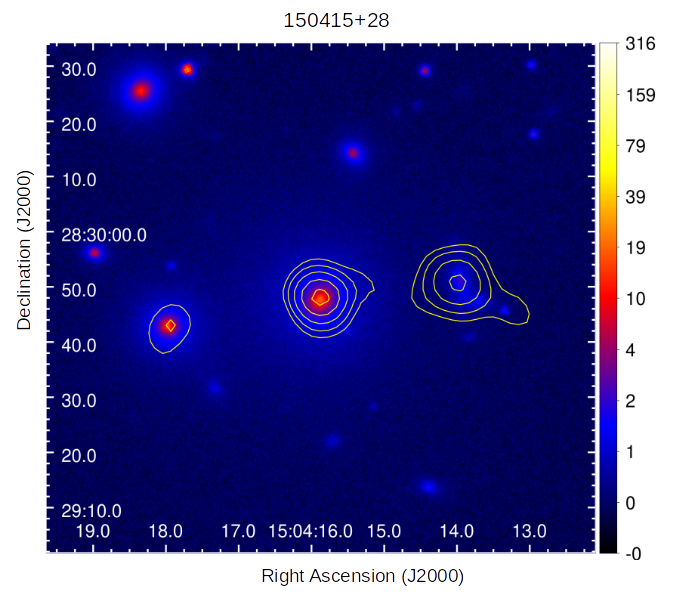}
\includegraphics[scale=0.35]{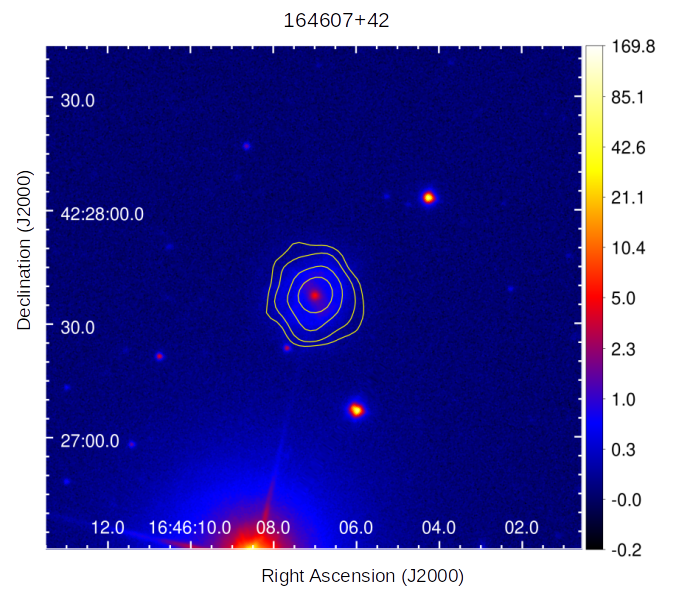}
\caption{The LOFAR 144 MHz images (yellow contours) overlaid on the SDSS i-band images for four objects. The presented contours have the following values: $\rm 1.5, 2.0, 2.5, 3.5, 4.5 \times 10^{-4} Jy\, beam^{-1}$ (source 095141$+$37); $\rm 1.5, 2.0, 3.5, 5.5, 7.5 \times 10^{-4} Jy\, beam^{-1}$ (source 112940$+$39); $\rm 2.4, 4.0, 6.5, 10.0, 14.0 \times 10^{-4} Jy\, beam^{-1}$ (source 150415$+$28); $\rm 2.5, 3.5, 5.5, 7.5 \times 10^{-4} Jy\, beam^{-1}$ (source 164607$+$42) .}  
\label{figure_appendix_lofar}
    \end{figure*}

Figure \ref{figure_appendix_lofar} shows the radio images at 144 MHz for four sources observed with LOFAR \citep{Shimwell}.

\clearpage
\section{Measurements and calculations made on the basis of VLBA observations}
Table \ref{table_vlba} presents measurements and calculations made on the basis of VLBA observations at the frequency of 8.7 GHz.

\begin{deluxetable}{l c r c c c c}[h!]
\tabletypesize{\footnotesize}
\tablecaption{Results of VLBA observations.}
\tablehead{
\multicolumn{1}{c}{Name} & \multicolumn{1}{c}{$\rm \theta_{maj} \times \theta_{min}$} & \multicolumn{1}{c}{$\rm S_{8.7\,GHz}$} & \multicolumn{1}{c}{$\rm T_B$}& Type & $\rm B_{eq}$& $\rm EV$\\
      & \multicolumn{1}{c}{[$\rm mas \times mas$]} & \multicolumn{1}{c}{[mJy]}   & \multicolumn{1}{c}{[K]}& &[mG]& [{\it c}]\\
\multicolumn{1}{c}{(1)} & \multicolumn{1}{c}{(2)} & \multicolumn{1}{c}{(3)}& \multicolumn{1}{c}{(4)}& (5)& (6) & (7)  
}
\startdata
024345$-$28 N & 3.6 $\times$ 1.0 & 19.6 ± 1.0 & 1.1 $\times 10^8 $& D & 22* & 0.76\\
\multicolumn{1}{r}{S}& 2.8 $\times$ 0.9 & 13.1 ± 0.7 & 1.0 $\times 10^8$& &$-$&$-$\\
024609$+$34 & 5.4 $\times$ 0.8  & 14.8 ± 0.7 & 7.1 $\times 10^7$& E & 37 & 1.47\\
031115$+$08 & 2.7 $\times$ 0.5 & 12.9 ± 0.6 & 1.7 $\times 10^8$& S & 20 & 0.30\\
053509$+$83 C& 3.3 $\times$ 0.3 & 7.1 ± 0.4 & 1.3 $\times 10^8$& Cj & 16 & 1.75\\
064001$+$28 C& 2.7 $\times$ 0.4 & 5.8 ± 0.3 & 1.1 $\times 10^8$& Cj & 19* & 0.16\\
\multicolumn{1}{r}{W} & 3.6 $\times$ 0.5 & 4.0 ± 0.2 & 4.0 $\times 10^7$& &$-$ &$-$\\
070837$+$32 & 4.9 $\times$ 0.3 & 5.9 ± 0.3 & 8.5 $\times 10^7$& E & $-$ & 2.39\\
071829$+$59 & 2.3 $\times$ 0.6 & 17.4 ± 0.9 & 3.1 $\times 10^8$& S & 26 & 1.37\\
095141$+$37 & 1.5 $\times$ 0.5 & 11.1 ± 0.6 & 3.2 $\times 10^8$& S & 34 & 0.70\\
101841$-$13 & 4.8 $\times$ 0.9 & 0.6 ± 0.1 & 2.6 $\times 10^6$& S & 14 & 0.36\\
105035$-$07 & 1.9 $\times$ 0.5 & 3.0 ± 0.2 & 6.2 $\times 10^7$& S & $-$ & 0.68\\
110239$-$06 N& 3.8 $\times$ 0.4 & 10.8 ± 1.1 & 3.1 $\times 10^8$& T & $-$ & 8.94\\
\multicolumn{1}{r}{C}& 3.8 $\times$ 0.5 & 11.5 ± 2.6 & 2.3 $\times 10^8$& &40* &$-$\\
\multicolumn{1}{r}{S}& 3.7 $\times$ 0.4 & 11.1 ± 3.3 & 2.9 $\times 10^8$& &$-$&$-$\\
112940$+$39 & 0.4 $\times$ 0.4 & 12.3 ± 0.6 & 2.2 $\times 10^9$& S & 41 & 0.20\\
114101$+$10 C& 3.3 $\times$ 1.0 & 4.2 ± 0.4 & 2.6 $\times 10^7$& Cj & 28 & 1.09\\
121619$+$12 & 2.1 $\times$ 0.5 & 5.2 ± 0.5 & 1.0 $\times 10^8$& S & 18 & 0.37\\
130400$-$11 & 3.6 $\times$ 0.4 & 6.7 ± 0.7 & 1.1 $\times 10^8$& S & 31 & 1.10\\
150415$+$28 & 4.2 $\times$ 0.9 & 7.0 ± 0.7 & 3.6 $\times 10^7$& E & 41 & 0.59\\
155847$+$14 & 3.4 $\times$ 0.5 & 20.6 ± 2.1 & 2.2 $\times 10^8$& S & 31 & 0.37\\
164607$+$42 & 0.9 $\times$ 0.3 & 3.5 ± 0.4 & 2.7 $\times 10^8$& S & 25 & 0.11\\
180940$+$24 & 3.1 $\times$ 1.1 & 1.7 ± 0.2 & 9.3 $\times 10^6$& S & $-$ & 0.13\\
183415$+$61 & 2.6 $\times$ 0.3 & 12.7 ± 1.3 & 3.6 $\times 10^8$& S & 36 & 0.17\\
195335$-$04 & 1.2 $\times$ 0.3 & 7.9 ± 0.8 & 4.2 $\times 10^8$& S & 54 & 0.13\\
203909$-$30 & 8.3 $\times$ 0.8 & 4.0 ± 0.4 & 1.2 $\times 10^7$& E & $-$ & 0.93\\
223933$-$22 E& 3.5 $\times$ 1.3 & 10.0 ± 1.0 & 4.5 $\times 10^7$& D & 71* & 0.78\\
\multicolumn{1}{r}{W} & 3.5 $\times$ 1.3 & 8.8 ± 0.9 & 3.8 $\times 10^7$& &$-$&$-$\\
233058$+$10 & 1.8 $\times$ 0.5 & 1.7 ± 0.1 & 3.4 $\times 10^7$& S & 36& 0.18\\
\enddata
\vspace{0.1 in}
\tablecomments{The columns are marked as follows: (1) source name and component markings: C - central and the brightest component, N - northern, S - southern, E - eastern, W - western; (2) deconvolved major and minor axis of the component; (3) the ﬂux density; (4) the brightness temperature; (5) radio morphology classification: Cj - Core-jet, D - double-lobed, E - extended structure, S - single component, T - triple; (6) equipartition magnetic field, * - in the case of more than one component the calculation was made for the brightest one; (7) lower limit on the expansion velocity given in speeds of light.}
\label{table_vlba}
\end{deluxetable}

\newpage
\section{Flux density measurements of 064001$+$28 components}
\label{sec:0640}

\begin{deluxetable}{c|r|l c c c |c}[h!]
\tabletypesize{\normalsize}
\tablecaption{Flux density measurements of the potential B and C components of 064001$+$28.}
\tablehead{
 Name & \multicolumn{1}{c|}{GMRT} & \multicolumn{4}{c|}{VLA}&\multicolumn{1}{c}{$\alpha_{thin}$}\\
 & \multicolumn{1}{c|}{648 MHz} & \multicolumn{1}{c}{1.5 GHz}& 3 GHz& 6 GHz& 10 GHz & \\
  & \multicolumn{1}{c|}{[mJy]} & \multicolumn{1}{c}{[mJy]} & \multicolumn{1}{c}{[mJy]}& \multicolumn{1}{c}{[mJy]}& \multicolumn{1}{c|}{[mJy]} & }
\startdata
B1 & 2.49$\pm$0.58 & 1.83$\pm$0.73 & 1.48$\pm$0.35 & 1.05$\pm$0.31 & 0.61$\pm$0.22 & -0.46$\pm$0.06\\
B2 & 9.90$\pm$0.62 & 3.36$\pm$1.70 &1.28$\pm$0.49& 0.26$\pm$0.08 & $-$ & -1.59$\pm$0.09\\
B3 & 7.24$\pm$0.93 &2.08$\pm$1.23 &0.96$\pm$0.23& 0.18$\pm$0.06 & $-$ & -1.57$\pm$0.13\\
C  & 10.74$\pm$0.64 &5.07$\pm$2.21&2.03$\pm$0.84&0.52$\pm$0.19 & $-$ & -1.30$\pm$0.10\\
\enddata
\vspace{0.1 in}
\tablecomments{The value of $\alpha_{thin}$ is the result of fitting a standard non-thermal power-law model of the form $S_{\nu} = a \nu^{\alpha_{thin}}$ to the measurements.}
\label{table_0640_comps_flux}
\end{deluxetable}

\section{Optical spectroscopy of 101841$-$13}
\label{sec:1018}

101841$-$13 was observed with the Palomar 200-inch Hale Telescope using Double Spectrograph (DBSP) on UT 2022 February 27. The on-source exposure times was $2 \times 300$~s and the observations were obtained through a slit with a width of 1.5~\arcsec, which was comparable to the seeing at the time. The night was not photometric, and data reductions followed standard procedures using IRAF.

\begin{figure*}[t!]
\centering  
\includegraphics[scale=0.18]{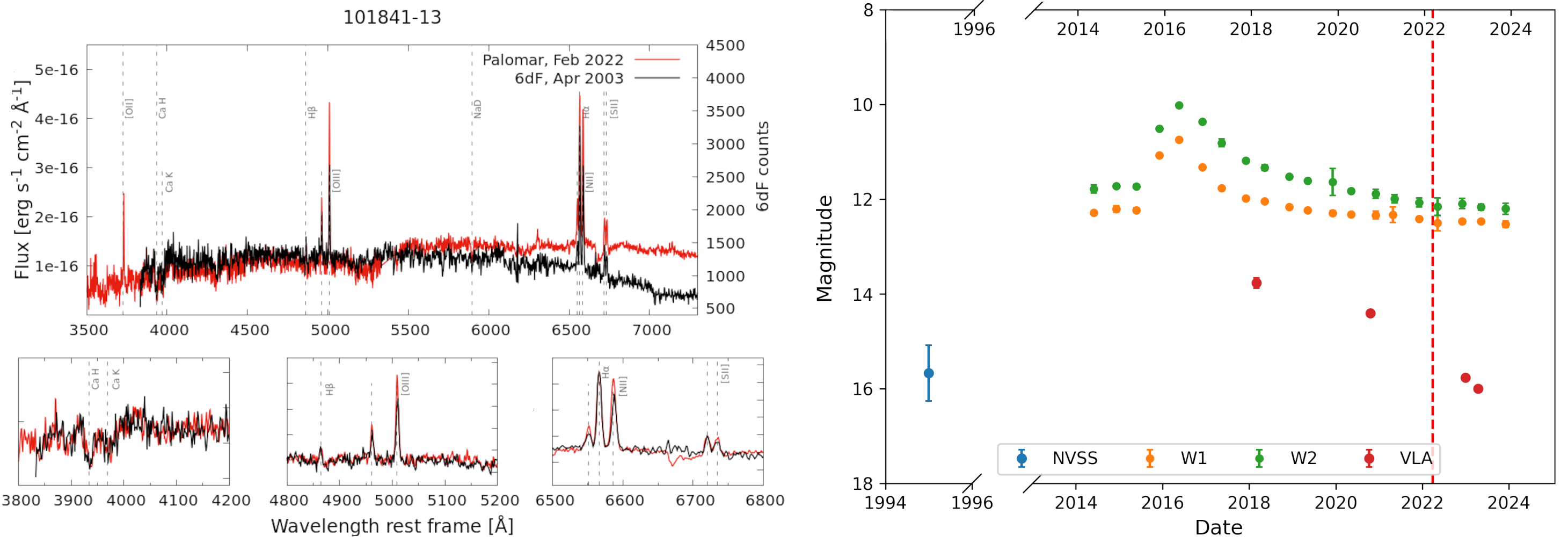}
\caption{(Left): Palomar (red) and archival 6dF (black) optical spectra of 101841$-$13 with strong emission lines indicated. The left axis shows the flux obtained from the (non-photometric) Palomar observations, and the right axis shows the uncalibrated flux in counts from the 6dF observations. Small panels below zoom in on key absorption and emission line regions:  Ca K and H lines (left), H$\beta$ and [OIII] lines (middle), and [NII], H$\alpha$ and [SII] lines (right).  Scaling of the 6dF spectrum was chosen as to fit the profiles of absorption and Balmer lines in Palomar spectrum. (Right): Radio 3~GHz light curve (red points), WISE W1 (yellow) and W2 (green) light curves and NVSS 1.4~GHz emission (blue point). The vertical red dashed line indicates the date of the Palomar spectroscopic observation.}  
\label{figure_spec_1018}
    \end{figure*}

We used the software package STARLIGHT\footnote{\url{www.starlight.ufsc.br}} to ﬁt the galaxy continuum, measure the stellar velocity dispersion, and measure emission line properties. STARLIGHT is an inverse stellar population synthesis code that recovers the stellar population of a galaxy by fitting a pixel-by-pixel model to the observed spectrum. The spectrum was corrected for Galactic extinction by adopting a \citet{cardelli1989} extinction law with $\rm R = 3.1$ and $A_V = 0.241$ mag. The intensities of the emission lines were measured by Gaussian fitting after subtracting the modeled stellar spectrum.
The optical spectrum of 101841$-$13 is presented in Figure \ref{figure_spec_1018} in red. The figure also includes the archival 2003 spectrum of this galaxy from the 6dF survey in black \citep{Jones2004, Jones2009}. Figure \ref{figure_spec_1018} also shows the radio and WISE infrared light curves with the date of our optical spectroscopic observation indicated.

\begin{figure*}[t!]
\centering  
\includegraphics[scale=0.29]{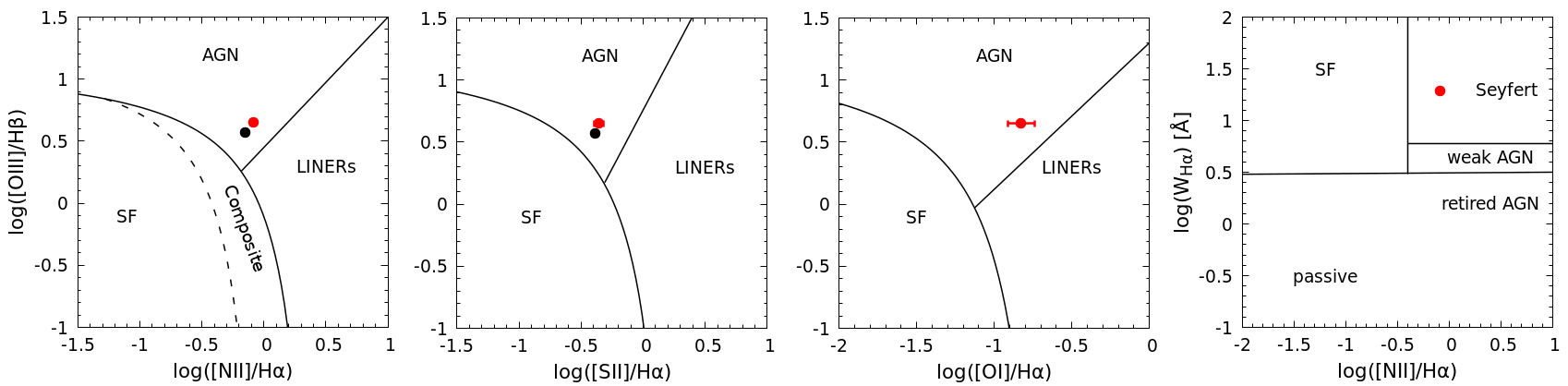}
\caption{BPT and WHAN diagrams \citep{Baldwin, Kewley2006, CidFernandes2011} for 101841$-$13. The red circle indicates values calculated based on measurements from the Palomar spectrum, and the black circle indicates upper limits estimated based on the 6dF spectrum.}  
\label{figure_bpt_1018}
    \end{figure*}

101841$-$13 does not show broad emission lines in either of these spectra. The narrow emission line flux ratios from the Palomar spectrum are as follows: $\rm log([O\,III]/H\beta) = 0.649\pm0.009$, $\rm log([N\,II]/H\alpha) = −0.086\pm0.018$, $\rm log([S\,II]/H\alpha) = -0.357\pm0.035$, $\rm log([O\,I]/H\alpha) = -0.827\pm0.046$ and $\rm log(W_{H\alpha})= 1.292\pm0.002$. Note that none of these values should be strongly affected by the non-photometric conditions of the observations. Ultimately, these values indicate accretion-driven photoionization. Therefore, based on the BPT and WHAN diagrams \citep{Baldwin, Kewley2006, CidFernandes2011}, 101841$-$13 can be classified as a Seyfert 2 galaxy (Figure \ref{figure_bpt_1018}).
The stellar velocity dispersion measured from the Palomar spectrum is $\rm \sigma_{*} = 104.3\pm5.7 km\,s^{-1}$. 
Using this value and the local $\rm M_{BH} - \sigma_{*}$ scaling relation \citep{Kormendy}, we estimate the black hole mass to be $\rm 2.7\pm0.7\times10^7\, M_{\odot}$. 
The 6dF spectrum of 101841$-$13 is not flux calibrated, so its analysis using the STARLIGHT code is not possible. Yet, based on the recorded counts, we estimate the line ratios from this spectrum as follows: $\rm log([O\,III]/H\beta) < 0.8$, $\rm log([N\,II]/H\alpha) < −0.15$, $\rm log([S\,II]/H\alpha) < -0.38$.  These values should be robust given the relatively stable sensitivity function over the small wavelength intervals for these pairs of lines. Next, we plot the line ratios of 101841$-$13 on the BPT diagrams (Figure \ref{figure_bpt_1018}), showing that the sources is clearly located in the area occupied by AGNs. However, since these estimates do not take into account the impact of extinction and stellar absorption in the Balmer lines, they should be treated with caution and as upper limits, though, based on the Palomar spectrum, we speculate that including these caveats would be unlikely to shift the source into the region of composite galaxies. 
We therefore infer that the clearly visible [O\,III], [N\,II] and [S\,II] lines in the 6dF spectrum of 101841$-$13 indicate the presence of underlying accretion already in the period before the radio and infrared brightening of the source.

\newpage
\section{High-frequency radio images}
\begin{figure*}[h!]
   \centering
   \includegraphics[scale=0.81]{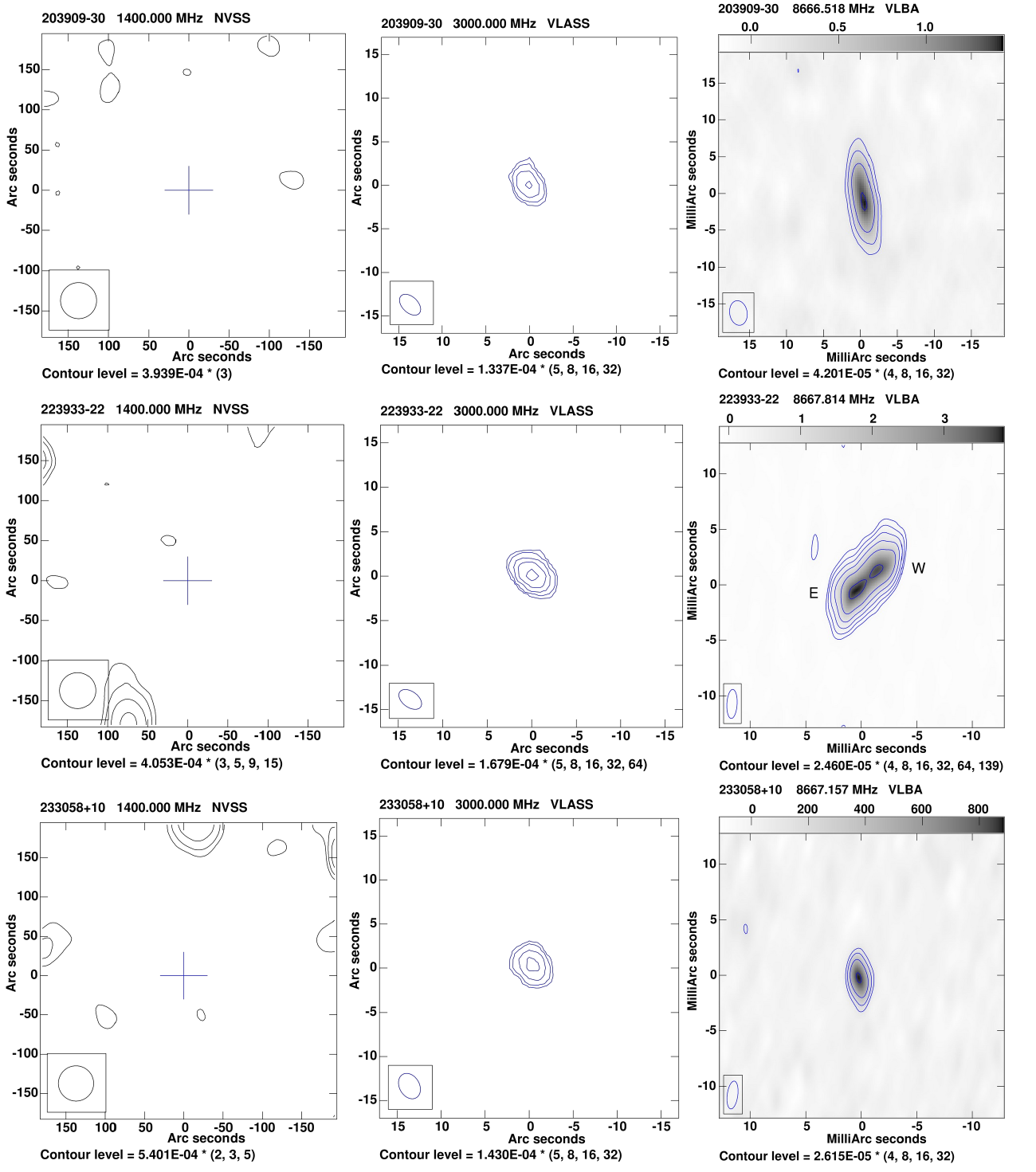}
    \caption{The following high-frequency images are presented from left to right: NVSS 1.4 GHz, VLASS 3 GHz (first epoch) and VLBA 8.7 GHz.}    
 \label{figure_appendix_images}
   \end{figure*}

Figure \ref{figure_appendix_images} is a continuation of the presentation of high-frequency images of our sources initiated in Figure \ref{figure_images}. They are, from left to right: NVSS 1.4 GHz, VLASS 3 GHz (first epoch) and VLBA 8.7 GHz. 

  \begin{figure*}[h!]
   \centering
   \includegraphics[scale=0.81]{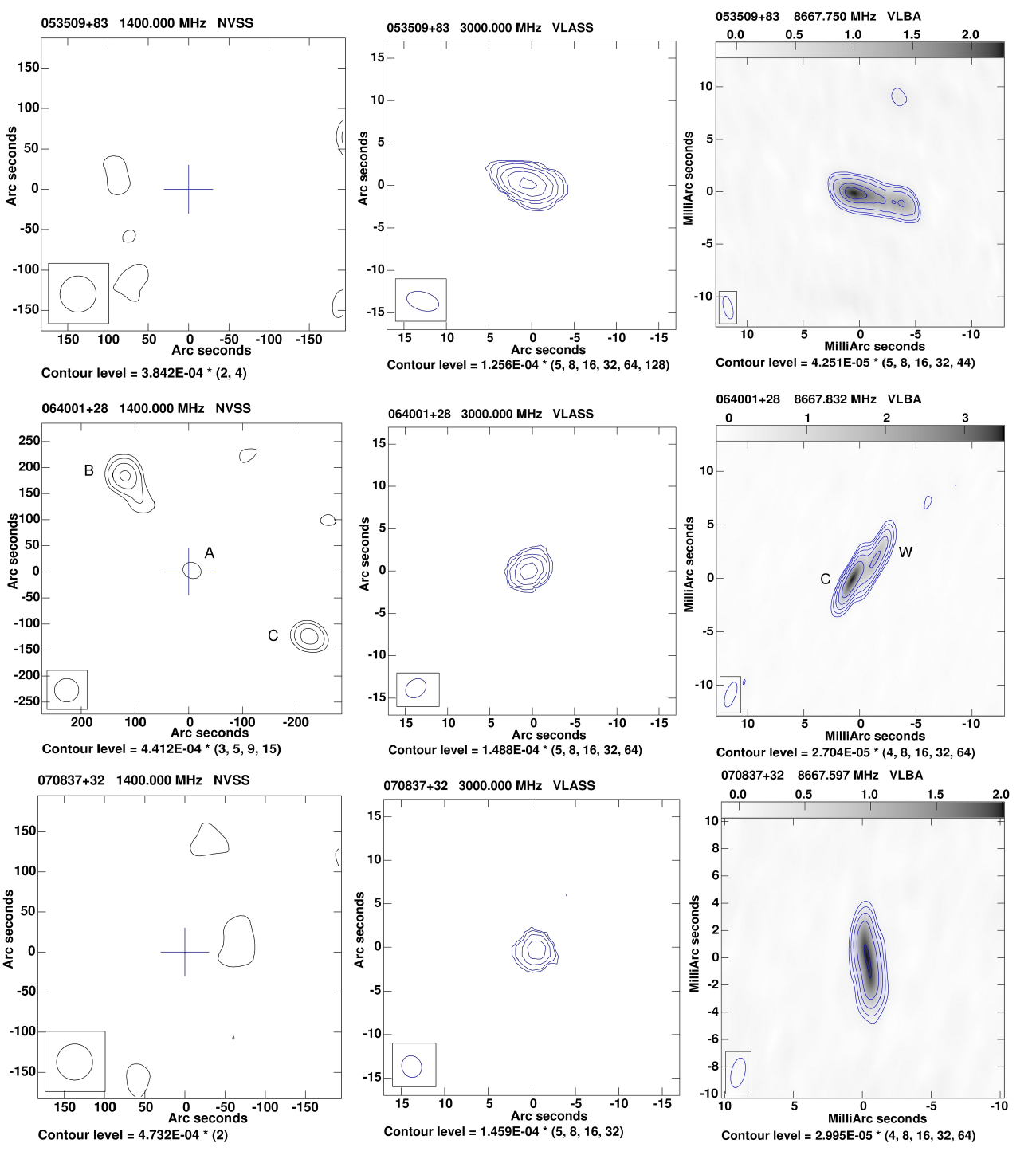}
   \end{figure*}

  \begin{figure*}[h!]
   \centering
   \includegraphics[scale=0.81]{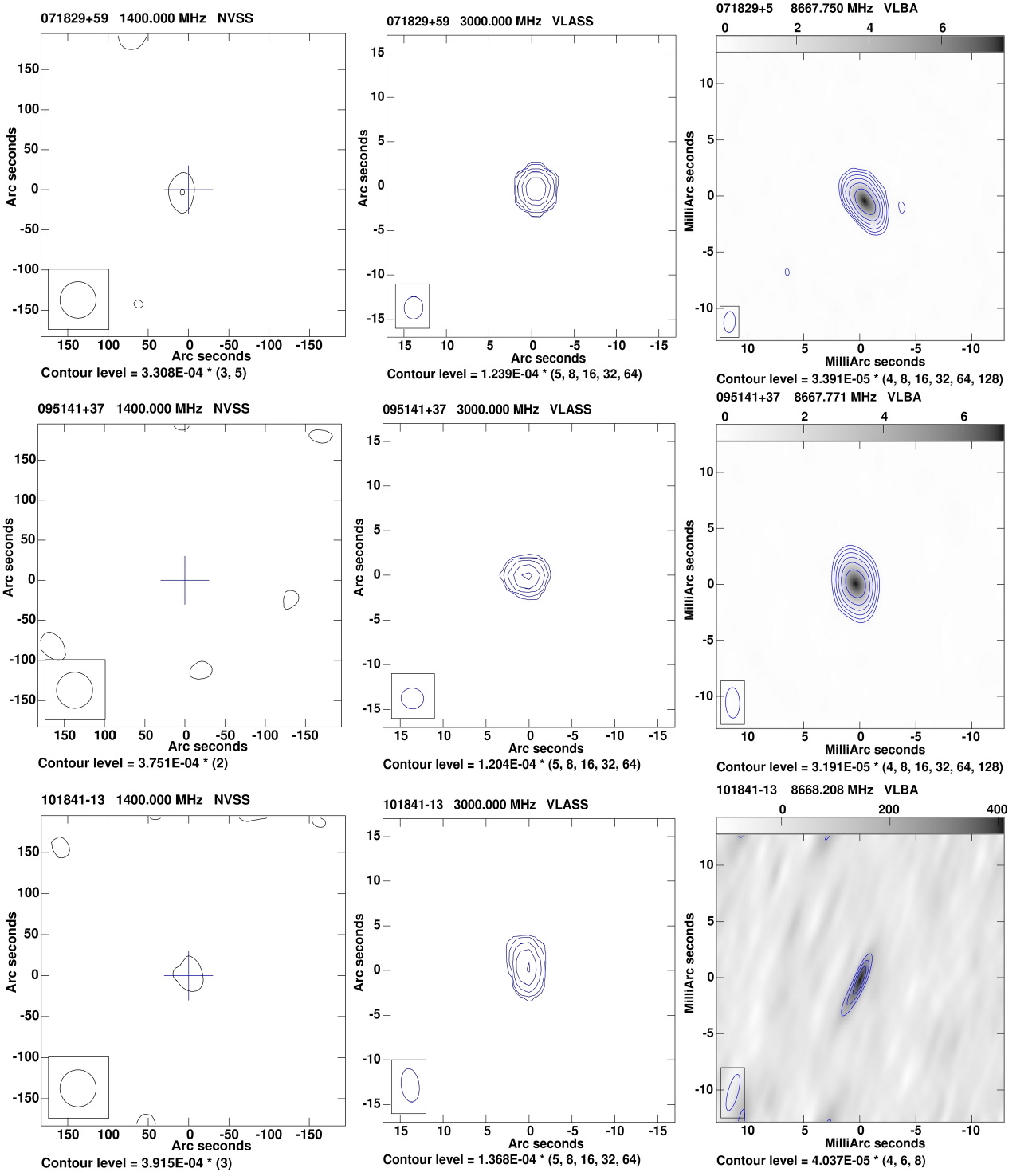}
   \end{figure*}

  \begin{figure*}[h!]
   \centering
   \includegraphics[scale=0.81]{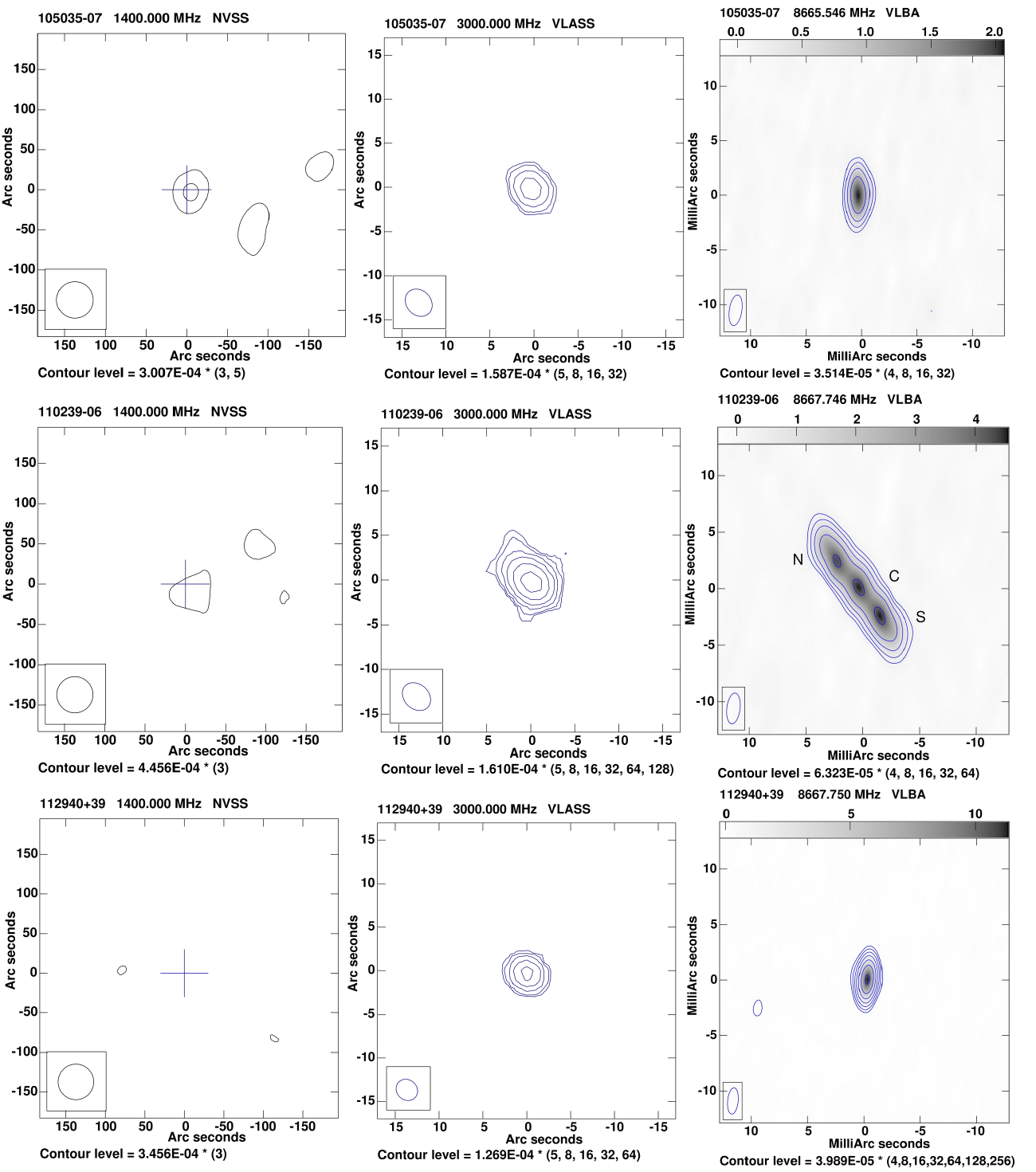}
   \end{figure*}

  \begin{figure*}[h!]
   \centering
   \includegraphics[scale=0.81]{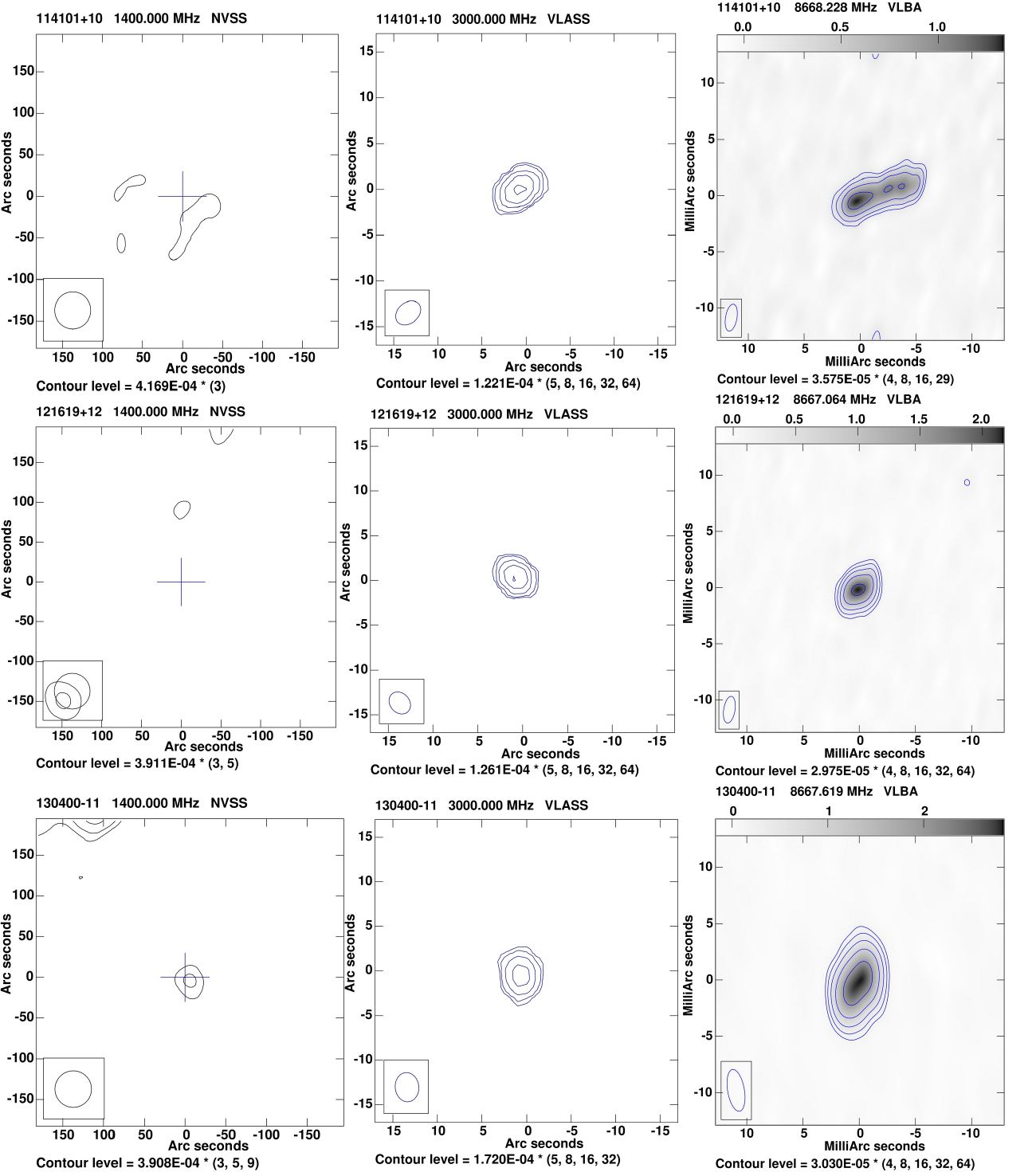}
   \end{figure*}

  \begin{figure*}[h!]
   \centering
   \includegraphics[scale=0.81]{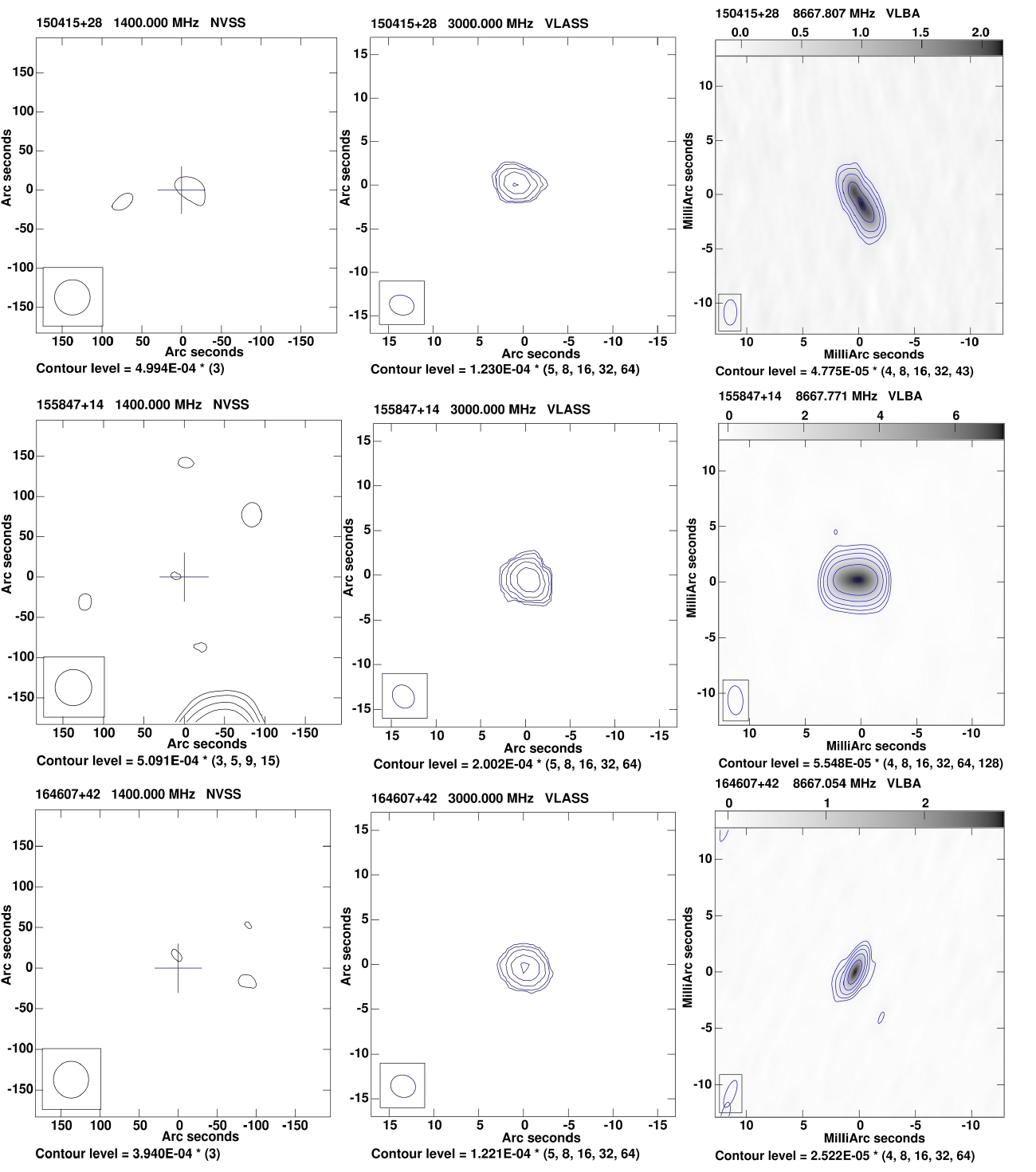}
   \end{figure*}

  \begin{figure*}[h!]
   \centering
   \includegraphics[scale=0.81]{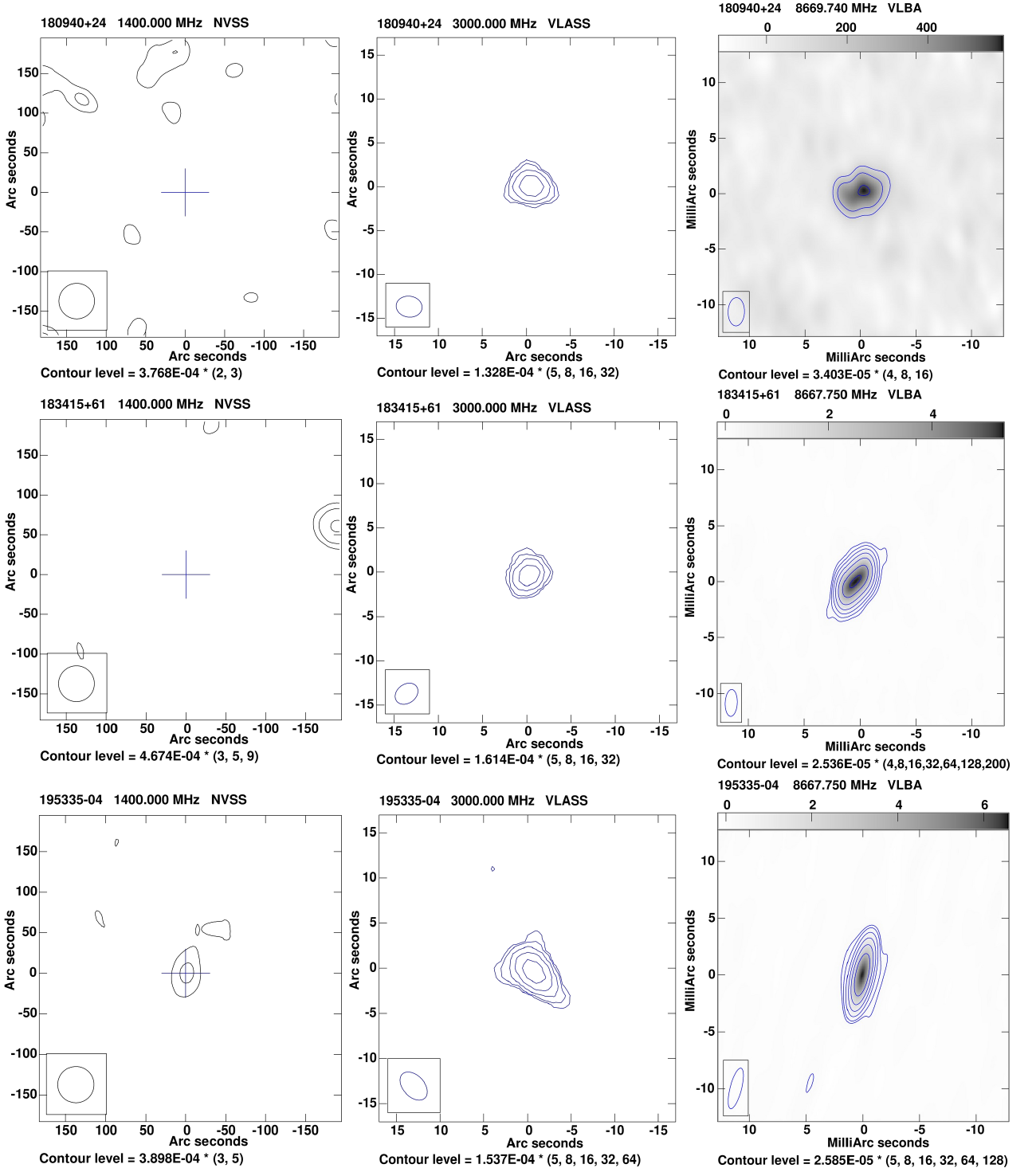}
   \end{figure*}

\end{document}